\documentclass[twocolumn,nofootinbib,prd,aps,tightenlines,floatfix]{revtex4}

\usepackage{slashed}
\usepackage{braket}
\usepackage{graphicx}
\usepackage{amssymb}

{\count255=\time\divide\count255 by 60 \xdef\hourmin{\number\count255}
  \multiply\count255 by-60\advance\count255 by\time
  \xdef\hourmin{\hourmin:\ifnum\count255<10 0\fi\the\count255}}

\def\vev#1{\left\langle #1 \right\rangle}
\def\li#1{ \text{Li}_2\left( #1 \right) }
\def\abs#1{ \left| #1 \right| }

\newcommand{\nn}{\nonumber \\ }
\newcommand{\e}{\epsilon}
\newcommand\eUV{\epsilon_{\text{UV}}}
\newcommand\eIR{\epsilon_{\text{IR}}}
\newcommand{\LL}{\mathsf{L}}

\newcommand{\Ls}{\mathsf{L}_{s}}
\newcommand{\Lt}{\mathsf{L}_{t}}
\newcommand{\Lu}{\mathsf{L}_{u}}
\newcommand{\Luts}{\mathsf{L}_{ut/s^2}}
\newcommand{\Lust}{\mathsf{L}_{us/t^2}}

\def\scamp{\mathsf{A}}

\def\lQ{\mathsf{L}_{Q}}
\def\lM{\mathsf{L}_{M}}

\def\lm{\mathsf{L}_{m}}
\def\Lst{\mathsf{L}_{s/t}}
\def\Ltu{\mathsf{L}_{t/u}}
\def\Lus{\mathsf{L}_{u/s}}

\def\darr#1{\raise1.5ex\hbox{$\leftrightarrow$}\mkern-16.5mu #1}
\def\rd{{{\rm d}}}
\def\aem{\alpha_{\text{em}}}
\def\sceth{ $\text{SCET}_{\text{EW}}$}
\def\scetl{ $\text{SCET}_{\gamma}$}

\def\pone{\textsf{\footnotesize CGKM1}}
\def\ptwo{\textsf{\footnotesize CGKM2}}

\begin{document}

\title{Electroweak Corrections using Effective Field Theory: Applications to the LHC}

\author{Jui-yu Chiu}
\affiliation{Department of Physics, University of California at San Diego,
  La Jolla, CA 92093}

\author{Randall Kelley}
\affiliation{Department of Physics, University of California at San Diego,
  La Jolla, CA 92093}

\author{Aneesh V.~Manohar}
\affiliation{Department of Physics, University of California at San Diego,
  La Jolla, CA 92093}
\date{\today\quad\hourmin}

\begin{abstract}
Electroweak Sudakov logarithms at high energy, of the form $(\alpha/\sin^2\theta_W)^n\log^m s/M_{Z,W}^2$, are summed using effective theory (EFT) methods. The exponentiation of Sudakov logarithms and factorization is discussed in the EFT formalism. Radiative corrections are computed to scattering processes  in the standard model involving an arbitrary number of external particles. The computations include non-zero particle masses such as the $t$-quark mass, electroweak mixing effects which lead to unequal $W$ and $Z$ masses and a massless photon, and Higgs corrections proportional to the top quark Yukawa coupling. The structure of the radiative corrections, and which terms are summed by the EFT renormalization group is discussed in detail. The omitted terms are smaller than 1\%. We give numerical results for the corrections to dijet production, dilepton production, $t \bar t$ production, and squark pair production. The purely electroweak corrections are significant --- about 15\% at 1~TeV, increasing to 30\% at 5~TeV, and they change both the scattering rate and angular distribution. The QCD corrections (which are well-known) are also computed with the EFT. They are much larger --- about a factor of four at 1~TeV, increasing to a factor of thirty at 5~TeV. Mass effects are also significant; the $q \bar q \to t \bar t$ rate is enchanced relative to the light-quark production rate by 40\%.

\end{abstract}

\maketitle
\tableofcontents

\section{Introduction}
\label{I}

Radiative corrections to high energy scattering processes have two powers of a large logarithm for each order in perturbation theory. These logarithms, referred to as Sudakov logarithms, lead to a breakdown of fixed order perturbation theory, and have to be summed to all orders. The Large Hadron Collider (LHC) has a center-of-mass energy of $\sqrt s =14$~TeV, and will be able to measure collisions with a partonic center-of-mass energy of several TeV, more than an order of magnitude larger than the masses of the electroweak gauge bosons.  Electroweak Sudakov corrections are not small at LHC energies, since $\alpha \log^2 s/M^2_{W,Z}/(4 \pi \sin^2 \theta_W) \sim 0.15$ at $\sqrt{s}=4$~TeV. In this paper, we will apply effective theory methods developed in two previous publications~\cite{cgkm1,cgkm2} to processes relevant for the LHC; in particular, we consider in detail dijet production, dilepton pair production, $t \bar t$ production, and squark pair production. In Refs.~\cite{cgkm1,cgkm2}, electroweak Sudakov corrections to the matrix element of an external current were found to be of order 10\%. Electroweak corrections to LHC cross-sections are about four times larger. Naively, one factor of two arises because scattering processes lead to four-particle operators, which have (approximately) twice the radiative correction of the two-particle current operator. The other factor of two arises in squaring the amplitude to obtain the cross-section. Thus purely electroweak corrections at the LHC are significant, and resummed contributions must be properly included to obtain a reliable prediction for the cross-section. There are, of course, QCD corrections which are even larger, and are also included.

There is an extensive literature on electroweak Sudakov effects~\cite{ccc,ciafaloni,fadin,kps,fkps,jkps,jkps4,beccaria,dp1,dp2,hori,beenakker,dmp,pozzorini,js,melles}. The computations use infrared evolution equations~\cite{fadin}, based on an analysis of the infrared structure of the perturbation theory amplitude and a factorization theorem for the Sudakov form factor~\cite{pqcd}. These summations have been  checked against one-loop~\cite{beccaria,dp1,dp2} and two-loop~\cite{hori,beenakker,dmp,pozzorini,js} computations.

The Sudakov logarithm $\log(s/M_{W,Z}^2)$ can be thought of as an infrared logarithm in the electroweak theory, since it diverges as $M_{W,Z}\to0$. By using an effective field theory (EFT), these infrared logarithms in the original theory can be converted to ultraviolet logarithms in the effective theory, and summed using standard renormalization group techniques. The effective theory needed is soft-collinear effective theory (SCET)~\cite{BFL,SCET1,SCET2,SCET3}, which has been used to study high energy processes in QCD~\cite{ira1}, and to perform Sudakov resummations arising from radiative gluon corrections.

This paper studies high energy electroweak corrections to processes relevant for the LHC, such as dijet production, dilepton pair production, $t \bar t$ production, and squark pair production, and expands on our previous works~\cite{cgkm1,cgkm2}, which will be referred to as \pone\ and \ptwo, respectively.  In \pone\ we showed how to compute $\log s/M^2_{W,Z}$ corrections to the Sudakov form factor for massless fermions using EFT methods. In \ptwo\ the results were generalized to massive fermions such as the top quark, including radiative corrections due to Higgs exchange.  The corrections were computed  without assuming that the Higgs and electroweak gauge bosons were degenerate in mass. The Higgs corrections when expanded to fixed order agree with previous results of Melles~\cite{melles}. The electroweak corrections to processes involving four external particles are computed in this paper. We will show that the results can be obtained by summing the Sudakov form-factor results of \ptwo\ over all pairs of external particles with appropriate group theoretic factors. We also show how the results can be generalized to processes involving an arbitrary number of external particles. 

There are different methods of counting the order of radiative corrections for the case of Sudakov corrections depending on whether one uses the amplitude or the logarithm of the amplitude. We discuss this issue in detail in Sec.~\ref{sec:exp}, where we also explain precisely which terms are summed in our computation. Roughly speaking, we use NLL running in QCD and LL running in the electroweak. The neglected terms are numerically less than 1\%.

The paper is organized as follows: the outline of the calculation and notation is given in Sec.~\ref{sec:outline}.  The general structure of Sudakov double-logarithms, exponentiation, and the log-counting rules we use are given in Sec.~\ref{sec:exp}. We also discuss the numerical convergence of the perturbation series. The SCET formalism we use for our calculation is described in Sec.~\ref{sec:wilson}, including the formalism for Wilson lines needed in multi-particle processes computed using an analytic regulator~\cite{bf,analytic}. The calculation of quark scattering and production is first calculated in a toy theory in section~\ref{sec:toy theory}. Results are also given for  massive quark production and squark production. The toy theory illustrates the theoretical tools needed for the standard model computation without the added complications of a chiral gauge theory with three gauge groups and particles in many different gauge representations. It also illustrates how one can compute the radiative corrections for theories with scalar particles, such as supersymmetric extensions of the standard model. Some observations on  the factorization of amplitudes are made in Sec.~\ref{sec:fact}. Radiative corrections in the standard model are given in Sec.~\ref{sec:SM}. There are a total of eighty different amplitudes that are needed, which are computed in this section. Detailed numerical results and plots are given in Sec.~\ref{sec:numerics}. Appendix~\ref{app:box} discusses the box graphs needed for the high scale matching computation, as well as the crossing matrix needed for the case of identical particles. The parameter integrals we require in Sec.~\ref{sec:SM} are tabulated in Appendix~\ref{app:integrals}. The top quark computation in \ptwo\ was incorrect, and the corrected result is given in Appendix~\ref{app:erratum}. The numerical values change by about 1\%.
  
\section{Outline of Calculation and Notation}
\label{sec:outline}

The Sudakov logarithms are summed by integrating the renormalization group equations in SCET. The formalism we use has been explained in detail in \ptwo. In this section, we outline the computation of four-particle processes; most of the results are well-known but will serve to define the notation we use in the rest of the paper. As in \ptwo, we first consider a toy gauge theory, a $SU(2)$ spontaneously broken gauge theory with coupling constant $\alpha$, where all gauge bosons have a common mass $M$.    This is the theory used in many previous computations~\cite{cgkm2, kps,fkps,jkps,jkps4,js}, and allows us to compare with previous results.  The results will then be generalized to the realistic case of the standard model. When extending the results of the toy theory to the standard model in Sec.~\ref{sec:SM}, Higgs exchange effects will be included as in \ptwo.

We consider two-to-two scattering at center-of-mass energies much larger than $M_Z$. We will generically use $Q \gg M_Z$ to denote the energetic scale, and work in the regime where $s,t,u$ are all of order $Q^2$, so that one has hard scattering kinematics. Our results apply to high energy scattering processes at fixed angles, such as jet production, but not to processes such as diffractive scattering. 

\begin{figure}
\begin{center}
\includegraphics[width=2.5cm]{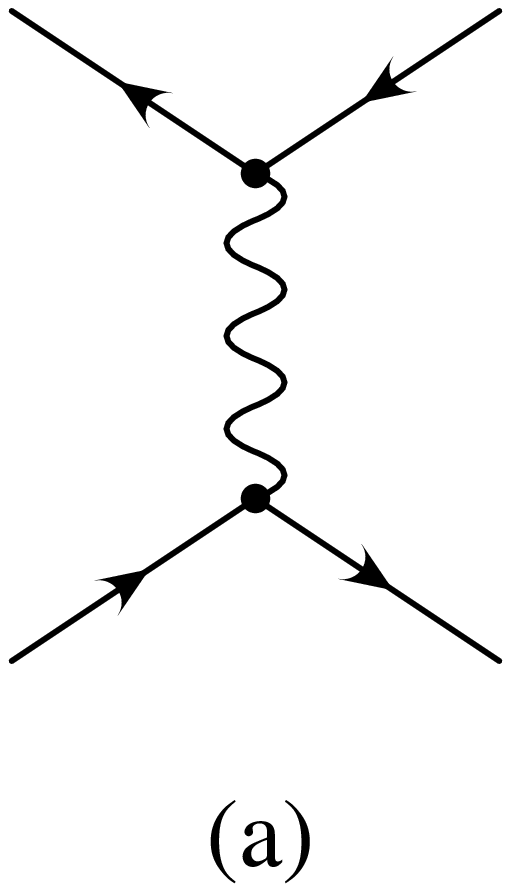} 
\hspace{2cm}
\includegraphics[width=2.5cm]{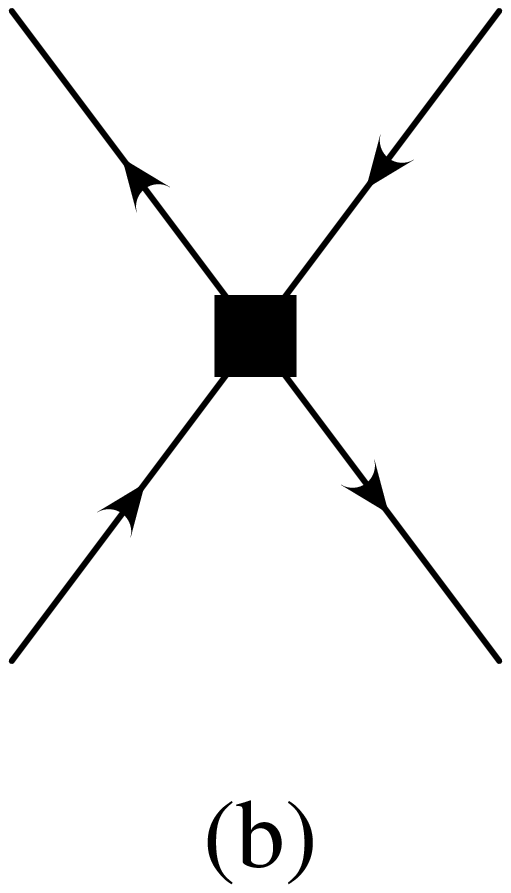}
\end{center}
\caption{\label{FgO2} Tree level matching onto the intermediate effective theory. The full theory amplitude in (a) turns into  scattering by a local operator in the effective theory, as shown in (b).}
\end{figure}

The scattering amplitude in the full theory arises from processes such as gauge boson exchange, as shown in Fig.~\ref{FgO2}a. 
The exchanged particle has virtuality of order $Q^2$. At the scale $\mu\sim Q$, we make a transition to SCET, which is an effective theory describing energetic particles with virtualities parameterically smaller than $Q^2$. The full theory process is treated in SCET as scattering by a set of local operators, as shown in the right-hand graph in Fig.~\ref{FgO2},
\begin{equation}
\label{EqO1}
iA_{\rm{full}} = \sum_i C_i \left(\mu\right) \bra{p_4 p_2} \mathcal{O}_i \left(\mu\right) \ket{p_1 p_3} ,
\end{equation}
where $\mathcal{O}_i$ are local SCET operators, and $C_i \left(\mu\right)$ are 
matching coefficients chosen so that the right-hand side reproduces the full theory amplitude up to power corrections of order $M^2/Q^2$. Power corrections can be systematically included by keeping higher dimension operators suppressed by powers of $Q^2$.  In our computation, we work to leading order in $M^2/Q^2$. The full and EFT have the same infrared  physics but different ultraviolet behavior, and so we must introduce a set of matching coefficients, $C_i(\mu)$ which correct for the different short distance properties of the two theories. The matching coefficients $C_i(\mu)$ are computed by comparing on-shell matrix elements in the full and effective theories at a scale $\mu \sim Q$. At this scale, infrared effects such as gauge boson and particle masses can be neglected, and so $C_i(\mu)$ can be computed using the unbroken gauge theory with massless particles.

The coefficients $C_i(\mu)$ are evolved from $\mu \sim Q$ down to the scale $\mu \sim M$ using the SCET anomalous dimensions. The evolution equation for the matching coefficients involves the (matrix) anomalous dimension, $\gamma_{ij}$,  and is 
\begin{equation}
\label{EqO7}
\mu \frac{d}{d\mu} C_i(\mu) = \gamma_{ij}(\mu) C_j(\mu).
\end{equation}
The anomalous dimension depends on the ultraviolet behavior of SCET, and is independent of particle masses. Like the matching at $Q$, it can be computed using the unbroken theory with massless particles. In SCET, the anomalous dimension matrix can depend on $\log Q^2/\mu^2$, so integrating Eq.~(\ref{EqO7}) sums the Sudakov double logarithms.
 
Once the coefficients $C_i(\mu)$ have been evolved down to a low scale of order $M$, we transition to a new effective theory, which is also SCET, but with the massive gauge bosons integrated out. In our toy example, this new theory has no gauge interactions, since all the gauge bosons are massive. In the standard model, the transition is from a theory with $SU(3)\times SU(2) \times U(1)$ gauge bosons which we call \sceth\ to a  new theory where the only gauge interactions are due to gluons and photons which we call \scetl. Operators $\mathcal{O}_i$ in \sceth\ are matched onto a set of operators $\widehat{ \mathcal{O}}_i$ in \scetl. A single $SU(3) \times SU(2) \times U(1)$ invariant operator $\mathcal{O}_i$ can break up into several operators $\widehat{ \mathcal{O}}_i$ which are $SU(3) \times U(1)_{\text{em}}$ invariant, but need not have full electroweak gauge invariance. The $\text{\sceth} \to \text{\scetl}$ matching requires treating massive gauge bosons in SCET, using the formalism developed in \pone, \ptwo.

The operators in \scetl\ are evolved down to a scale set by the experimental observables of interest, and then used to compute the desired observables. For example, if one is interested in jet production, then the operators would be scaled down to $\mu$ of order the typical jet invariant mass. The operators can then be used to compute jet observables. This paper focuses on electroweak corrections, and we will not discuss this final step of the computation, since it is performed as discussed in earlier work~\cite{2jet}. In our numerical results, we will choose this low energy scale to be 30~GeV. The electroweak corrections are not very sensitive to this scale, since the only effects below $M_Z$ are electromagnetic. The QCD corrections are scale dependent; the $\mu$ dependence in the SCET running cancels the $\mu$ dependence of the jet matrix elements to the order of the computation. We have not analyzed this in detail since we concentrate on electroweak effects in this paper. In Sec.~\ref{sec:numerics}, only Fig.~\ref{fig:num1} and Fig.~\ref{fig:angnum1} have significant $\mu$ dependence.

The bulk of the paper discusses the computation of the anomalous dimensions in \sceth\ and \scetl, and the matching between \sceth\ and \scetl, which require SCET operators involving four-particles. We introduce the notation necessary to deal with an arbitrary number of particles. Most of the notation is standard to SCET, and we only discuss those features which are necessary for the extension to $r$-particles.

The $r$ energetic particles are described by SCET fields $\xi_{n_i,p_i}$ labelled by momentum $p_i$ and light-cone direction $n_i$, $i=1,\ldots,r$. There are $r$ light-cone directions $n_i$, $n_i^2=0$, where $n_i^\mu = (1, \mathbf{n}_i)$, with $\mathbf{n}_i$ a unit vector near the direction of motion of particle $i$. We will also define $r$ light-cone directions $\bar n_i$ by reversing the sign of space components of $n_i$, i.e.\ by applying parity to $n_i$, $\bar n_i^\mu = (1, -\mathbf{n}_i)$. Note that $\bar n_i \cdot n_i =2$. The momentum of any particle can be written as
\begin{eqnarray}
p_i^\mu &=& \frac 1 2 n_i^\mu \left(\bar n_i \cdot p_i\right) + \frac 1 2 \bar n_i^\mu \left(n_i \cdot p_i\right) + p_{i,\perp}^\mu.
\label{3}
\end{eqnarray}
If $\mathbf{n}_i$ is chosen to be exactly along the direction of $p_i$, then $p_{i,\perp}^\mu=0$. The particles are energetic, with $\bar n_i \cdot p_i \sim Q$. In the case of only two energetic particles, one can work  in the Breit frame where the particles are back-to-back, with $\bar n_1=n_2$ and $\bar n_2=n_1$, so that one only deals with two null vectors $n_1$ and $\bar n_1$, conventionally called $n$ and $\bar n$.

\begin{figure}
\begin{center}
\includegraphics[width=2.5cm]{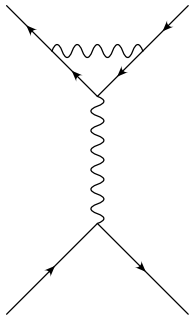}
\end{center}
\caption{\label{fig:radcor} Vertex correction to the scattering amplitude in the full theory.}
\end{figure}

\begin{figure}
\begin{center}
\includegraphics[width=3cm]{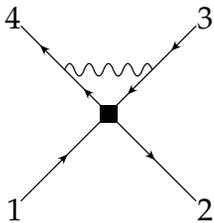}
\end{center}
\caption{\label{fig:vertex} Vertex correction in SCET.}
\end{figure}

Consider a radiative correction graph to the tree-level process Fig.~\ref{FgO2}, such as the vertex correction shown in Fig.~\ref{fig:radcor} in the full theory.
The gauge boson exhanged between the two fermion lines still has virtuality of order $Q^2$, and so the diagram behaves like the graph in Fig.~\ref{fig:vertex}, with the highly virtual gauge boson shrunk to a point.
As is well-know, there are several different momentum regions which contribute to the loop integral in Fig.~\ref{fig:radcor}. If the components of the gauge boson loop momentum are of order $Q$, then the gauge boson has virtuality of order $Q^2$. This contribution is not present in SCET, and is included in the one-loop matching coefficients at the scale $Q$. The other regions, which are included in SCET, are when the gauge boson is collinear to particle~1 ($n_1$-collinear gauge boson), to particle~2 ($n_2$-collinear gauge bosons), or is ultrasoft. The SCET theory thus contains $n_i$-collinear gauge  bosons for each particle direction, $i=1,\ldots,r$,  with momenta scaling like $p_i$, denoted by $A_{n_i,p_i}$ with labels, as well as ultrasoft gauge bosons  denoted by $A$, with no labels, which couple to all the particles, analogous to the soft and ultrasoft fields introduced in NRQCD~\cite{lmr}. We work in the regime where the kinematic variables such as $s,t$ are of order $Q^2$, and the invariant masses of the final states are much smaller than $Q^2$. The SCET power counting parameter is $\lambda=M/Q$. The formalism is valid for observables that can be constructed out of variables in the effective theory, for which the reduction to effective theory vertices such as in Fig.~(\ref{fig:vertex}) is valid. In particular, it is valid for jet observables and top decay observables at the LHC.

Notation: We use the abbreviations
\begin{eqnarray}
\lM = \log \frac{M^2}{\mu^2},\ \lm = \log \frac{m^2}{\mu^2},\ \lQ = \log \frac{Q^2}{\mu^2} \nn
\Ls = \log \frac{-s}{\mu^2},\ \Lt = \log \frac{-t}{\mu^2},\ \Lu = \log \frac{-u}{\mu^2} \nn
\Lst = \log \frac{s}{t} = \log(-s)-\log(-t),\nn
\Ltu = \log \frac{t}{u}= \log(-t)-\log(-u),\nn
\Luts=\log \frac{ut}{s^2}= \log(-u)+\log(-t)-2\log(-s).
\label{abbrev}
\end{eqnarray}
For scattering kinematics, $s>0$, $t<0$, and $u<0$. All logarithms arise in the form $\log(-x-i0^+)$ for $x=s,t,u$, so that $\log(-s-i0^+)=\log s - i \pi$. Similarly,
$\Lst=\log(-s)-\log(-t)=\log(-s/t)-i \pi$, and $\LL_{t/s}=\log(-t)-\log(-s)=\log(-t/s)+i \pi$. This procedure can be used to find the branch cut of logarithms with negative argument which occur in the subsequent formul\ae.

\section{Exponentiation and Log-Counting}
\label{sec:exp}

The exponentiation properties of Sudakov logarithms, and the relation between the renormalization group results and those obtained by exponentiating fixed order computations was discussed in \ptwo. This section summarizes the results we need for our standard model calculation.

The scattering amplitude $\scamp$ has an expansion of the form\footnote{For multi-particle scattering, $A$ is actually a matrix of amplitudes, and matrix ordering is important. We discuss the simpler case of the Sudakov form factor, where $A$ is a number. This is sufficient to study the exponentiating and log power-counting we need. The matrix case is discussed in Sec.~\ref{sec:fact}.}
\begin{eqnarray}
\scamp &=& \left( \begin{array}{cccccccccc} 1 \\[5pt]
\alpha \LL^2 & \alpha \LL & \alpha  \\[5pt]
\alpha^2 \LL^4 & \alpha^2 \LL^3 & \alpha^2 \LL^2 & \alpha^2 \LL & \alpha^2  \\[5pt]
\alpha^3 \LL^6 & \multicolumn{4}{c}{\ldots} \\[5pt]
\vdots
\end{array}\right)
\label{x1}
\end{eqnarray}
where $\alpha$ represents a gauge coupling constant ($\alpha_1$, $\alpha_2$ or $\alpha_s$), $M$ is an electroweak gauge boson mass ($M_W$ or $M_Z$) and $Q\gg M$ is of order the center of mass energy of the scattering process, and $\LL=\log Q/M$ is the large logarithm. Each entry in Eq.~(\ref{x1}) has a numerical coefficient, and the total amplitude is given by summing all the terms. The first row is the tree-level result, the second row is the one-loop contribution, etc.\ The $\alpha^n$ contribtion has logarithms up to power $\LL^{2n}$.

The logarithm of the scattering amplitude has an expansion of the form~\cite{collins,mueller,sen}
\begin{eqnarray}
\log \scamp &=& \left( \begin{array}{cccccccccc} 
\alpha \LL^2 & \alpha \LL & \alpha  \\[5pt]
\alpha^2 \LL^3 & \alpha^2 \LL^2 &  \alpha^2 \LL & \alpha^2  \\[5pt]
\alpha^3 \LL^4 & \alpha^3 \LL^3 & \alpha^3 \LL^2 &  \alpha^3 \LL & \alpha^3  \\[5pt]
\alpha^4 \LL^5 & \multicolumn{4}{c}{\ldots} \\[5pt]
\vdots
\end{array}\right)
\label{x2}
\end{eqnarray}
where the $\alpha^n$ contribtion now has logarithms only up to power $\LL^{n+1}$, and the amplitude has been normalized so that its tree-level  value is unity. The $n^{\text{th}}$ row can be computing using perturbation theory at $n$ loops. There are far fewer coefficients in Eq.~(\ref{x2}) than Eq.~(\ref{x1}), so the form Eq.~(\ref{x2}) for $\log \scamp$ is highly non-trivial. Equation~(\ref{x2}) is referred to as the exponentiated form of the amplitude, since $\scamp$ is given by exponentiating the r.h.s. The first column gives the leading-log (LL) series, the second gives the next-to-leading-log (NLL) series, etc.\footnote{The LL, NLL, etc.\ counting used here is different from that used in fixed order calculations. The relation between the two is explained in \ptwo, and after Eq.~(\ref{10x}) in this section.}  

The EFT computation naturally gives the scattering amplitude in exponentiated form. In general, there are several possible gauge invariants that contribute to the scattering amplitude, so that $\scamp$ is a matrix. The EFT computation gives the proper matrix ordering to be used for the exponentiated form of $\scamp$. The difference between different matrix orderings can be computed using the Baker-Cambell-Hausdorff theorem. If $X$ and $Y$ are matrices, then
\begin{eqnarray}
e^Z &=& e^X e^Y \nn
 Z &=& X+Y+\left[X,Y\right]+\frac{1}{12}\left[X,\left[X,Y\right]\right]+
\frac1{12}\left[Y,\left[Y,X\right]\right]+\ldots\nn
\label{bch}
\end{eqnarray}
where all the higher order terms are multiple commutators of $X$ and $Y$. If $X$ and $Y$ represent contribution to $\log \scamp$ of the form Eq.~(\ref{x2}), then $X$ and $Y$ are of order $\alpha^{n_{X,Y}} \LL^{m_{X,Y}}$ where $m_{X,Y} \le n_{X,Y}+1$. Thus one could in principle generate terms in $\log \scamp$ of the form $\alpha^n \LL^m$ with $m > n+1$ by reordering a matrix product using Eq.~(\ref{bch}). This does not occur, because, as discussed in Sec.~\ref{sec:fact}, the leading Sudakov series $\alpha^n \LL^{n+1}$ is proportional to the unit matrix, and so drops out of the commutators in Eq.~(\ref{bch}), so that the form Eq.~(\ref{x2}) is preserved independent of the matrix ordering.

When $\LL$ is large, fixed order perturbation theory breaks down,  and one needs to sum the logarithmically enhanced higher order corrections. There are two interesting regimes relevant for the standard model, in which resummation is necessary. The first is the leading-log (LL) regime in which $\alpha \LL$ is of order unity.\footnote{Including loop factors of $4\pi$.} This is the regime in TeV scale scattering for strong interaction corrections, where $\alpha \to \alpha_s$. Using $\LL \sim 1/\alpha$, the various terms in Eqs.~(\ref{x1}) are of order
\begin{eqnarray}
\scamp &=& \left( \begin{array}{cccccccccc} 1 \\[5pt]
\frac{1}{\alpha} & 1 & \alpha  \\[5pt]
\frac{1}{\alpha^2} & \frac{1}{\alpha} & 1 & \alpha & \alpha^2  \\[5pt]
\frac{1}{\alpha^3} & \multicolumn{4}{c}{\ldots} \\
\vdots
\end{array}\right).
\label{x3}
\end{eqnarray}
Clearly the fixed order perturbation expansion breaks down, and higher order terms grow with inverse powers of $\alpha$. To obtain a reliable value for the amplitude requires summing all terms along and below the diagonal, i.e.\ all terms of order unity or larger. The first super-diagonal gives the order $\alpha$ correction, the second super-diagonal gives the order $\alpha^2$ correction, etc.

The terms in the exponentiated form Eq.~(\ref{x2}) are of order
\begin{eqnarray}
\log \scamp &=& \left( \begin{array}{cccccccccc} 
\frac{1}{\alpha} & 1 & \alpha  \\[5pt]
\frac{1}{\alpha} & 1 &  \alpha & \alpha^2  \\[5pt]
\frac{1}{\alpha} & 1 & \alpha &  \alpha^2 & \alpha^3  \\[5pt]
\frac{1}{\alpha} & \multicolumn{4}{c}{\ldots} \\
\vdots
\end{array}\right).
\label{x4}
\end{eqnarray}
The expression for $\log \scamp$ has already achieved a partial summation of higher order terms. The largest terms are order $1/\alpha$, and there are no terms with higher powers of $1/\alpha$. To obtain $\log \scamp$ requires summing the first column (the LL series) and the second column (the NLL series). The NNLL series gives order $\alpha$ corrections, the N${}^3$LL series gives the order $\alpha^2$ corrections, and so on. While the NLL series is suppressed by one power of $\alpha$ relative to the LL series, it cannot be considered as a correction to the scattering amplitude $\scamp$, since we have to exponentiate $\log \scamp$. If we write $f_n$ for the N${}^n$LL contribution to $\log \scamp$, then
\begin{eqnarray}
\log \scamp &=& \frac{1}{\alpha} f_0 + f_1 + \alpha f_2 + \ldots\nn
&=& \frac{1}{\alpha}\left[ f_0 + \alpha f_1 + \alpha^2 f_2 + \ldots \right]
\end{eqnarray}
so that $f_1$ and $f_2$ are corrections to $\log \scamp$. However,
\begin{eqnarray}
 \scamp &=& \exp\left[\frac{1}{\alpha} f_0 + f_1 + \alpha f_2 + \ldots\right]\nn
 &=&e^{\frac{1}{\alpha} f_0} \times e^{f_1} \times e^{\alpha f_2} \times \ldots
 \label{10x}
\end{eqnarray}
and $\exp f_1$ can make a large change in $\scamp$. Only $f_2$ and higher can be considered as corrections to $\scamp$.

\newpage

The counting discussed above is consistent with that used in renormalization group improved perturbation theory computations. In much of the literature, it is more common to use a different counting, which we denote by the subscript $\text{FO}$. The $\text{LL}_{\text{FO}}$ terms are those in $\scamp$ (\emph{not} $\log \scamp$) of the form $\alpha^n \LL^{2n}$, the $\text{NLL}_{\text{FO}}$ terms are those in $\scamp$ of the form $\alpha^n \LL^{2n-1}$, and in general, the $\text{N}^k\text{LL}_{\text{FO}}$ terms are those in $\scamp$ of the form $\alpha^n \LL^{2n-k}$. 
In terms of fixed-order counting, Eq.~(\ref{x2}) can be written as
\begin{widetext}
\begin{eqnarray}
\log \scamp &=& \left( \begin{array}{cccccccccc} 
\alpha \LL^2 \sim \text{LL}_{\text{FO}} & \alpha \LL \sim \text{NLL}_{\text{FO}} & \alpha \sim \text{N}^2\text{LL}_{\text{FO}} \\[5pt]
\alpha^2 \LL^3 \sim \text{NLL}_{\text{FO}} & \alpha^2 \LL^2 \sim \text{N}^2\text{LL}_{\text{FO}} &  \alpha^2 \LL \sim \text{N}^3\text{LL}_{\text{FO}} & \alpha^2  \sim \text{N}^4\text{LL}_{\text{FO}}\\[5pt]
\alpha^3 \LL^4 \sim \text{N}^2\text{LL}_{\text{FO}} & \alpha^3 \LL^3 \sim \text{N}^3\text{LL}_{\text{FO}}& \alpha^3 \LL^2 \sim \text{N}^4\text{LL}_{\text{FO}}&  \alpha^3 \LL \sim \text{N}^5\text{LL}_{\text{FO}} & \alpha^3 \sim \text{N}^6\text{LL}_{\text{FO}} \\[5pt]
\alpha^4 \LL^5 \sim \text{N}^3\text{LL}_{\text{FO}}& \multicolumn{4}{c}{\ldots} \\[5pt]
\vdots
\end{array}\right)
\label{Nx2}
\end{eqnarray}
\end{widetext}
and terms in $\scamp$ obtained by exponentiating are given by combining the powers of $\text{N}$.

Note that with this counting, terms of a given series grow at higher order in perturbation theory, e.g.\ the $\text{N}^3\text{LL}_{\text{FO}}$ terms are $\alpha^2 \LL$, $\alpha^3 \LL^3$, $\alpha^4 \LL^5$, \ldots, $\alpha^n \LL^{2n-3}$, which in the leading-log regime are of order $\alpha$, $1$, $1/\alpha$, \ldots, $1/\alpha^{n-3}$, and grow at higher orders. One can see this clearly from Eq.~(\ref{10x}) --- $f_{k+1}$ is of order $\alpha^k$, and is small for $k \ge 1$, as are all terms in the expansion of $\exp \alpha^k f_{k+1}$. However, the perturbation expansion for $\scamp$ contains the prefactor $\exp f_0/\alpha$, and the terms $(f_0/\alpha)^n$ in the expansion of this prefactor for $n > k$ mutiply the small terms in the expansion of $\exp \alpha^k f_{k+1}$ to produce terms which are larger than unity, with a series of large contributions of alternating sign (since $f_0$ is negative). The problem is that the tree-level value $\scamp=1$ is not close to the true result for $\scamp$; the leading contribution $\exp f_0/\alpha$ has an essential singularity at $\alpha=0$ in the perturbation expansion. The second term $\exp f_1$ also is not small. Only after these two contributions are factored out and properly exponentiated does one have a reliable perturbation expansion. Summing all terms up to order $\text{N}^k\text{LL}_{\text{FO}}$ does not give a reliable calculation, because $\text{N}^{k+1} \text{LL}_{\text{FO}}$ terms at order $\alpha^r$, $r \ge k+1$ are larger than unity. It is essential to properly exponentiate the $f_0$ and $f_1$ contributions to get a reliable expansion. Once this done, the higher order contributions are a small correction to the full amplitude $\scamp$. The amplitude $\scamp$ can be very different from the tree-level amplitude (a factor of 100 in our problem), and still be reliably computed in perturbation theory.

The second regime we consider is the leading-log-squared (LL${}^2$) regime in which $\alpha \LL^2$ is of order unity. This is the regime in TeV scale scattering for electroweak corrections, with $\alpha \to \alpha_{1,2}$.  Using $\LL \sim 1/\alpha^{1/2}$, the various terms in Eqs.~(\ref{x1}) are of order
\begin{eqnarray}
\scamp &=& \left( \begin{array}{cccccccccc} 1 \\[5pt]
1 & \alpha^{1/2} & \alpha  \\[5pt]
1 & \alpha^{1/2} & \alpha & \alpha^{3/2} & \alpha^2  \\[5pt]
1 & \multicolumn{4}{c}{\ldots} \\[5pt]
\multicolumn{5}{c}{\vdots}
\end{array}\right)
\label{x5}
\end{eqnarray}
and in Eq.~(\ref{x2}) are of order
\begin{eqnarray}
\log \scamp &=& \left( \begin{array}{cccccccccc} 
1 & \alpha^{1/2} & \alpha  \\[5pt]
\alpha^{1/2} & \alpha &  \alpha^{3/2} & \alpha^2  \\[5pt]
\alpha & \alpha^{3/2}  & \alpha^2 &  \alpha^{5/2} & \alpha^3  \\[5pt]
\alpha^2 &  \multicolumn{4}{c}{\ldots} \\[5pt]
\vdots
\end{array}\right).
\label{x6}
\end{eqnarray}
The computation of $\scamp$ requires summing the first column, the Sudakov double-logs of order $\alpha^n \LL^{2n}$. The remaining terms can be treated in a perturbative expansion. The second column gives the correction of order $\alpha^{1/2}$, the third column the order $\alpha$ correction, etc.\ The exponentiated form $\log \scamp$ can be computed to order unity from the  $\alpha \LL^2$ term. The first correction, of order $\alpha^{1/2}$, is from the $\alpha^2 \LL^3$ and $\alpha \LL$ terms, the order $\alpha$ correction is from the $\alpha^3 L^4$, $\alpha^2 L^2$, and $\alpha$ terms, etc.\ We will refer to these as the  LL${}^2$ (leading-log-squared), NLL${}^2$, NNLL${}^2$, etc.\ contributions to $\log \scamp$.

The scattering amplitude  in the EFT computation has the form~\cite{cgkm1,cgkm2}
\begin{eqnarray}
&& \scamp =  \exp\left[ D_0(\alpha(M))  +  D_1 (\alpha(M)) \log\frac{Q^2}{M^2}\right] \nn
&&\times  \exp\left\{- \int_{M}^{Q}\frac{{\rm d}\mu}{\mu}  \left[A(\alpha(\mu)) \log\frac{\mu^2}{Q^2}+ B (\alpha(\mu))
\right]\right\}\nn
&&\times \exp C(\alpha(Q))
\label{13a}
\end{eqnarray}
Here $\exp C(\alpha(Q))$ is the high scale matching coefficient at $Q^2$, $\gamma(\mu)=A(\alpha(\mu)) \log(\mu^2/Q^2)+B(\alpha(\mu))$ is the SCET anomalous dimension between $Q$ and $M$, $\exp D(\alpha(M))$, $D(\alpha(M))=D_0(\alpha(M)) +  D_1 (\alpha(M)) \log{Q^2}/{M^2}$ is the low scale matching coefficient at $M$, $\alpha$ the gauge coupling constant ($\alpha_1$, $\alpha_2$ or $\alpha_s$), $M$ is the electroweak gauge boson mass ($M_W$ or $M_Z$) and $Q\gg M$ is of order the center of mass energy of the scattering process. $A$ is called the cusp anomalous dimension, and is linear in $\log Q$ to all orders in perturbation theory~\cite{dis,Bauer:2003pi}. The low-scale matching $\exp D$ has a single-log term $D_1$ to all orders in perturbation theory~\cite{cgkm1,cgkm2}. The LL series is given by the one-loop cusp anomalous dimension, the NLL series by the two-loop cusp anomalous dimension, the one-loop value of $B$ and the one-loop value of $D_1$, the NNLL series by the three-loop cusp, two-loop $B$ and $D_1$, and one-loop $D_0$ and $C$, and the N${}^n$LL series by the $n+1$ loop cusp, the $n$-loop $B$ and $D_1$, and the $n-1$ loop $D_0$ and $C$. Eq.~(\ref{13a}) for the standard model, which we study in this paper, sums the QCD and electroweak corrections, including cross terms such as $\alpha_s \alpha_{1,2}$, $\alpha_s g_t^2$, or $\alpha_{1,2} g_t^2$ which depend on mixed products of the Yukawa, strong and electroweak coupling constants.

\subsection{Absence of some terms in the Sudakov expansion}

In Eq.~(\ref{x2}), we wrote the generic expansion for $\log \scamp$. In the standard model, one gets the form Eq.~(\ref{x2}) where $\alpha^n$ can be a product of the gauge or Yukawa couplings. It is interesting to note that not all possible terms are present. The leading Sudakov series in $\log \scamp$ of the form $\alpha^n \LL^{n+1}$ is given by integrating the one-loop cusp anomalous dimension with the leading order $\beta$-function. The one-loop cusp anomalous dimension $\Gamma(\mu)$ is trivially a sum over the different gauge groups, since there can be no mixed terms like $\alpha_s \alpha_{1,2}$ at one-loop, and because there is no Yukawa contribution to the cusp anomalous dimension (see \pone). The one-loop gauge $\beta$-function also does not mix different gauge couplings. Thus the leading Sudakov series is a sum of independent terms for each gauge group, with no mixed contributions, i.e.\ there are terms of the form $\alpha_s^n \LL^{2n}$, $\alpha_1^n \LL^{2n}$ and $\alpha_2^n \LL^{2n}$, but no terms of the form $\alpha_s^n \alpha_{1,2}^m \LL^{2n+2m}$ for $n,m\not=0$. 

The first contribution to the cusp anomalous dimension which involves couplings from two-different gauge groups, and so cannot be written as the sum of contributions over individual groups, arises at four-loop order.\footnote{i.e., $\Gamma(\alpha_s,\alpha_1,\alpha_2)=\Gamma_s(\alpha_s)+\Gamma_1(\alpha_1)+\Gamma_2(\alpha_2)$ up to three-loop order. We would like to thank Z.~Bern and L.~Dixon for helpful correspondence on this point.} The two-loop $\beta$-function also has contributions from two different gauge couplings. Thus at LL, the running strong coupling $\alpha_s$ only gets modified by terms of the form $\alpha_s (\alpha_s \LL)^n$, but at NLL, one can have terms of the form $\alpha_s(\alpha_{1,2} \alpha_s \LL)(\alpha_s \LL)^n (\alpha_{1,2} \LL)^m$. The $(\alpha_{1,2} \alpha_s \LL)$ factor comes from one insertion of the two-loop $\beta$-function in the renormalization group integration, and the other factors come from using the leading-order $\beta$-functions for the remaining integration of $\alpha_s$ and $\alpha_{1,2}$. 

Using the above, and noting that the matching conditions and non-cusp anomalous dimensions have all allowed terms, one finds that one can get all possible terms in Eq.~(\ref{x2}) for the $\text{N}^2\text{LL}$ and higher series (third column and beyond). For the LL series (first column), all terms have a single gauge coupling. For the NLL (second column),  all terms can occur with the exception of the $\alpha^2 \LL^2$ contribution, which can only have a single gauge coupling, so that terms such as $\alpha_s \alpha_1 \LL^2$ are absent.

\subsection{Terms included in the computation}\label{sec:terms}

In the standard model, the radiative corrections involve the strong coupling $\alpha_s$ and the electroweak couplings $\alpha_{1,2}$. For log-counting purposes, we assume that the strong coupling is in the leading-log regime, and the electroweak couplings are in the leading-log-squared regime. Let $a$ be the log-counting parameter. Then $\LL \sim 1/a$, $\alpha_{1,2} \sim a^2$, $\alpha_s \sim a$. The top-quark Yukawa coupling is also treated as the same order as the electroweak couplings, $g_t^2 \sim \alpha_{1,2} \sim a^2$. The terms in $\log \scamp$ are given in Eq.~(\ref{x2}), but now each $\alpha$ can be either a strong coupling $\alpha_s$ of order $a$ or an electroweak coupling $\alpha_{1,2}$ of order $a^2$. The order of terms with all couplings equal to $\alpha_s$ are given by Eq.~(\ref{x4}) with $\alpha \to a$, those with one coupling $\alpha_{1,2}$ and the rest $\alpha_s$ are given by Eq.~(\ref{x4}) with $a \times (\alpha \to  a)$, etc.\ The leading terms of order $1/a$ in $\log \scamp$ are given by summing the $\alpha_s^n \LL^{n+1}$ terms, i.e.\ the leading-log QCD series. The order $1$ terms are given by summing the $\alpha_s^n \LL^n$ and $\alpha_{1,2}\alpha_s^{n-1} \LL^{n+1}$ terms, i.e.\ the NLL QCD series and the LL series with one power of the electroweak coupling. The order $a$ corrections are given by summing the $\alpha_s^n \LL^{n-1}$, $\alpha_{1,2}\alpha_s^{n-1} \LL^n$ and $\alpha_{1,2}^2 \alpha_s^{n-2} \LL^{n+1}$ terms, etc.\ In the exponentiated form Eq.~(\ref{x2}), one only needs to include electroweak corrections at low orders, so that summing terms to order unity only require one-loop electroweak computations, to order $a$ only requires two-loop electroweak corrections, etc.\ In contrast, the unexponentiated form Eq.~(\ref{x1}) of fixed-order computations requires electroweak corrections of arbitrarily high order to sum all terms of order unity or larger.

In the numerical results of Sec.~\ref{sec:numerics}, we include the one-loop QCD, electroweak and Higgs corrections, as well as the two-loop QCD anomalous dimension~\cite{aybat} and two-loop running of the gauge coupling constants. This includes the entire one-loop correction to the scattering amplitude, as well as all higher order corrections which are formally of order $1/a$ or $a^0$. The terms we neglect are order $a$ or higher in the log-counting, and at least second order in the gauge couplings constants $\alpha_{s,1,2}$.  The error due to the neglected terms is numerically less than 1\% in the rate.

In terms of the commonly used fixed order counting, we have included all $\text{LL}_{\text{FO}}$ and $\text{NLL}_{\text{FO}}$ terms for both the QCD and electroweak corrections. In addition we have included all $\text{NNLL}_{\text{FO}}$ of the form $\alpha_s^n \LL^{2n-2}$ and $\alpha_s^{n-1}\alpha_{1,2}\LL^{2n-2}$. Using the counting that $\alpha_s \sim a$ and $\alpha_{1,2} \sim a^2$, and counting $a^n \LL^{2n-k}$ as $\text{N}^k\text{LL}_{\text{FO}}$, we have summed all terms of order 
$\text{N}^3\text{LL}_{\text{FO}}$. In terms of the exponentiated form Eq.~(\ref{x2}), which is the form given by SCET and used for the numerics, we have included
\begin{eqnarray}
\log \scamp &=& \left( \begin{array}{cccccccccc} 
\surd & \surd & \surd  \\[5pt]
\surd & \text{not}\ \alpha_{1,2}^2\LL^2 &  \text{only}\ \alpha_s^2 \LL & \times  \\[5pt]
\surd & \text{not}\ \alpha_{1,2}^3\LL^3 & \times &  \times  & \times  \\[5pt]
\surd & \text{not}\ \alpha_{1,2}^4\LL^4 & \multicolumn{3}{c}{\ldots} \\[5pt]
\vdots & \vdots &
\end{array}\right)
\label{incx2}
\end{eqnarray}
where $\surd$ means all terms have been included, $\times$ means no terms have been included. The largest terms omitted are $\alpha_{1,2}^2\LL^2$, $\alpha_s^3 \LL^2$ and $\alpha_s \alpha_{1,2}\LL$, and are estimated to be $( \alpha/(\pi \sin^2 \theta_W)^2\LL^2 \sim 0.006$, $(\alpha_s/\pi)^3 \LL^2 \sim 0.003$ and $ \alpha_s \alpha/(\pi^2 \sin^2 \theta_W)\LL \sim 0.003$ using $\LL \sim \log (4 \text{TeV})^2/M_Z^2 \sim 7$. This gives a sub-1\% error. The $\alpha_{1,2}^2\LL^2$ term arises from the two-loop electroweak cusp anomalous dimension, and the $\alpha_s^3 \LL^2$ term from the three-loop QCD cusp anomalous dimension. These are known, and could be easily included in the computation. We have checked that these change the rates by less than 1\%.

\section{SCET Formalism and Wilson Lines}
\label{sec:wilson}

In SCET, $n_1$ collinear gauge bosons can interact with particle~1, or with the other particles in the process. The coupling of $n_1$-collinear gauge bosons to particle~1 is included explicitly in the SCET Lagrangian. The particle-gauge interactions are identical to those in the full-theory, and there is no simplification on making the transition to SCET. However, if an $n_1$-collinear gauge boson interacts with a particle other than 1 (pick particle~2 for definiteness), then particle~2 becomes off-shell by an amount of order $Q$, and the intermediate particle~2 propagators can integrated out, giving a Wilson line interaction in SCET. The form of these operators was derived in Ref.~\cite{SCET2}. We will use the definitions
\begin{eqnarray}
W^{(n_2)}_{n_1} &=& \left[\exp\left(-g \frac{1}{\overline{\mathcal{P}}} n_2 \cdot A_{n_1,q}^A T^A \right)\right]
\label{wilson}
\end{eqnarray}
which is the expression given in Ref.~\cite{SCET2} with the replacement $n \to n_1$, $\bar n \to n_2$. The gauge generators $T^A$ are in the representation $\frak{R}_2$ of particle~2. The subscript $n_1$ is a reminder that the Wilson line contains $n_1$-collinear gauge fields, and the superscript ${(n_2)}$ is a reminder that the integration path is directed along $n_2$, and that the gauge generators are in the representation of particle~2. 

$W^{(n_2)}_{n_1}$ is a $d_2 \times d_2$ matrix where $d_2$ is the dimension of  $\frak{R}_2$, and transforms under $n_1$-collinear gauge transformations as
 \begin{equation}
\label{EqA3j}
\left[W^{(n_2)}_{n_1}\right]_{ab} \to  U_{ac}^{(2)} \left[W^{(n_2)}_{n_1}\right]_{cb}
\end{equation}
where $U^{(2)}$ is the gauge transformation matrix in the $\frak{R}_2$ representation. One can similarly define $W^{(n_j)}_{n_i}$ for any pair $ij$ of particles, with $i\not=j$. It is convenient to treat all gauge indices as incoming, i.e.\ an outgoing fermion line in the gauge representation $\frak{R}$ will be treated as an incoming fermion in the representation $\bar\frak{R}$.

A generic gauge invariant local operator in the full theory can be written as the gauge invariant product of fields,
\begin{eqnarray}
\mathcal{O} &=& \sum_{\left\{ a_i \right\}} c\left(\left\{ a_i \right\}\right)\prod_i \chi_{i,a_i}(0)
\label{6}
\end{eqnarray}
where $\chi_{i,a_i}$ is $\psi_{i,a_i}$ for incoming particles, $\chi_{i,a_i}=\psi^\dagger_{i,a_i}$ for outgoing particles, and $c$ is a Clebsch-Gordan coefficient. $\chi_i$ transforms as $\frak{R}_i$ for incoming particles, and as $\bar\frak{R}_i$ for outgoing particles. The indices $a_i$ are gauge indices, and $c\left(\left\{ a_i \right\}\right)\equiv c(a_1,\ldots,a_r)$ is the Clebsch-Gordan coefficient for combining the product of fields into a gauge singlet. For $n_1$-collinear gauge couplings, the field $\chi_{1}$ in Eq.~(\ref{6}) can be replaced by the SCET field $\xi_{n_1,p_1}$, and the other fields are replaced by Wilson lines.
Collinear gauge invariance implies that the operator Eq.~(\ref{6}) in the effective theory
is
\begin{eqnarray}
\mathcal{O} &=& \sum_{\left\{ a_i \right\}} c\left(\left\{ a_i \right\}\right)\prod_i \left[ W_{n_i}^{(\bar n_i)\dagger}\xi_{n_i,p_i}\right]_{a_i}
\label{13}
\end{eqnarray}
which is gauge invariant under  collinear gauge transformations. The sum of all graphs in the full theory with $n_1$-collinear gauge emission off any of the particles $1,\ldots, r$ in the full theory operator Eq.~(\ref{6}) is equivalent to $n_1$-collinear emission from $\xi_{n_1,p_1}$, or from the Wilson line $W_{n_1}^{\bar n_1}$~\cite{SCET2} in the operator Eq.~(\ref{13}).

\begin{figure}
\begin{center}
\includegraphics[width=2.5cm]{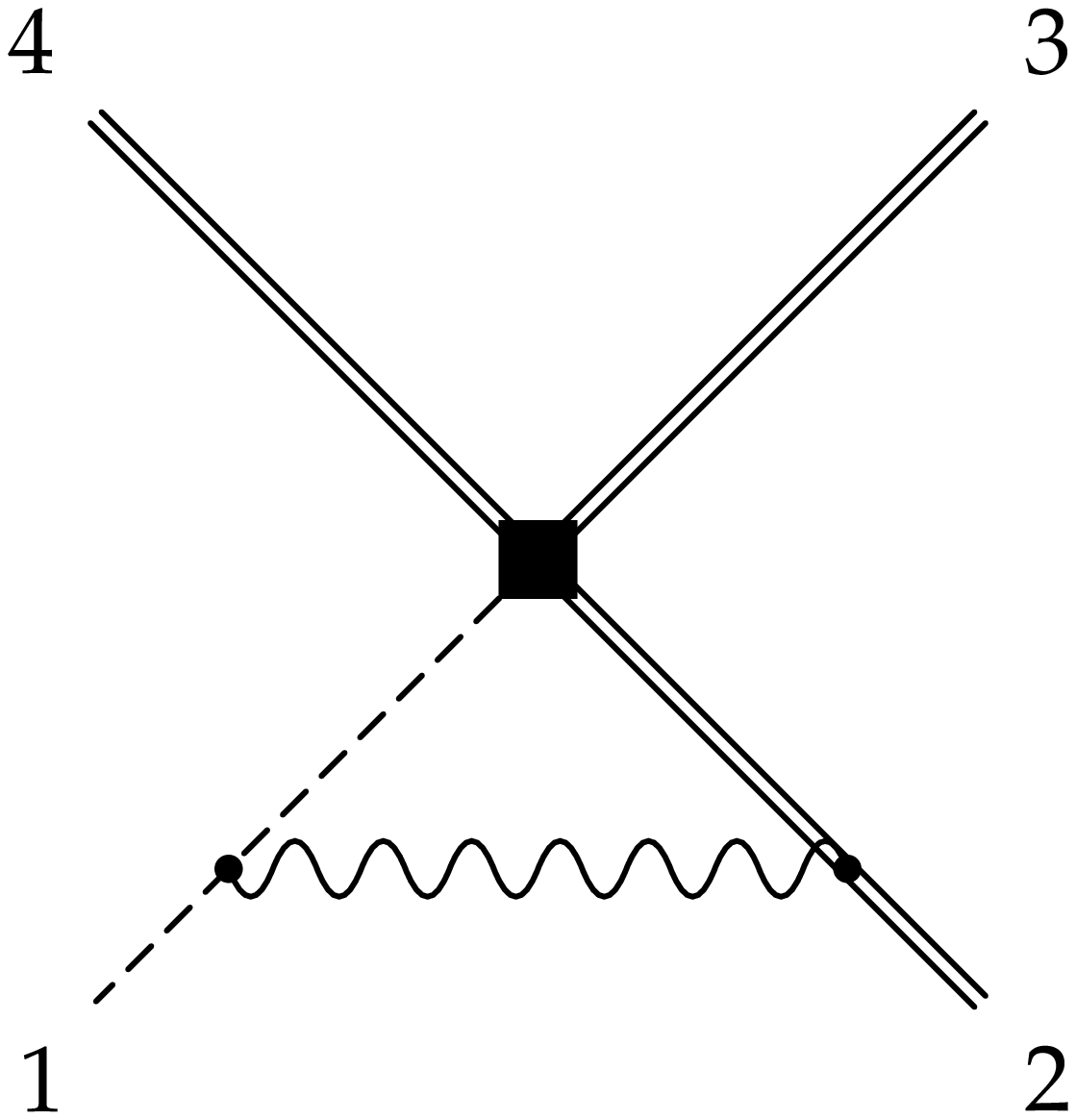}\quad
\includegraphics[width=2.5cm]{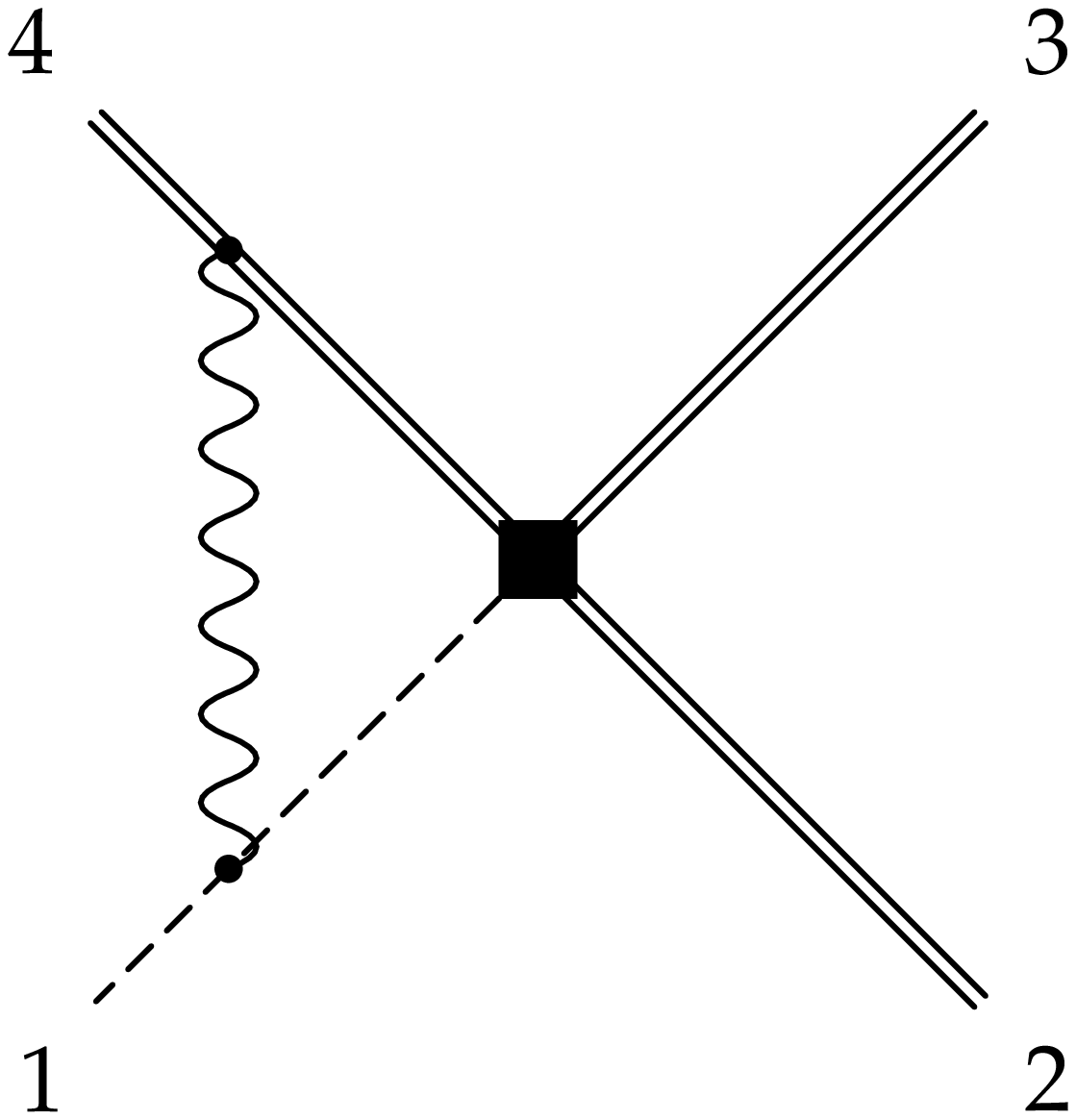}\quad
\includegraphics[width=2.5cm]{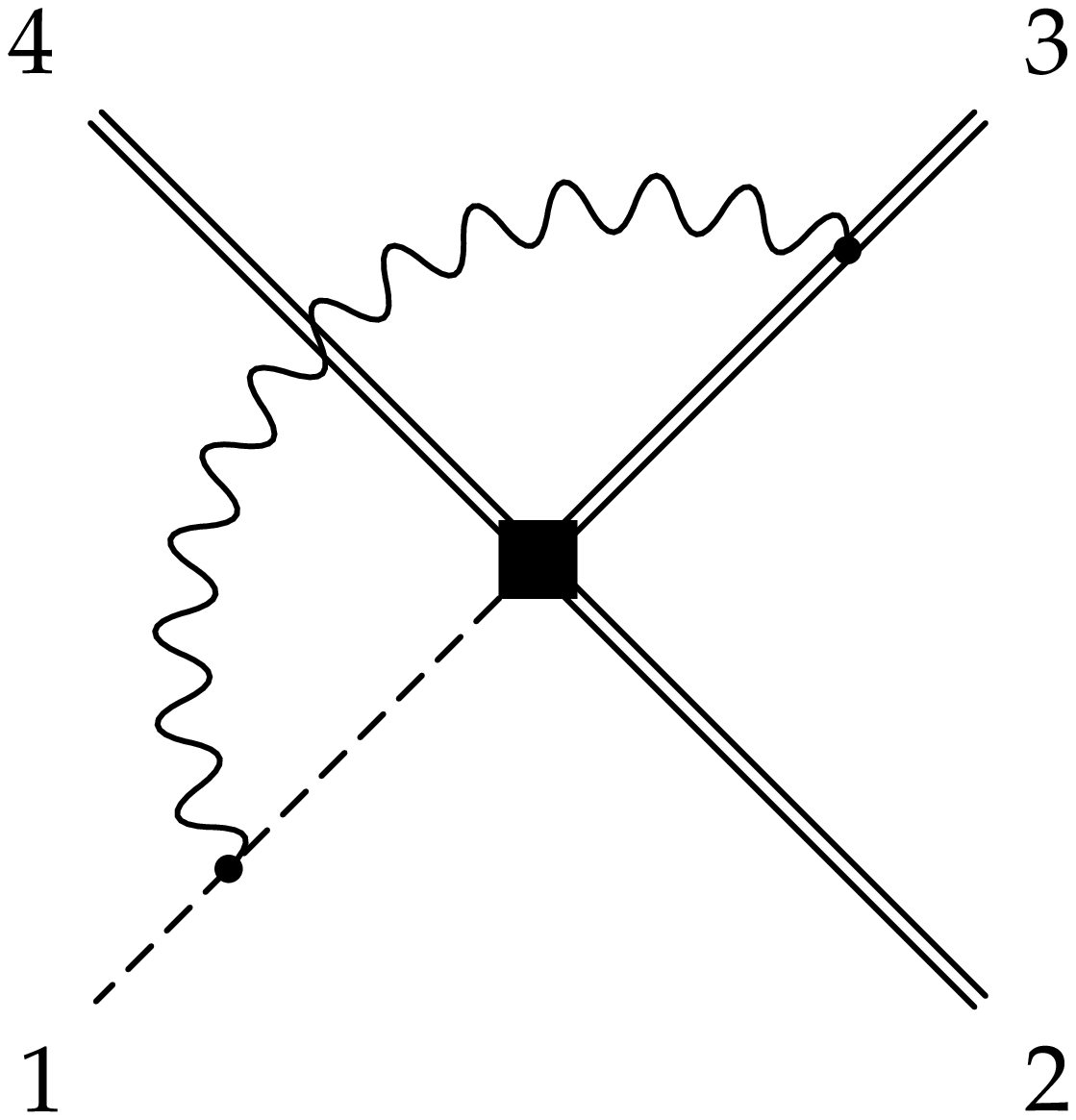}
\end{center}
\caption{\label{fig:n1scet} Graphical representation of $n_1$-collinear interactions in SCET.}
\end{figure}

The structure Eq.~(\ref{13}) is non-trivial, and requires combining terms with gluon emission from all the particles, and using the fact that the operator is a gauge singlet.
The Feynman rules for multiple gauge emission of $n_1$-collinear gluons from particle $i$ gives factors of the form
\begin{eqnarray}
\frac{\epsilon \cdot n_i}{k \cdot n_i}.
\end{eqnarray}
The $n_1$-collinear gauge field has momentum $k$ and polarization $\epsilon$ in the $n_1$-direction at leading order in SCET power counting, so the above expression can be replaced by
\begin{eqnarray}
\frac{\epsilon \cdot n_i}{k \cdot n_i} &\to & \frac{n_1 \cdot n_i}{n_1 \cdot n_i} 
\frac{\epsilon \cdot \bar n_1}{k \cdot \bar n_1} =\frac{\epsilon \cdot \bar n_1}{k \cdot \bar n_1} 
\label{11}
\end{eqnarray}
using the leading (first) term in Eq.~(\ref{3}) for the decomposition of both $k$ and $\epsilon$. This expression is independent of $n_i$. This means that one can change the direction $n_i$, provided $n_i \cdot n_1$ remains leading order in the power counting, i.e.\ $n_i$ does not become almost parallel to $n_1$. One can thus move all the $n_i$ labels so that they all point in a common direction, which can conveniently be chosen to be $\bar n_1$. \emph{This choice only makes reference to particle~1, and has no information about the directions of the other particles.} This is the basis for soft-collinear factorization.

In this paper, we will use the analytic regulator~\cite{bf,analytic} used in \pone, \ptwo. With analytic regularization, Eq.~(\ref{11}) becomes
\begin{eqnarray}
\frac{\epsilon \cdot n_i}{\left(k \cdot n_i\right)^{1+\delta}} &\to & \frac{n_1 \cdot n_i}{\left(n_1 \cdot n_i\right)^{1+\delta}} 
\frac{\epsilon \cdot \bar n_1}{\left(k \cdot \bar n_1\right)^{1+\delta}} \nn
&=& \frac{1}{\left(n_1 \cdot n_i\right)^{\delta}} 
\frac{\epsilon \cdot \bar n_1}{\left(k \cdot \bar n_1\right)^{1+\delta}}
\label{reg}
\end{eqnarray}
and the $n_i$ dependence no longer cancels. Thus the identities which allowed one to combine all the $n_1$-collinear emissions into a single Wilson line in the $\bar n_1$ direction no longer hold. This is a big drawback of the analytic regulator. It is possible to use other regulators which do not have this problem~\cite{hoang2}, but then there are other subtleties which must be addressed, related to zero-bin subtractions~\cite{zerobin}, which are necessary for soft-collinear factorization~\cite{idilbi1,idilbi2,lee,collins:fact}. With the analytic regulator, $n_1$-colllinear interactions cannot be encoded in a single Wilson line in the $\bar n_1$ direction; instead one needs to include Wilson lines along the directions of all the other particles. In the scattering case, this means that $n_1$-collinear interactions at one loop are given graphically by Fig.~\ref{fig:n1scet}. This is equivalent to evaluating the collinear graphs in the full theory using the method of regions with an analytic regulator. We have followed this procedure because it allows for a direct comparison of our intermediate results with previous work.

\section{Sudakov Corrections to Scattering Processes}
\label{sec:toy theory}

In this section we use the toy model to calculate the amplitudes for $qq \to qq$, $q\bar{q} \to q\bar{q}$, $q \bar q \to t \bar t$, and $q \bar{q} \to \tilde{q} \tilde{q}^c$, where $\tilde q$ denotes a colored scalar particle such as a squark. We will call the gauge symmetry color and the particles quarks. The corresponding results in the standard model are given in Sec.~\ref{sec:SM}. 

An interesting result is that the SCET $S$-matrix elements are given by summing the results for the two-particle case, the on-shell Sudakov form-factor given in \pone, \ptwo\ over all pairs of particles. We first compute the $q\bar{q} \to q\bar{q}$ amplitude explicitly by summing the diagrams, and show how the answer can be written as a sum over two-particle $S$-matrix elements. The general proof is given in Sec.~\ref{sec:pairs}.

In this section, as in \ptwo\, we use the decomposition
\begin{eqnarray}
C &=& C^{(0)}+ \frac{\alpha}{4\pi} C^{(1)} \ldots
\label{15}
\end{eqnarray}
of coefficients and anomalous dimensions into their tree-level and one-loop values. In the next section on the standard model, we will explicitly include the $\alpha/(4 \pi)$ factor in the defintion of $C^{(1)}$, since there are several different gauge coupling constants.

\subsection{Light Quark Production}
\label{ssec:qq_scattering}

\begin{figure}
\begin{center}
\includegraphics[width=4cm]{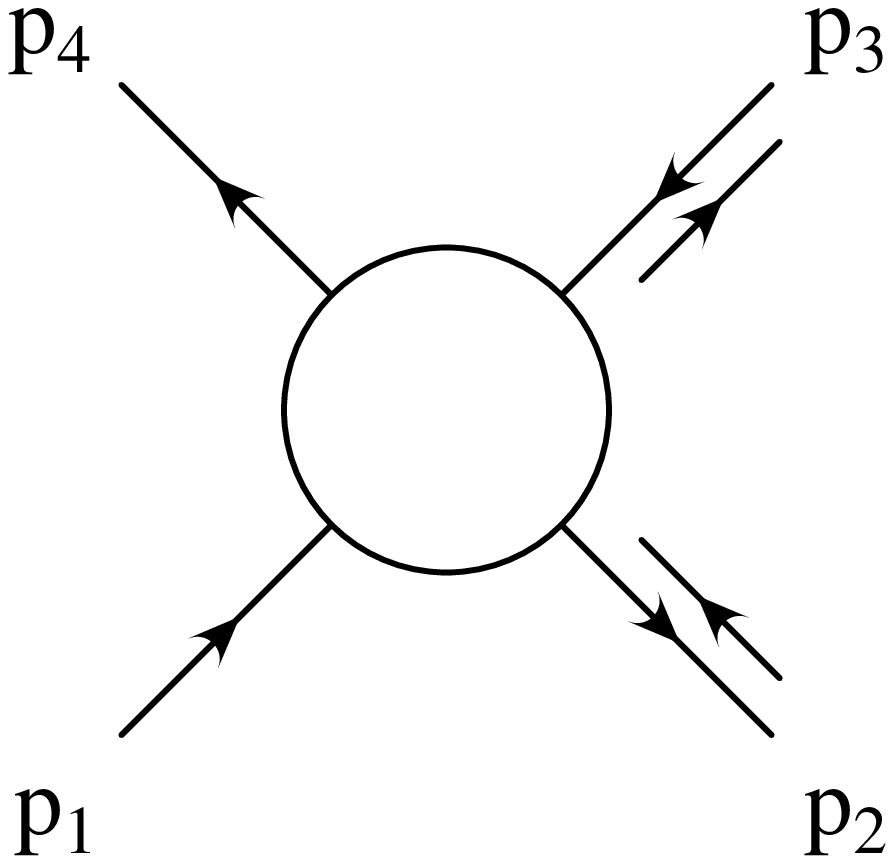}
\end{center}
\caption{\label{FgC1} Pair production $q(p_1) + \bar{q}(p_2) \to  q^\prime(p_4)+\bar{q}^\prime(p_3)$. Time runs vertically. }
\end{figure}

We start with light quark pair-production, $q\bar q\to q^\prime  \bar q^\prime$. The kinematics for $q\bar q\to q^\prime  \bar q^\prime$ is illustrated schematically in Fig.~\ref{FgC1} where the incoming and outgoing particles have momenta $p_1$, $p_2$  and $p_3$, $p_4$, respectively, and we work in the limit $s,t,u \gg M^2 \gg m_i^2$. The external particles are all on-shell $(p_i^2$ = $m^2_i)$.  The Mandelstam variables are $s=(p_1+p_2)^2$, $t=(p_4-p_1)^2$ and $u=(p_3-p_1)^2$. We assume $q$ and $q^\prime$ are different flavors, so that only the $s$-channel annihilation graphs contribute. Identical flavors are discussed in Appendix~\ref{app:box}.

At the scale $\mu \sim Q$ the full theory is matched onto SCET, and the full theory amplitude at leading order in the power counting is expressed as a sum of local operator matrix elements, as in Eq.~(\ref{EqO1}). The gauge-invariant operators in the effective theory are
\begin{eqnarray}
\mathcal{O}_{1 {L \atop R} {L \atop R} } &=&  [\bar{\xi}_4 W_4] t^a \gamma^\mu P_{L \atop R} [W_3^\dagger \xi_3]\, [\bar{\xi}_2 W_2] t^a \gamma_\mu P_{L \atop R} [W_1^\dagger \xi_1]\nn[5pt]
\mathcal{O}_{2 {L \atop R} {L \atop R} } &=&[\bar{\xi}_4 W_4]  \gamma^\mu P_{L \atop R} [W_3^\dagger \xi_3]\, [\bar{\xi}_2 W_2]  \gamma_\mu P_{L \atop R} [W_1^\dagger \xi_1].\nn
\label{16}
\end{eqnarray}
There are only two operators which contribute because the fermions are in the fundamental representation of the gauge group. For other representations, there can be more invariants which contribute, e.g., for isospin one fermions, there are three invariant amplitudes in the $I=0,1,2$ channels.

At tree-level, 
\begin{eqnarray}
C_{1LL}^{(0)}&=&C_{1LR}^{(0)}=C_{1RL}^{(0)}=C_{1RR}^{(0)}= \frac{4 \pi \alpha}{s}
\nn[10pt]
C_{2LL}^{(0)}&=&C_{2LR}^{(0)}=C_{2RL}^{(0)}=C_{2RR}^{(0)}=0 
\label{16t}
\end{eqnarray}
from the graph in Fig.~(\ref{FgO2}).

\begin{figure}
\begin{center}
\includegraphics[width=2.5cm]{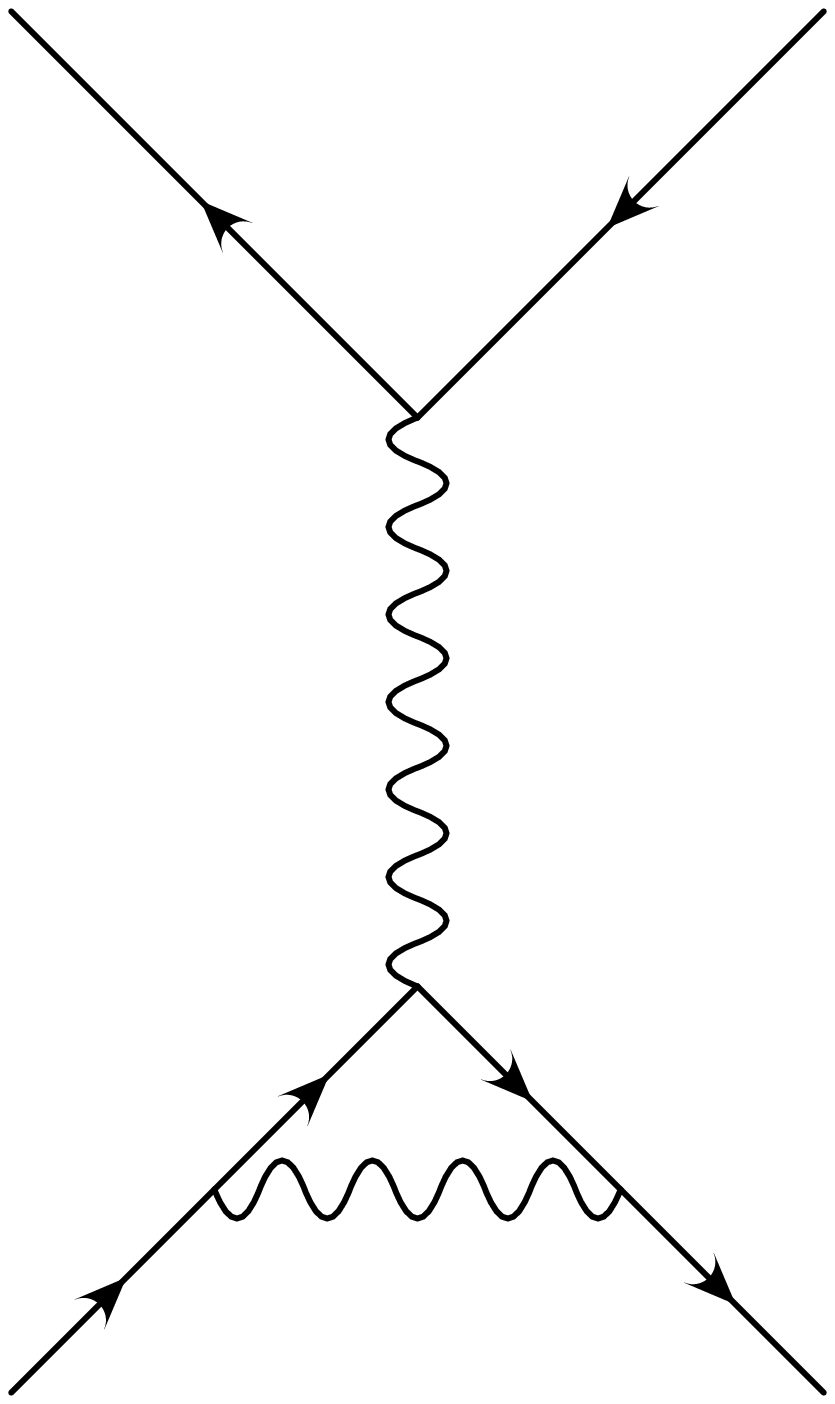}\qquad
\includegraphics[width=2.5cm]{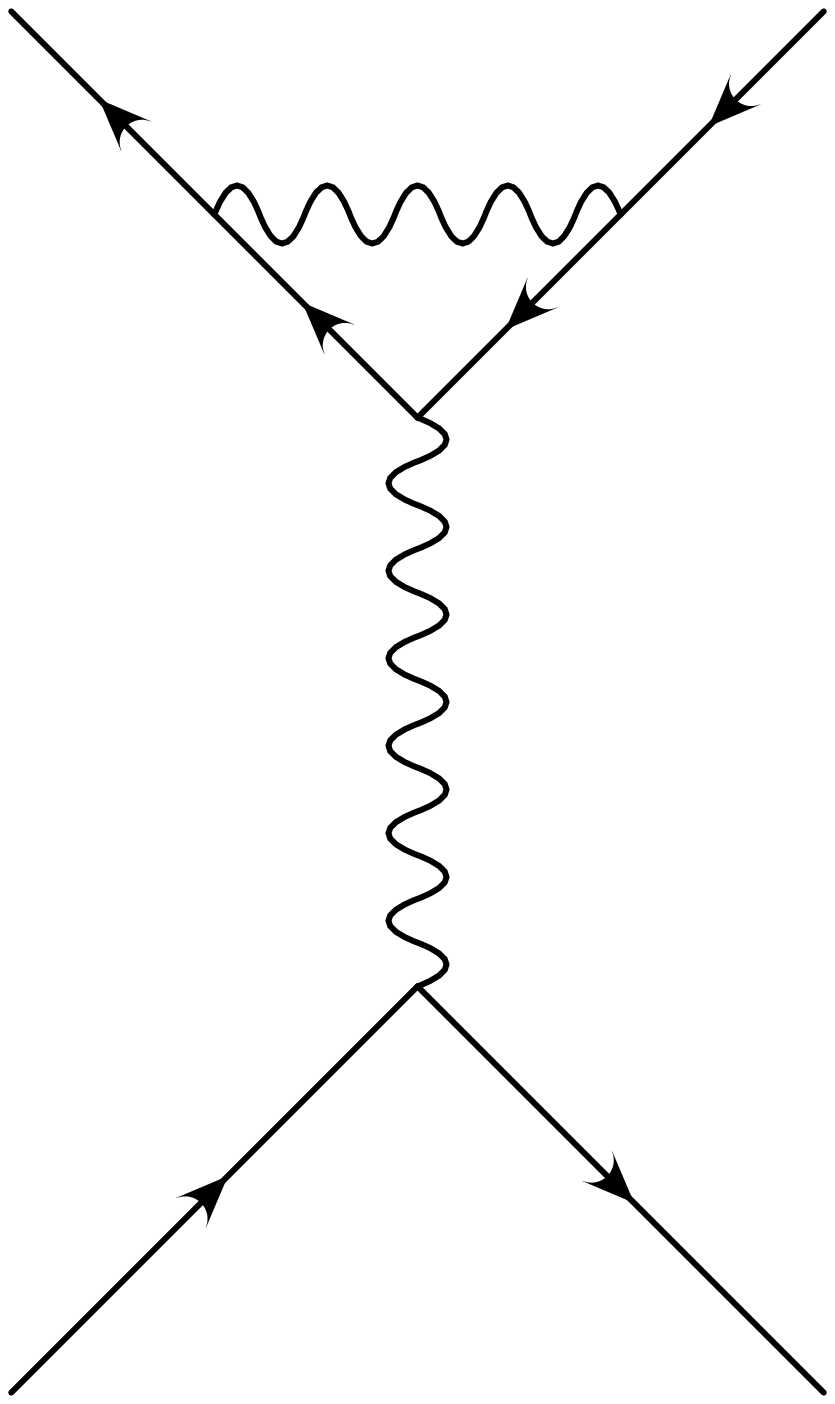}\\ \vspace{5mm}
\includegraphics[width=2.5cm]{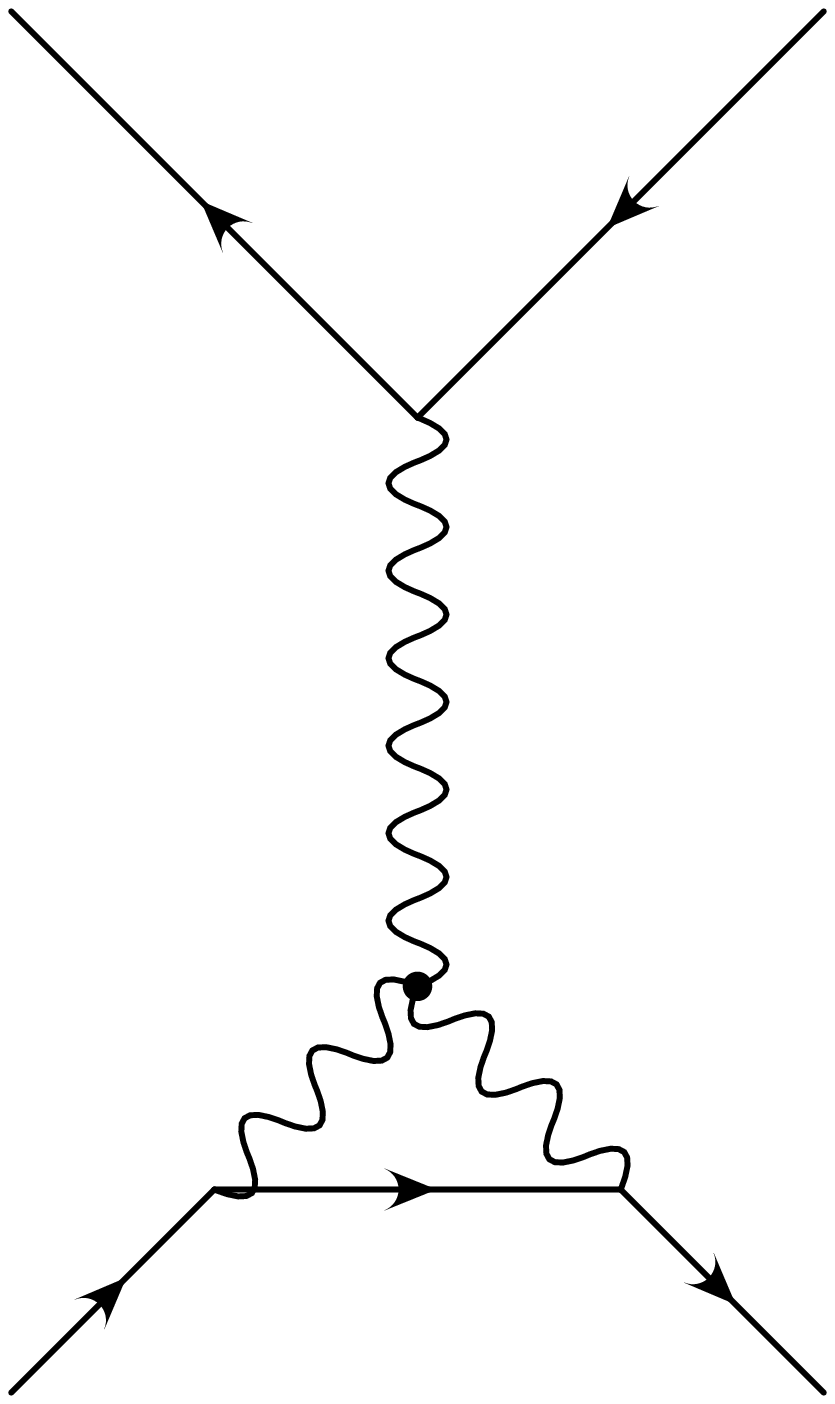}\qquad
\includegraphics[width=2.5cm]{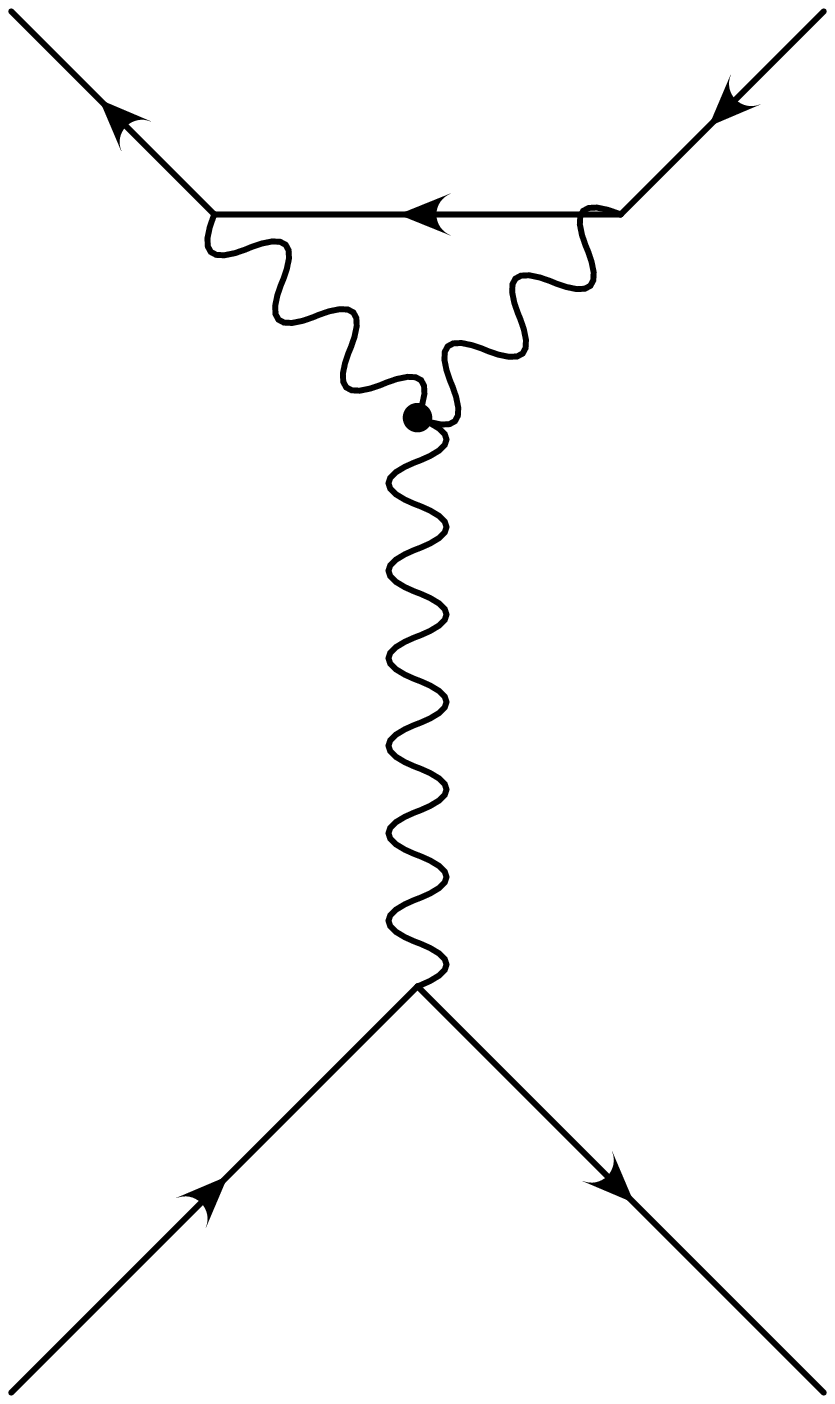}\\ \vspace{5mm}
\includegraphics[width=2.5cm]{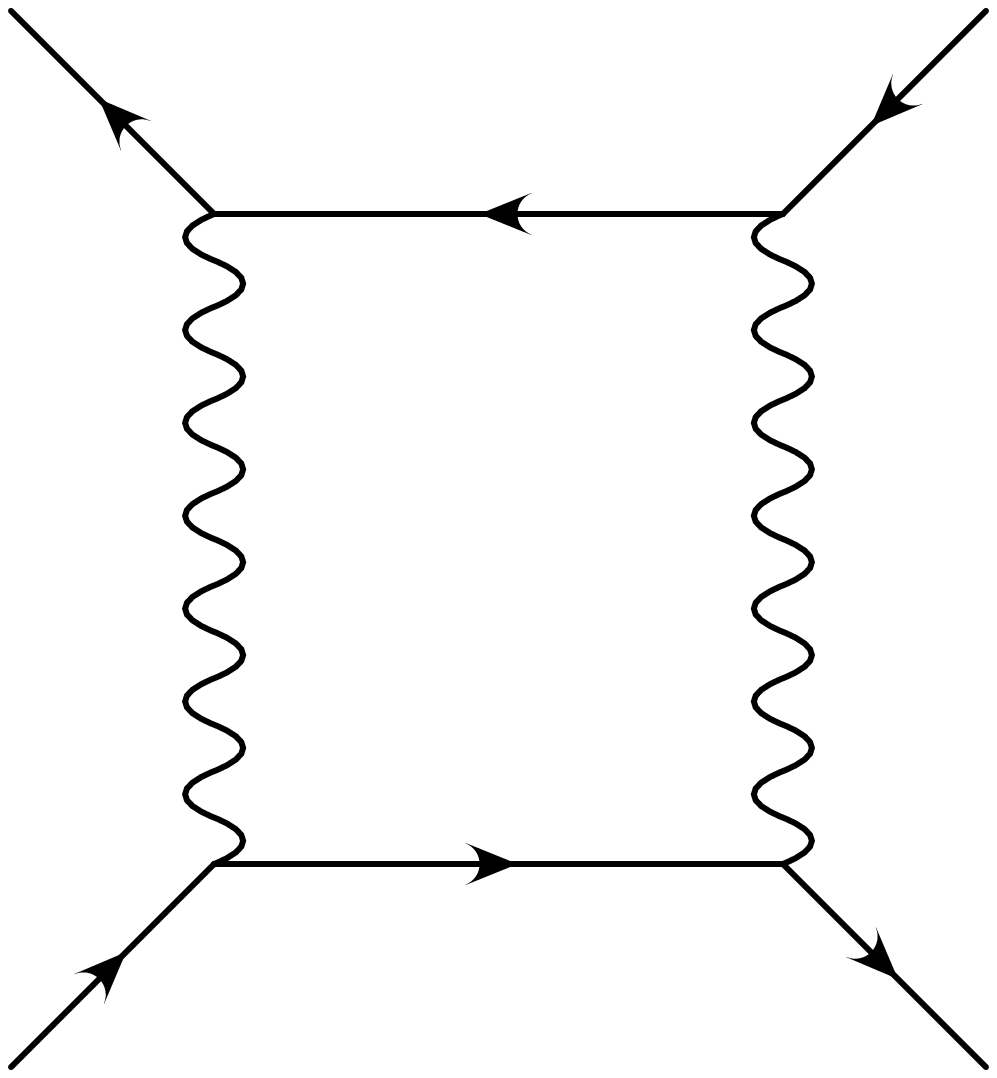}\qquad
\includegraphics[width=2.5cm]{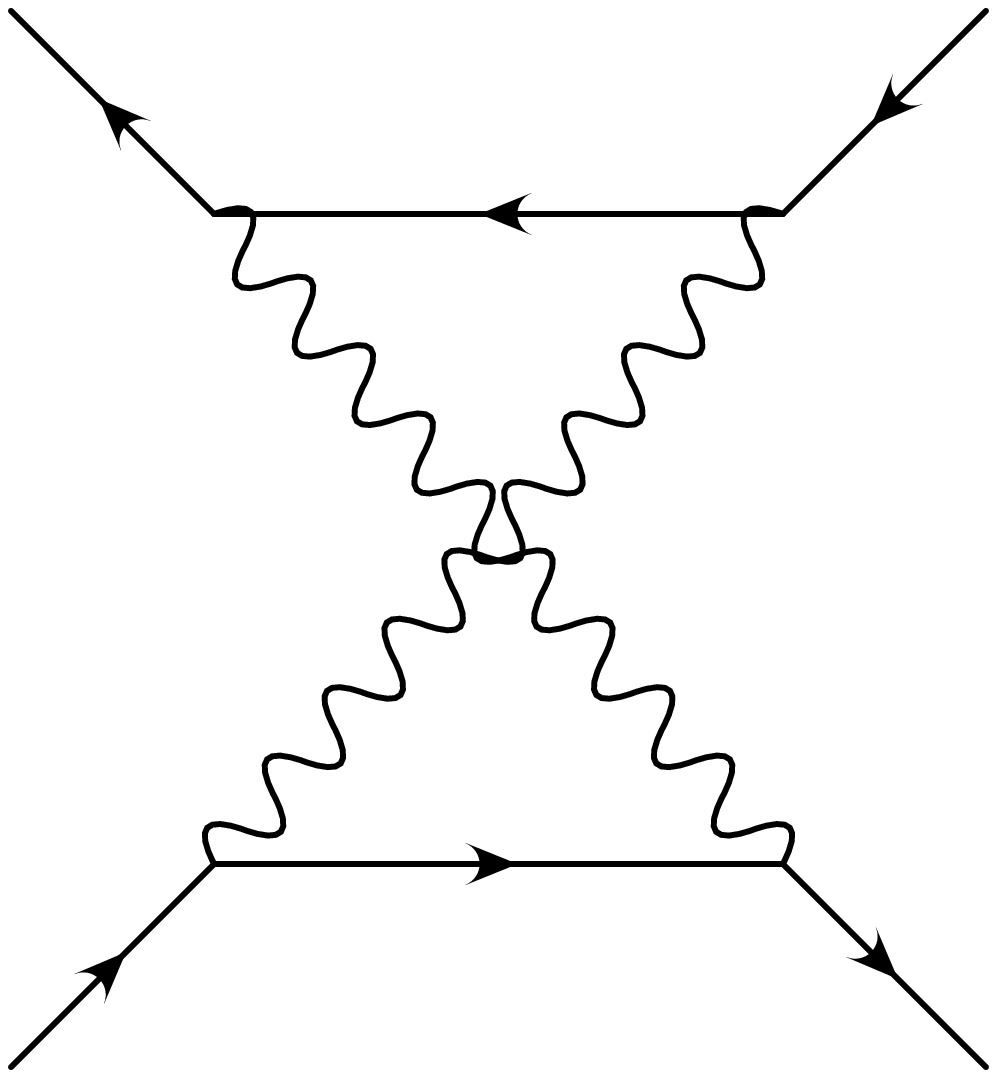}
\end{center}
\caption{\label{FgA1} One loop corrections to pair production in the full theory. Wavefunction and vacuum polarization graphs are not shown}
\end{figure}

The one-loop corrections in the full theory are given by the diagrams in Fig.~\ref{FgA1}, as well as vacuum polarization and wavefunction graphs. 
The one-loop corrections in the effective theory are given by computing radiative corrections to the matrix elements of the four-fermi operators $\mathcal{O}_i$ with tree-level coefficients, and the one-loop matching corrections $C_i^{(1)}$ are given by the difference of the two computations. The graphs in the effective theory vanish on-shell in dimensional regularization, so the one-loop matching coefficients are given by the full theory graphs computed on-shell~\cite{dis,hqet,eft}.  In the full theory matching computation, infrared scales such as the gauge boson mass $M$ and fermion masses $m_i$, which are all much smaller than $Q$, can be set to zero. Thus the coefficients $C_i$ are given by the graphs in Fig.~(\ref{FgA1}) with all masses set to zero. The computation is summarized in Appendix~\ref{app:box}, and agrees with previous calculations~\cite{jkps4, Bohm, Roth}. The one loop coefficients are (removing an overall $\alpha/(4\pi)$, see Eq.~(\ref{15})):
\begin{eqnarray}
C^{(1)}_{1LL} &=& C^{(1)}_{1RR} =   \frac{4\pi\alpha}{s} \biggl[  X(s,t) -\frac{(C_d+C_A)}{4}  \tilde  f(s,t)  \biggr] \nn[10pt]
C^{(1)}_{2LL} &=&  C^{(1)}_{2RR} =  -\frac{4\pi\alpha}{s}\tilde  f(s,t) C_1 \nn[10pt]
C^{(1)}_{1LR} &=&  C^{(1)}_{1RL} = \frac{4\pi\alpha}{s} \biggl[ X(s,u) +  \frac{(C_d-C_A)}{4}  \tilde f(s,u) \biggr] \nn[10pt]
C^{(1)}_{2LR}  &=& C^{(1)}_{2RL} = \frac{4\pi\alpha}{s}\tilde f(s,u) C_1 
\label{17}
\end{eqnarray}
where
\begin{eqnarray}
X(s,t) &=& 2C_F  \left(-\Ls^2+3\Ls+\frac{\pi^2}{6}-8 \right ) \nn
&&+ C_A  \biggl(2\Ls^2-2\LL_{-s-t} \Ls-\frac{11}{3} \Ls+\pi ^2+\frac{85}{9} \biggr)\nn
&&  + \left(\frac43\Ls-\frac{20}{9}\right)  T_F n_F
+ \left(\frac13\Ls-\frac{8}{9}\right)  T_F n_s \nn[10pt]
\tilde f(s, t) &=&  -\frac{2s}{s+t} \LL_{t/s} + \frac{s(s+2t)}{(s+t)^2}\biggl(\LL^2_{t/s} + \pi^2 \biggr)\nn
&&+4 \Ls \LL_{t/(-s-t)} \,.\label{20}
\end{eqnarray} 
Here $n_F$ and $n_S$ are the number of Dirac fermions and complex scalars.
The group theory invariants $C_d$ and $C_1$ are defined in Eq.~(\ref{31}), (\ref{31a}) below. The high scale matching is the only piece of the computation which cannot be obtained by summing the Sudakov form-factor results over all pairs of particles. 

If the initial and final quark flavors are identical, then there are also $t$-channel graphs which contribute to the matching (see App.~\ref{app:box}).

The next step is to compute the anomalous dimension in SCET between $Q$ and $M$, and the matching corrections in SCET at $M$ when the gauge bosons are integrated out. Both results can be obtained simultaneously by computing the on-shell matrix elements of $\mathcal{O}_i$ in SCET. The finite part of the graph gives the matching correction, and the infinite part gives the anomalous dimension. The SCET diagrams are $n_i$-collinear diagrams and ultrasoft graphs. As in \pone, \ptwo\,  the ultrasoft graphs vanish on-shell with the analytic regulator, so the only graphs which contribute are the collinear graphs.

The one-loop $n_i$-sector graphs are given in Fig.~(\ref{fig:n1scet}). Particle~$i$ is given by the field $\xi_i$, and the remaining particles are represented by Wilson lines.  The computations are done using the same analytic regularization method used in Refs.~\pone, \ptwo. The regulated $n_i$-collinear propagator denominator is
\begin{eqnarray}
\frac{1}{(p_i+k)^2} &\to& \frac{(-\nu_i^2)^{\delta_i}}{\left[(p_i+k)^2\right]^{1+\delta_i}}.
\end{eqnarray}
The propagator denominator for particle $j$ interacting with $n_i$-collinear gluons becomes
\begin{eqnarray}
\frac{1}{(p_j+k)^2} &\to& \frac{(-\nu_j^2)^{\delta_j}}{\left[(p_j+k)^2\right]^{1+\delta_j}}
\to \frac{(-\nu_j^2)^{\delta_j}}{\left[2 p_j \cdot k\right]^{1+\delta_j}}.
\label{19}
\end{eqnarray}
At leading order in SCET power counting, $p_i$ and $k$ are $n_i$-collinear, so  $p_i^\mu = n_i^\mu \left(\bar n_i \cdot p_i\right)/2$, $k^\mu = n_i^\mu \left(\bar n_i \cdot k\right)/2$ and
\begin{eqnarray}
\frac{1}{(p_j+k)^2} &\to& \frac{(-\nu_j^2)^{\delta_j}}{\left[\frac12 (\bar n_j \cdot  p_j) (n_j \cdot n_i)\right]^{1+\delta_j}}.
\label{19a}
\end{eqnarray}
Thus the analytic continuation of the Wilson line propagator arising from particle $j$ is
\begin{eqnarray}
\frac{1}{\bar n_i \cdot k } &\to&  \frac{(- \nu_{i}^{(j)})^{\delta_j}}{(\bar n_i \cdot k)^{1+\delta_j} }\nn
\nu_{i}^{(j)} &=&  \frac{\nu_j^2}{\left[\frac12 (\bar n_j \cdot  p_j) (n_j \cdot n_i)\right]}.
\label{21}
\end{eqnarray}
The key observation is that the $\nu_j$ regulator parameter when particle $j$ is the $n_j$-collinear field $\xi_{n_j,p_j}$ is related to the $\nu^{(j)}_i$ regulator parameter when particle $j$ interacts with $n_i$-collinear gluons as a Wilson line. This feature was already studied in \pone, \ptwo\, and leads to a calculable logarithmic violation of factorization, as discussed further in Sec.~\ref{sec:fact}.

The $n_i$ collinear graph with the  particle~$j$ Wilson line is then identical to the $n_1$ collinear graph interacting with the $n_2$ Wilson line result in \pone, \ptwo\ with the replacement $\nu_1 \to \nu_i$ for the collinear particle regulator, and $\nu_2^+ \to \nu_{i}^{(j)}$ for the Wilson line regulator. The regulator variables $\nu_i, \nu_{i}^{(j)}$ only appear in logarithms, and $
\nu_{i}^{(j)}$ only appears in the boost-invariant combination
\begin{eqnarray}
\frac{\nu_{i}^{(j)}}{\bar n_i \cdot p_i} &=&  \frac{\nu_j^2}{\frac12 (\bar n_j \cdot  p_j) (n_j \cdot n_i)(\bar n_i \cdot p_i)}= \frac{\nu_j^2}{2 p_i \cdot p_j}.
\label{22}
\end{eqnarray}
In the Sudakov form-factor results in \pone, \ptwo\,  $2p_1 \cdot p_2=Q^2$, and Eq.~(\ref{22}) was the origin of the $\log Q^2$ terms in SCET. Here $2 p_i \cdot p_j$ depends on the kinematic variables, and gives a dependence on $\log s$, $\log t$ and $\log u$.

\begin{figure}
\begin{center}
\includegraphics[width=2.5cm]{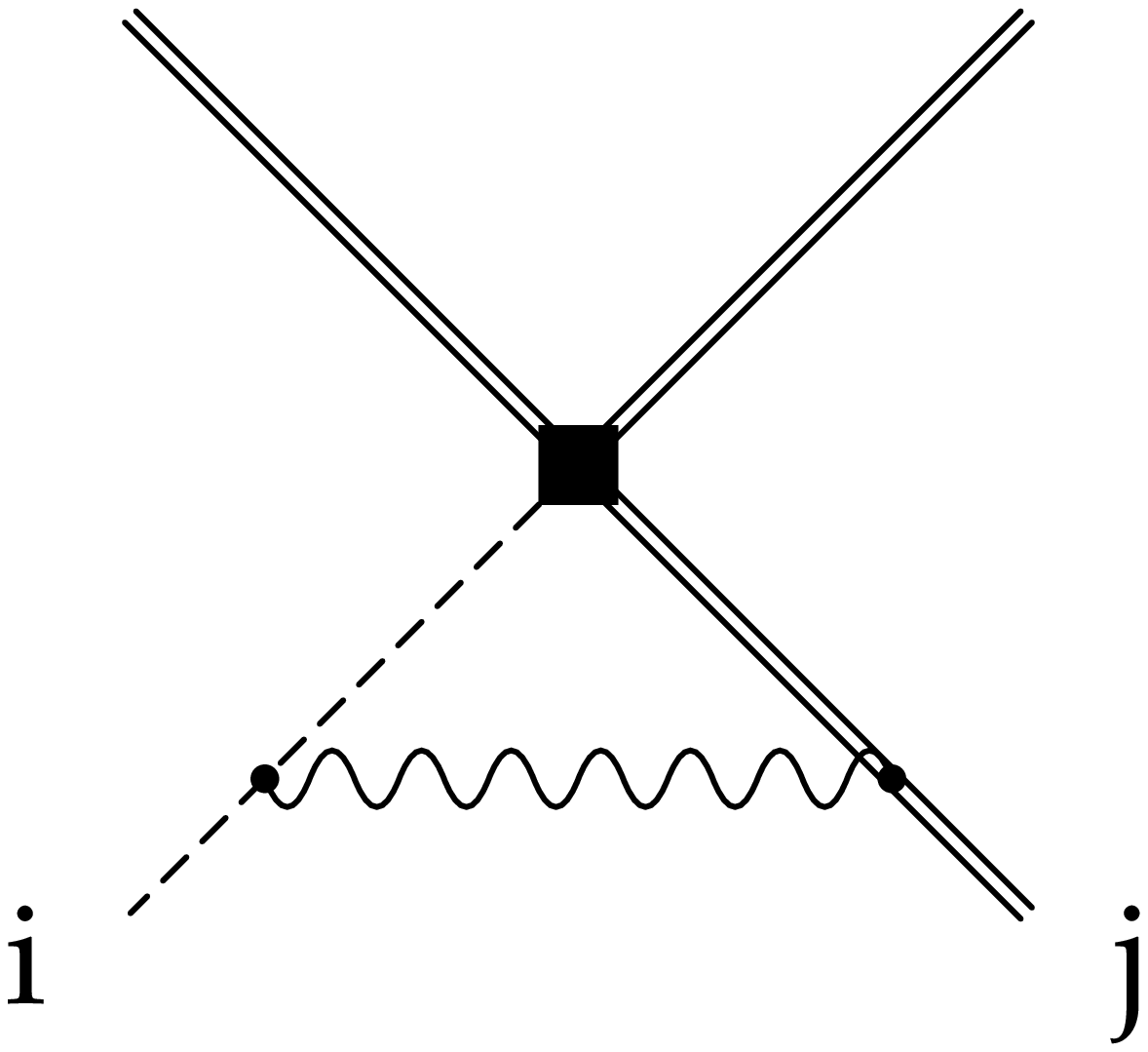}\qquad
\includegraphics[width=2.5cm]{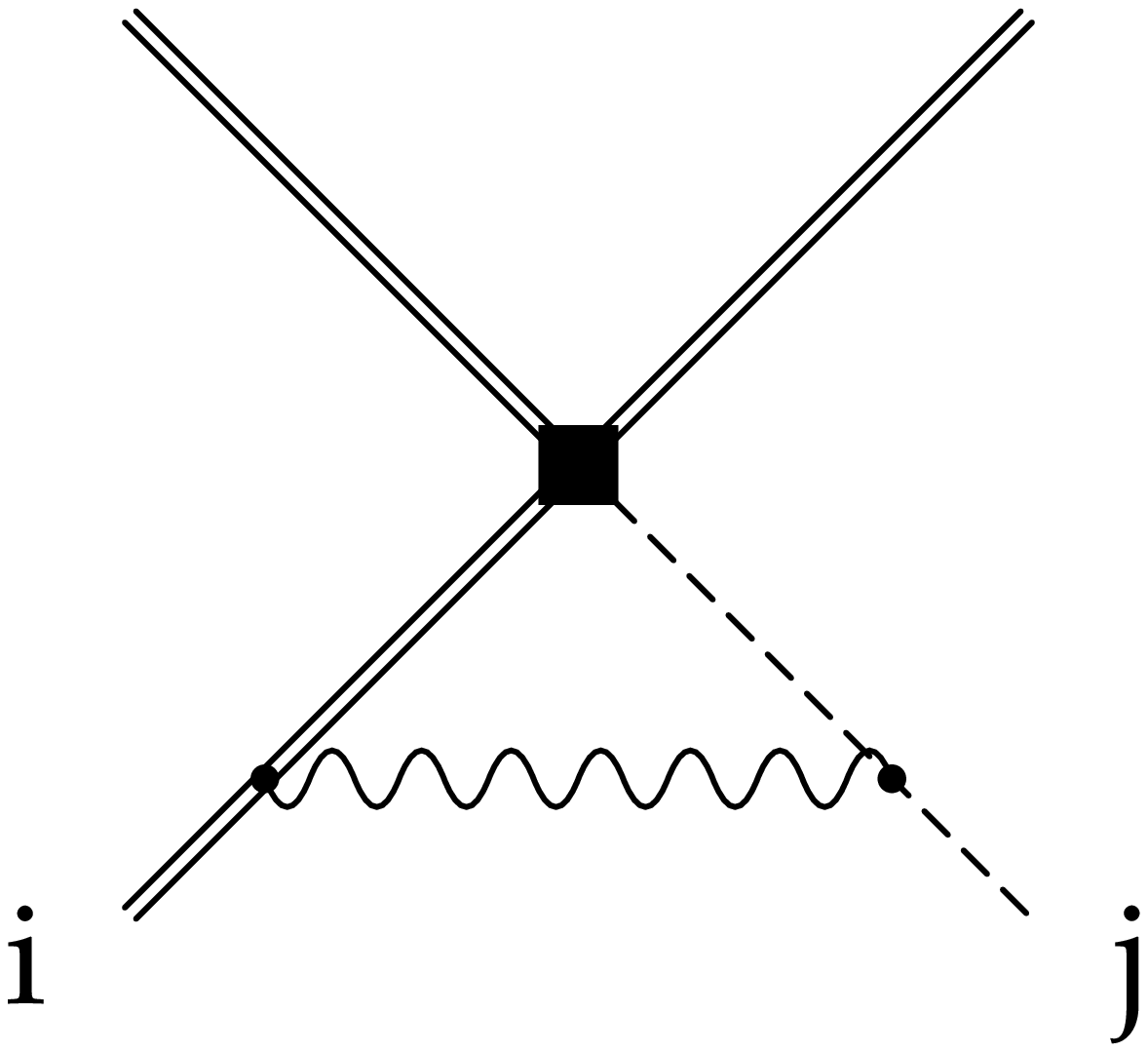}
\end{center}
\caption{Collinear graphs involving particles $i$ and $j$ which are related. \label{fig:B}}
\end{figure}

In the Sudakov form-factor computation, there was a non-trivial cancellation between the $n$-collinear and $\bar n$-collinear graphs, so that the sum of the graphs was independent of the analytic regulator parameters $\nu_i$. There is a similar cancellation here. There are two graphs which are related to each other: graphs with gauge boson exchange between $i$ and $j$ in which $i$ is $n_i$-collinear and $j$ is a Wilson line, and in which $i$ is a Wilson line and $j$ is $n_j$-collinear (see Fig.~\ref{fig:B}). 
These graphs have identical color factors. The regulator cancellation depended on two identities given in Appendix~A in \ptwo. The corresponding relations here are
\begin{eqnarray}
&&\left(\log \frac{ \nu_i^{(j)}}{\bar n_i \cdot p_i} + \log \frac{\nu_i^2}{\mu^2}\right)
-\left(\log \frac{ \nu_j^{(i)}}{\bar n_j \cdot p_j} + \log \frac{\nu_j^2}{\mu^2}\right)\nn
&=& \log \frac{2\nu_i^2\nu_j^2}{\mu^2\left( \bar n_j \cdot p_j \right)\left( \bar n_i \cdot p_i \right) 
\left(n_i \cdot n_j \right)}- \left( i \leftrightarrow j \right)\nn
&=& 0\nn
&&\left(\log \frac{\nu_i^{(j)}}{\bar n_i \cdot p_i} - \log \frac{\nu_i^2}{\mu^2}\right)
+\left(\log \frac{\nu_j^{(i)}}{\bar n_j \cdot p_j} - \log \frac{\nu_j^2}{\mu^2}\right)\nn
&=& \log \frac{2\mu^2\nu_j^2}{\nu_i^2\left( \bar n_j \cdot p_j \right) \left( \bar n_i \cdot p_i \right)\left(n_i \cdot n_j \right)}+ \left( i \leftrightarrow j \right)\nn
&=&  2\log \frac{\mu^2}{\left( \bar n_j \cdot p_j \right) \left( \bar n_i \cdot p_i \right)\frac 12 
\left(n_i \cdot n_j \right)}\nn
&=&  2\log \frac{\mu^2}{2 p_i \cdot p_j}
\end{eqnarray}
which follow from Eq.~(\ref{21}), so the $\nu$ cancellation continues to hold. Thus the collinear graphs are obtained by the collinear graphs in the Sudakov form-factor case with the replacement $Q^2 \to 2 p_i \cdot p_j$, and summing over pairs with the appropriate group theory factor. The ultrasoft graphs vanish on-shell, as in the Sudakov form-factor case, so the complete answer is given by adding the wavefunction renormalization graphs to the collinear contribution.

The terms which depend on $\log( p_i \cdot p_j)$ arise from the regularization of Wilson lines using the analytic regulator. They depend on the momenta of both particles, so it is clear that in $n_i$-collinear graphs, it is not possible to combine the Wilson lines for the other particles into a single Wilson line, as that would lose information on the $p_j$ dependence. 

Note that the $r$-particle result obtained by combining the Sudakov form-factors over all pairs of particles is valid even if all the momenta flowing into the operator do not add to zero, i.e.\ even if there is some momentum inserted at the vertex. In the case of 2-particle scattering, we are interested in operator insertions at zero momentum, and the six $p_i \cdot p_j$ invariants can be written in terms of two independent Mandelstam variables.

The SCET graphs do not change the Lorentz or chiral structure of the operators, and only cause rearragements of the gauge indices. Thus $\mathcal{O}_{1LL}$ can mix only with $\mathcal{O}_{2LL}$. Furthermore, the mixing matrix for $\mathcal{O}_{1hh^\prime}$, $\mathcal{O}_{2hh^\prime}$ is independent of the chirality labels $h,h^\prime$. To keep track of the gauge
indices, it is convenient to denote $\mathcal{O}_{1hh^\prime}$, $\mathcal{O}_{2hh^\prime}$ by
\begin{eqnarray}
\mathcal{O}_1 &=& t^a \otimes t^a \nn
\mathcal{O}_2 &=& 1 \otimes 1.
\end{eqnarray}
The SCET graphs are then a $2\times2$ matrix in $\mathcal{O}_{1,2}$ space, and a unit matrix in chirality ($h,h^\prime)$ space.

The sum of the $n$-collinear and $\bar n$-collinear vertx graphs with the gauge factor $C_F$ omitted is
\begin{eqnarray}
\Gamma\left(Q^2\right) &=&\frac{\alpha}{4\pi} \Biggl[\frac{2}{\e^2} + \frac{4}{\e} - \frac{2}{\e} \lQ   -\lM^2+2\lM \lQ \nn
&&- 4\lM+4-\frac{5\pi^2}{6} \Biggr] \,.
\label{EqB7} 
\end{eqnarray}
The wavefunction renormalization, omitting group theory factors is\footnote{These are Eq.~(43) with the wavefunction correction removed and  Eq.~(40) of \ptwo. Since we work on-shell, there is no need to introduce infrared modes whose virtuality is governed by the off-shellness~\cite{dorsten}.}
\begin{eqnarray}
\delta Z^{-1} &=&\frac{\alpha}{4\pi} \Biggl[ \frac{1}{\e}  - \lM-\frac12 \Biggr]
\end{eqnarray}

The sum of graphs in Fig.~(\ref{fig:B}) which connect particles~1 and 2 is thus
\begin{eqnarray}
\Gamma_{12}\!\left(-2p_1 \cdot p_2\right) t^b t^a t^b \otimes t^a 
\end{eqnarray}
if the operator at the vertex is $\mathcal{O}_1$, and
\begin{eqnarray}
\Gamma_{12}\!\left(-2p_1 \cdot p_2\right) t^b t^b \otimes 1 
\end{eqnarray}
if the operator at the vertex is $\mathcal{O}_2$. The minus signs relative to Eq.~(\ref{EqB7}) arise because both momenta $p_{1,2}$ are incoming, whereas in Eq.~(\ref{EqB7}) computed in \ptwo, $p_1$ was incoming, $p_2$ was outgoing, and $Q^2=2p_1 \cdot p_2$. It is useful to add subscripts to $\Gamma$ denoting the particles involved in the diagram.

The group theory factors can be simplified using
\begin{eqnarray}
t^a t^a &=& C_F\ 1 \nn
t^a t^b t^a &=& \left(C_F-\frac12 C_A\right) t^b \nn
t^a t^b \otimes t^a t^b &=& C_1\ 1 \otimes 1 +\frac14\left(C_d-C_A\right)t^a \otimes t^a\nn
t^a t^b \otimes t^b t^a &=& C_1\ 1 \otimes 1 +\frac14\left(C_d+C_A\right)t^a \otimes t^a
\label{31}
\end{eqnarray}
in the notation of Ref.~\cite{Manohar:2000hj}. For an $SU(N)$ gauge theory,
\begin{eqnarray}
C_A &=& N\nn
C_F &=& \frac{N^2-1}{2N}\nn
C_d &=& \frac{N^2-4}{N}\nn
C_1 &=& \frac{N^2-1}{4N^2}
\label{31a}
\end{eqnarray}
so $C_F=4/3$, $C_A=3$, $C_d=5/3$, $C_1=2/9$ for $SU(3)$ and
$C_F=3/4$, $C_A=2$, $C_d=0$, $C_1=3/16$ for $SU(2)$. 

The matrix element of $\mathcal{O}_1$ is
\begin{eqnarray}
&&\left(C_F-\frac12C_A\right)\left(t^a \otimes t^a\right) \Gamma_{12}(-2 p_1 \cdot p_2)\nn
&&+\left(C_F-\frac12C_A\right)\left(t^a \otimes t^a\right) \Gamma_{34}(-2 p_3 \cdot p_4)\nn
&&+\left(C_1 1 \otimes 1 + \frac14\left(C_d+C_A\right)t^a \otimes t^a\right)\Gamma_{14}(2 p_1 \cdot p_4)\nn
&&+\left(C_1 1 \otimes 1 + \frac14\left(C_d+C_A\right)t^a \otimes t^a\right)\Gamma_{23}(2 p_2 \cdot p_3)\nn
&&-\left(C_1 1 \otimes 1 + \frac14\left(C_d-C_A\right)t^a \otimes t^a\right)\Gamma_{13}(2 p_1 \cdot p_3)\nn
&&-\left(C_1 1 \otimes 1 + \frac14\left(C_d-C_A\right)t^a \otimes t^a\right)\Gamma_{24}(2 p_2 \cdot p_4)\nn
&&-\frac12\left(\delta Z_{1}^{-1}+\delta Z_{2}^{-1}+\delta Z_{3}^{-1}+\delta Z_{4}^{-1}\right)C_F \left(t^a \otimes t^a\right) \,.\nn
\label{30}
\end{eqnarray}
The terms are given by summing over the six possible choices of particle pairs, and  including the wavefunction contribution for each particle. The terms from gluon exchange between $13$ or $24$ have minus signs, from charge conjugation, since both lines have color flowing into the vertex.\footnote{Equation~(\ref{30})  is true even if there is non-zero momentum inserted at the operator vertex, so that $p_1 + p_2 \not=p_3+p_4$.}

For $\mathcal{O}_2$, one has instead
\begin{eqnarray}
&&C_F\left(1 \otimes 1\right) \Gamma_{12}(-2 p_1 \cdot p_2)+C_F\left(1 \otimes 1\right) \Gamma_{34}(2 p_3 \cdot p_4)\nn
&&+\left(t^a \otimes t^a\right)\Gamma_{14}(-2 p_1 \cdot p_4)+\left(t^a \otimes t^a\right)\Gamma_{23}(2 p_2 \cdot p_3)\nn
&&-\left(t^a \otimes t^a\right)\Gamma_{13}(2 p_1 \cdot p_3)-\left(t^a \otimes t^a\right)\Gamma_{24}(2 p_2 \cdot p_4)\nn
&&-\frac12\left(\delta Z_{1}^{-1}+\delta Z_{2}^{-1}+\delta Z_{3}^{-1}+\delta Z_{4}^{-1}\right)C_F \left(1 \otimes 1\right) \,.
\label{32}
\end{eqnarray}
Equations~(\ref{30}), (\ref{32}) can be written in matrix form, by defining the matrix
\begin{eqnarray}
\mathcal{R} &=& \widetilde{\mathcal{R}}\, \openone+\mathcal{R}_S\nn
\widetilde{\mathcal{R}}&=&C_F \Bigl[ \Gamma_{12}(-2p_1\cdot p_2)-\frac12 \delta Z_{1}^{-1} -\frac12 \delta Z_{2}^{-1}\Bigr]\nn
&&+C_F \Bigl[ \Gamma_{34}(-2p_3\cdot p_4)-\frac12 \delta Z_{3}^{-1} -\frac12 \delta Z_{4}^{-1}\Bigr]\nn[10pt]
\mathcal{R}_S &=&\left[ \begin{array}{cc} 
\frac14 C_d r_1 + \frac14 C_A r_2
& r_1 \\[10pt]
 C_1r_1 & 0 \end{array}\right]\nn[10pt]
r_1 &=& \Gamma_{14}(2p_1 \cdot p_4)+\Gamma_{23}(2p_2 \cdot p_3)\nn
&&-\Gamma_{13}(2p_1 \cdot p_3)-\Gamma_{24}(2p_2 \cdot p_4)\nn
r_2 &=& \Gamma_{14}(2p_1 \cdot p_4)+\Gamma_{23}(2p_2 \cdot p_3)\nn
&&+\Gamma_{13}(2p_1 \cdot p_3)+\Gamma_{24}(2p_2 \cdot p_4)\nn
&& -2 \Gamma_{12}(-2p_1\cdot p_2)-2 \Gamma_{34}(-2p_3\cdot p_4)\,.
 \label{34a}
\end{eqnarray}
Equation~(\ref{34a}) has an interesting structure --- It has a diagonal piece $\widetilde{\mathcal{R}}$, which is the sum of the on-shell Sudakov form factor graphs  (including wavefunction factors) for $1 \to 2$ and $3 \to 4$, and a term $\mathcal{R}_S$, which depends on the amplitude linear combinations $r_1$ and $r_2$. $\mathcal{R}_S$ contains differences of $\Gamma_{ij}$. One can include wavefunction factors in $\mathcal{R}_S$ by the replacement
\begin{eqnarray}
\Gamma_{ij} &\to& \mathcal{S}_{ij} \equiv \Gamma_{ij}-\frac12\delta Z_{i}^{-1}-\frac12 \delta Z_{j}^{-1}
\end{eqnarray}
without changing $r_1$ and $r_2$. We will thus use Eq.~(\ref{34a}) in the form
\begin{eqnarray}
\mathcal{R} &=& \widetilde{\mathcal{R}}\,\openone+\mathcal{R}_S\nn
\widetilde{\mathcal{R}}&=& C_F \mathcal{S}_{12}(-2p_1\cdot p_2)+C_F  \mathcal{S}_{34}(-2p_3\cdot p_4) \nn[10pt]
\mathcal{R}_S &=&\left[ \begin{array}{cc} 
\frac14 C_d r_1 + \frac14 C_\mathcal{S} r_2
& r_1 \\[10pt]
 C_1r_1 & 0 \end{array}\right]\nn[10pt]
r_1 &=& \mathcal{S}_{14}(2p_1 \cdot p_4)+\mathcal{S}_{23}(2p_2 \cdot p_3)\nn
&&-\mathcal{S}_{13}(2p_1 \cdot p_3)-\mathcal{S}_{24}(2p_2 \cdot p_4)\nn[5pt]
r_2 &=& \mathcal{S}_{14}(2p_1 \cdot p_4)+\mathcal{S}_{23}(2p_2 \cdot p_3)\nn
&&+\mathcal{S}_{13}(2p_1 \cdot p_3)+\mathcal{S}_{24}(2p_2 \cdot p_4)\nn
&& -2 \mathcal{S}_{12}(-2p_1\cdot p_2)-2 \mathcal{S}_{34}(-2p_3\cdot p_4)
 \label{34}
\end{eqnarray}
where $\mathcal{S}$ is the on-shell Sudakov form-factor including wavefunction corrections, i.e.\ an $S$-matrix element, without any color factors. The $r_1$ and $r_2$ terms contain differences of Sudakov form-factors, and so do not contain Sudakov double-logs, which are universal, don't depend on particle type, and cancel in the difference.

The on-shell matrix element of the effective Lagrangian $C_i \mathcal{O}_i$ including wavefunction factors is
\begin{eqnarray}
\left[ 
\begin{array}{cc}
\vev{ {\mathcal{O}_1} }^{(0)} & \vev{ {\mathcal{O}_2}}^{(0)}  
\end{array}\right] \left(1+\mathcal{R}\right) \left[ 
\begin{array}{c}
C_1 \\ C_2 
\end{array}\right]
\label{35}
\end{eqnarray}
where $C_i$ are the operator coefficients and $\mathcal{O}^{(0)}$ are the tree-level matrix elements.

Equations~(\ref{34}), (\ref{35}) are master equations we will use for our scattering computations.  For example, to compute the matching correction when the massive gauge bosons are integrated out, we use
\begin{eqnarray}
\left[ 
\begin{array}{cc}
\tilde C_1 \\ \tilde C_2 
\end{array}\right] &=& \left(1+R\right) \left[ 
\begin{array}{cc}
C_1 \\ C_2 
\end{array}\right]
\label{35a}
\end{eqnarray}
where $R$ is the finite part of $\mathcal{R}$, and $C$ and $\tilde C$ are the coefficients in the high-energy theory with gauge bosons and the low-energy theory without gauge bosons, respectively. Similarly, the anomalous dimension matrix is
\begin{eqnarray}
\mu \frac{\rd}{\rd \mu} \left[ 
\begin{array}{cc}
C_1 \\ C_2 
\end{array}\right] &=& \gamma \left[ 
\begin{array}{cc}
C_1 \\ C_2 
\end{array}\right]
\label{35b}
\end{eqnarray}
where $\gamma$ is the anomalous dimension computed using the $1/\epsilon$ terms in $\mathcal{R}$, i.e. $-2$ times the $1/\epsilon$ terms in $\mathcal{R}$ at one loop. The matching conditions and anomalous dimensions are given by Eq.~(\ref{34}) with $\mathcal{S}_{ij}$ replaced by the corresponding Sudakov form-factor matching correction and anomalous dimension computed in \ptwo\, without any additional Feynman graph computations.

We now apply the master formula to the SCET anomalous dimension for $q \bar q \to q^\prime \bar q^\prime$ in the region $Q > \mu > M$, and to the matching condition at $M$. The anomalous dimension is given using Eq.~(\ref{34}) with $\mathcal{S}$ replaced by the SCET anomalous dimension for the Sudakov form factor, i.e.\ by $\gamma^{(1)}$ for the bifermion operators in Table~I of \ptwo, $\mathcal{S}  \to 4 \lQ - 6$. The anomalous dimension matrix is
\begin{eqnarray}
\gamma^{(1)} &=& \widetilde{\gamma}^{(1)}\, \openone +\gamma_S^{(1)}\nn
\widetilde{\gamma}^{(1)} &=& 2C_F\left(4 \log\frac{-s}{\mu^2}-6\right)\nn
\gamma_S^{(1)} &=& \left[ \begin{array}{cc} 
2 C_d \log \frac{t}{u}+2 C_A \log \frac{ut}{s^2}
& 8 \log \frac{t}{u} \\[5pt]
 8C_1 \log \frac{t}{u} & 0
\end{array}\right] 
\label{39}
\end{eqnarray}
or using the notation defined in Eq.~(\ref{abbrev}),
\begin{eqnarray}
\gamma^{(1)} &=& \widetilde{\gamma}^{(1)}\, \openone +\gamma_S^{(1)}\nn
\widetilde{\gamma}^{(1)} &=& 2C_F\left(4 \Ls-6\right)\nn
\gamma_S^{(1)} &=& \left[ \begin{array}{cc} 
2 C_d \Ltu +2 C_A \Luts
& 8 \Ltu \\[5pt]
 8C_1 \Ltu & 0
\end{array}\right] \,.
\label{39c}
\end{eqnarray}
All logarithms of negative argument are defined by the branch $\log (-s-i0^+)$, and $\log(ut/s^2)\equiv \log(-u -i0^+)+\log(-u-i0^+)-2\log(-s-i0^+)$, etc.\ as discussed earlier.   The off-diagonal terms vanish at $t=u$, i.e.\ when the center-of-mass scattering angle is $\pi/2$. $\gamma_S$  is  called the soft anomalous dimension. We will see explicitly that the soft anomalous dimension and the soft-matching $R_S$ are universal, and independent of the external states, i.e. they are the same for fermions and scalars, and independent of the particle masses. In our computation using the analytic regulator, $\gamma_S$ arises from \emph{collinear} graphs; the ultrasoft graphs all vanish. 

The anomalous dimension $\widetilde \gamma$ is twice the anomalous dimension for the Sudakov form-factor, and contains a $\log (-s/\mu^2)$ term which produces double logs in the amplitude on integration. The soft anomalous dimension $\gamma_S$ does not contain any parameterically large logarithms, since $s, t, u$ are all formally of order $Q^2$.

The matching matrix is given by replacing $\mathcal{S}$ by the matching $D^{(1)}$ for bifermion operators  in Table~I of \ptwo\, $\mathcal{S} \to -\lM^2+2\lM \lQ -3\lM+9/2-5\pi^2/6$,
\begin{eqnarray}
R^{(1)} &=& \widetilde{R}^{(1)}\,\openone+R_S^{(1)}\nn
\widetilde{R}^{(1)} &=&2 C_F \Bigl[   -\lM^2+2\lM\Ls- 3\lM+\frac92-\frac{5\pi^2}{6}\Bigr] \nonumber\\[10pt]
R_S^{(1)} &=& \lM \left[ \begin{array}{cc} 
 C_d  \Ltu +C_A  \Luts 
&  4 \Ltu \\[10pt]
4 C_1 \Ltu & 0 \end{array}\right]\,.
\label{38}
\end{eqnarray}
Note that there is a non-trivial low-scale matching correction. At $\mu=M$, $\lM=0$, and 
$R_S^{(1)}$ vanishes. This is an accident in the toy model at one-loop. In the standard model, $R_S^{(1)}$ does not vanish, and has terms of the form $\log M_W^2/M_Z^2$.

This completes the computation for quark production. The matching and anomalous dimensions are combined to give the final amplitude in the usual way, and give the exponentiated SCET form for the Sudakov logarithms discussed in \ptwo. The matrices $\gamma_S$ and $R_S$ are universal, and have the same values for heavy quark production and for squark production, as we see explicitly below.

\subsection{Light Quark Scattering}
\label{ssec:pair_production}

\begin{figure}
\begin{center}
\includegraphics[width=3cm]{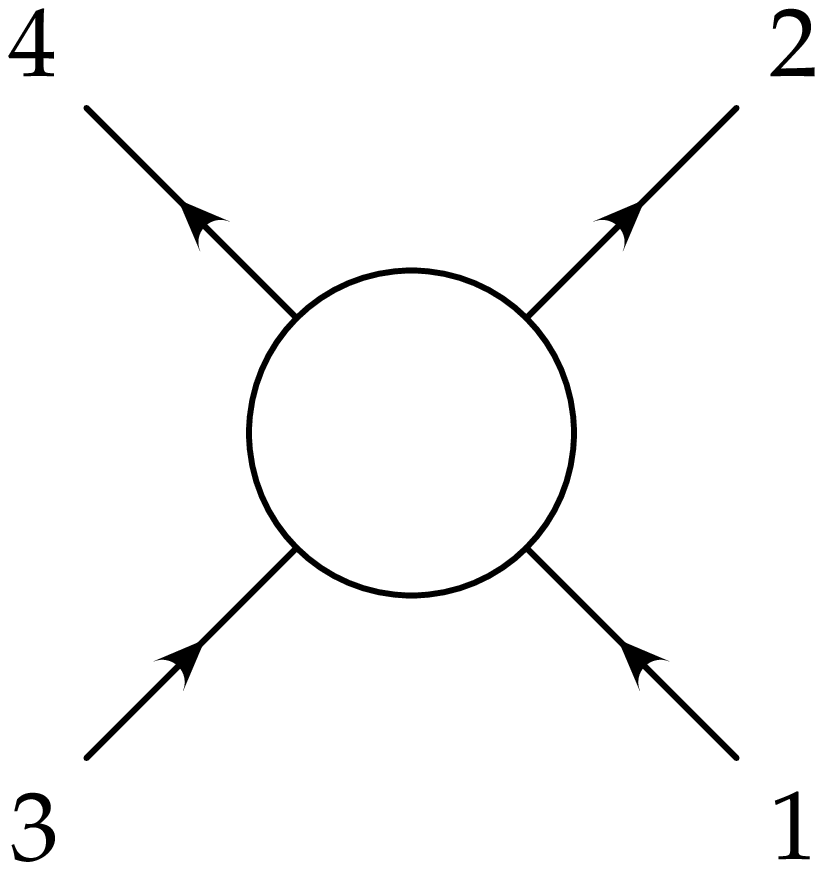}
\end{center}
\caption{\label{FgA0} Quark scattering $q(p_1)+q^\prime(p_3) \to q(p_2) + q^\prime(p_4)$. Time runs vertically.}
\end{figure}

The next process we consider is light quark production, $q(p_1)+q(p_3) \to q^\prime(p_2) + q^\prime(p_4)$ (see Fig.~\ref{FgA0}) with $q\not=q^\prime$, which is related to quark scattering in Fig.~\ref{FgC1} by crossing symmetry, with the replacements $p_2 \to -p_2$, $p_3 \to -p_3$. The Mandelstam variables for quark scattering are $s=(p_1+p_3)^2$, $t=(p_2-p_1)^2$ and $u=(p_4-p_1)^2$, so the amplitudes are obtained from those in the previous section by the replacement $s \to t$, $t \to u$, $u \to s$. Identical flavors are discussed in Sec.~\ref{app:box}.

The anomalous dimension matrix is
\begin{eqnarray}
\gamma^{(1)} &=& \widetilde{\gamma}^{(1)}\, \openone +\gamma_S^{(1)}\nn
\widetilde{\gamma}^{(1)} &=& 2C_F\left(4 \Lt-6\right)\nn
\gamma_S^{(1)} &=& \left[ \begin{array}{cc} 
2 C_d \Lus+2 C_A \Lust & 8 \Lus \\[5pt]
 8C_1\Lus & 0
\end{array}\right] 
\label{39a}
\end{eqnarray}
and the matching matrix is
\begin{eqnarray}
R^{(1)} &=& \widetilde{R}^{(1)}\, \openone +R_S^{(1)}\nn
\widetilde{R}^{(1)} &=&2 C_F \Bigl[  -\lM^2+2\lM \Lt - 3\lM+\frac92-\frac{5\pi^2}{6}\Bigr] \nn
R_S^{(1)} &=& \lM \left[ \begin{array}{cc} 
 C_d \Lus +C_A  \Lust
&  4\Lus \\[10pt]
4 C_1 \Lus & 0 \end{array}\right]\,.
\label{38a}
\end{eqnarray}

\subsection{Heavy Quark Production}
\label{sec:top_production}

Consider the annihilation of a light quark-antiquark pair to produce a heavy quark-antiquark pair, suggestively labeled $t\bar t$,  via the process $q(p_1) + \bar{q}(p_2) \to t(p_4) + \bar{t}(p_3)$.  The kinematics and Mandelstam variables are the same as Sec.~\ref{ssec:pair_production}; the only difference is that the final particles have mass $m$ which is not negligible compared with the gauge boson mass $M$, but is much smaller than $Q$, so that  $s,t,u \sim Q^2  \gg m^2,\, M^2$.

The first step is to match the full theory onto SCET at $\mu \sim Q$.  The fields $\xi_{n_3}$ and $\xi_{n_4}$ are now taken to have mass $m$~ \cite{ira2, llw}. The matching condition at $Q$ can be computed by from the full theory graphs with all scales much smaller than $Q$ set to zero, so the matching at $Q$ is the same as for the light quark case.

The second step is to run SCET operators in the effective theory from $Q$ to $m$.  The SCET anomalous dimension is independent of low mass scales and again gives the same result as in the massless case, Eq.~(\ref{39a}).

The third step is to switch at the scale $\mu \sim m$ to an effective theory where the heavy quarks are described by heavy quark effective theory (HQET) fields $t_{v_3}$ and $t_{v_4}$ \cite{book}. The four-fermi SCET operators of Eq.~(\ref{16}) are matched onto the SCET/HQET operators:
\begin{eqnarray}
\label{EqD1}
{\mathcal{O} }_1 \to {\mathcal{O}}'_1 &=&\bar{t}_{v_4} t^a \gamma^\mu P_{L \atop R}  t_{v_3}\, [\bar{\xi}_{n_2} W_{n_2}]  t^a \gamma_\mu P_{L \atop R} [W_{n_1}^\dagger \xi_{n_1}] \nn
{\mathcal{O}}_2 \to  {\mathcal{O} }'_2 &=&  \bar{t}_{v_4} \gamma^\mu  P_{L \atop R} t_{v_3}\,[\bar{\xi}_{n_2} W_{n_2}] \gamma_\mu P_{L \atop R}  [W_{n_1}^\dagger \xi_{n_1}] \,.
\end{eqnarray}
The HQET fields do not transform under a collinear gauge transformation; therefore, there is no factor analogous to the $W^\dagger_n$ Wilson line that goes along with $\xi_{n}$.  The heavy fields $t_{v_i}$ still couple to ultrasoft gauge bosons. 

The matching condition at $\mu \sim m$ is given by computing the difference of the graphs in the theory where particles 3 and 4 are described by SCET fields, and the same graphs computed when the two particles are described by HQET fields. Particles 1 and 2 continue to be described by SCET fields. The group theory and kinematic factors for each pair of particles remain unchanged as we switch from SCET to HQET, so the matching condition is given by Eqs.~(\ref{34}) with each $\Gamma$ being replaced by the difference of the corresponding graph in the two theories. Thus one can use
\begin{eqnarray}
\mathcal{S}_{12} &\to& 0\nn
\mathcal{S}_{34} &\to& R_{hh} \nn
\mathcal{S}_{ij}(x) &\to& R_{hl} \qquad ij=13,14,23,24
\label{42}
\end{eqnarray}
where $R_{hh}$ is the matrix element for the Sudakov form-factor in going from two SCET to two HQET fields, and $R_{hl}$ is the matrix element for the transition from two SCET fields to one SCET and one HQET field, dropping any overall group theory factors. The matching coefficients can be read off from Eq.~(80) and Eq.~(85) in \ptwo\,
\begin{eqnarray}
\mathcal{S}_{12} &\to& 0\nn
\mathcal{S}_{34} &\to& R+T\nn
\mathcal{S}_{ij}(x) &\to& R \qquad ij=13,14,23,24\nn
R &=& T =\frac12 \lm^2-\frac12\lm+\frac{\pi^2}{12}+2
\label{43}
\end{eqnarray}
where we use the entries from the first rows of Tables~II and IV of \ptwo. Thus
\begin{eqnarray}
\mathcal{R}^{(1)} &=& \widetilde{\mathcal{R}}^{(1)}\,\openone+\mathcal{R}_S^{(1)}\nn
\widetilde{\mathcal{R}}^{(1)} &=& C_F \left(R+T\right)\nn
&=& C_F \left(\lm^2-\lm+\frac{\pi^2}{6}+4\right)\nn
\mathcal{R}_S^{(1)} &=& 0
\label{46}
\end{eqnarray}
using Eq.~(\ref{34}) for the matching. $\mathcal{R}_S^{(1)}$ vanishes since $r_1=2R-2R=0$ and $r_2=4R-2(R+T)=0$.

The anomalous dimension below $m$ is given by using Eq.~(\ref{34}) with the replacement analogous to Eq.~(\ref{43}) for the anomalous dimension,
\begin{eqnarray}
\mathcal{S}_{12}(-2p_1\cdot p_2) &\to& \gamma_{1}(-s)\nn
\mathcal{S}_{34}(-2p_3 \cdot p_4) &\to&\gamma_{3}(-s) \nn
\mathcal{S}_{13}(2p_1\cdot p_3) &\to& \gamma_{2}(-u)\nn
\mathcal{S}_{14}(2p_1\cdot p_4) &\to& \gamma_{2}(-t)\nn
\mathcal{S}_{23}(2p_2\cdot p_3) &\to& \gamma_{2}(-t)\nn
\mathcal{S}_{24}(2p_2\cdot p_4)&\to& \gamma_{2}(-u)
\label{43a}
\end{eqnarray}
where $\gamma_{1,2,3}(Q^2)$ are the entries from the first rows of Tables~I, II and IV of \ptwo. They are the anomalous dimensions for $ll$, $hl$ and $hh$ currents, respectively. The anomalous dimension matrix in the HQET/SCET theory is
\begin{eqnarray}
\gamma^{(1)} &=& \widetilde{\gamma}^{(1)}\,\openone+\gamma_S^{(1)}\nn
\widetilde{\gamma}^{(1)} &=& C_F\left(\gamma_{1}(-s)+\gamma_{3}(-s)\right)\nn
r_1 &\to& 2\gamma_{2}(-t)-2\gamma_{2}(-u)\nn
r_2 &\to&2\gamma_{2}(-t)+2\gamma_{2}(-u)-2\gamma_{1}(-s)-2\gamma_{3}(-s)\nn
\gamma_1(Q^2) &=& 4 \lQ-6\nn
\gamma_2(Q^2) &=& 4 \lQ-2 \lm-5\nn
\gamma_3(Q^2) &=& 4 \left[w r(w)-1\right]\nn
r(w) &=&\frac{ \log \left(w + \sqrt{w^2-1}\right)}{\sqrt{w^2-1}}\nn
w&=& 1+\frac{Q^2}{2m^2}
\label{48}
\end{eqnarray}
Since we are working in the limit $Q^2 \gg m^2$, $w r(w) - 1\to \log(2w) -1 \to
\log(Q^2/m^2) -1$ up to power corrections. This gives
\begin{eqnarray}
\gamma^{(1)} &=& \widetilde{\gamma}^{(1)}\,\openone+\gamma_S^{(1)}\nn
\widetilde{\gamma}^{(1)} &=& C_F\left(8 \Ls -4\lm -10\right)\nn
\gamma_S^{(1)} &=& \left[ \begin{array}{cc} 
2 C_d  \Ltu +2 C_A \Luts
& 8  \Ltu \\[5pt]
8 C_1  \Ltu & 0
\end{array}\right] \,.
\label{48a}
\end{eqnarray}

The last step is to integrate out the gauge boson at $\mu \sim M$ and transition to the theory with no gauge bosons. The matching is given by Eq.~(\ref{34}) where $\mathcal{S}_{ij}$ are replaced by the corresponding results for the Sudakov form-factor matching,
\begin{eqnarray}
\mathcal{S}_{12}(-2p_1\cdot p_2) &\to& D(-s)\nn
\mathcal{S}_{34}(-2p_3 \cdot p_4) &\to& U(-s)\nn
\mathcal{S}_{13}(2p_1\cdot p_3) &\to& S(-u)\nn
\mathcal{S}_{14}(2p_1\cdot p_4) &\to& S(-t)\nn
\mathcal{S}_{23}(2p_2\cdot p_3) &\to& S(-t)\nn
\mathcal{S}_{24}(2p_2\cdot p_4) &\to& S(-u)
\label{43b}
\end{eqnarray}
where $D,S,U$ are given in the first rows of Tables~I, II and IV, respectively, of \ptwo. The matching is
\begin{eqnarray}
R^{(1)} &=& \widetilde{R}^{(1)}\,\openone+R_S^{(1)}\nn
\widetilde{R}^{(1)} &=& C_F\left(D(-s)+U(-s)\right)\nn
r_1 &\to& 2S(-t)-2S(-u)\nn
r_2 &\to&2S(-t)+2S(-u)-2D(-s)-2U(-s)
\label{50}
\end{eqnarray}
so that
\begin{eqnarray}
R^{(1)} &=& \widetilde{R}^{(1)}\,\openone+R_S^{(1)}\nn
\widetilde{R}^{(1)}  &=& C_F\Bigl(-\lM^2+4\lM \Ls -2\lM \lm \nn
&&-5\lM+\frac92-\frac{5\pi^2}{6}\Bigr)\nn
R_S^{(1)} &=& \lM \left[ \begin{array}{cc} 
C_d \Ltu +  C_A \Luts
& 4\Ltu \\[5pt]
4C_1 \Ltu & 0
\end{array}\right] \,.
\label{50a}
\end{eqnarray}

In summary, the computation proceeds as follows: (a) Match at $\mu \sim \sqrt{s}$ using Eq.~(\ref{16}), (\ref{17}) (b) Run between $\sqrt{s}$ and $m$ using Eq.~(\ref{39})
(c) Matching at $m$ using Eq.~(\ref{46}) (d) Run between $m$ and $M$ using Eq.~(\ref{48a}) (e) Match at $M$ using Eq.~(\ref{50a}).

If the fermion mass is not much larger than $M$, as is the case for the top-quark, one can replace (c), (d) and (e) by a single step, ($c^\prime$) Integrate out the fermion and gauge bosons simultaneously at $\mu \sim m \sim M$, as in Secs.~VIII~D,G of \ptwo. In this case, the matching is given by Eq.~(87), (91) of \ptwo:
\begin{eqnarray}
\mathcal{S}_{12}(-2p_1\cdot p_2) &\to& D(-s)\nn
\mathcal{S}_{34}(-2p_3 \cdot p_4) &\to& D(-s)+2 f_F(z)-h_F(z)\nn
\mathcal{S}_{13}(2p_1\cdot p_3) &\to& D(-u)+f_F(z)-h_F(z)/2\nn
\mathcal{S}_{14}(2p_1\cdot p_4) &\to& D(-t)+f_F(z)-h_F(z)/2\nn
\mathcal{S}_{23}(2p_2\cdot p_3) &\to& D(-t)+f_F(z)-h_F(z)/2\nn
\mathcal{S}_{24}(2p_2\cdot p_4) &\to& D(-u)+f_F(z)-h_F(z)/2 \nn
z &=& \frac{m^2}{M^2}
\label{43c}
\end{eqnarray}
where the functions $f_F$ and $h_F$ are given in Appendix~B of \ptwo. They are the change in the matching condition due to the quark mass. The matching matrix becomes
\begin{eqnarray}
R^{(1)} &=& \widetilde{R}^{(1)}\, \openone +R_S^{(1)}\nn
\widetilde{R}^{(1)} &=& C_F\left(2D(-s)+2 f_F(z)-h_F(z)\right)\nn
&=& 2C_F\Bigl(-\lM^2+2\lM\Ls -3\lM+\frac92\nn
&&-\frac{5\pi^2}{6} +f_F(z)-h_F(z)/2\Bigr)\nn
R_S^{(1)} &=& \lM \left[ \begin{array}{cc} 
C_d \Ltu +  C_A \Luts
& 4\Ltu \\[5pt]
4C_1\Ltu & 0
\end{array}\right]\label{51a}
\end{eqnarray}
The $f_F$ and $h_F$ terms cancel in $R_S$.

\subsection{Squark Pair Production}
\label{sec:squark_production}

\begin{figure}
\begin{center}
\includegraphics[width=2.5cm]{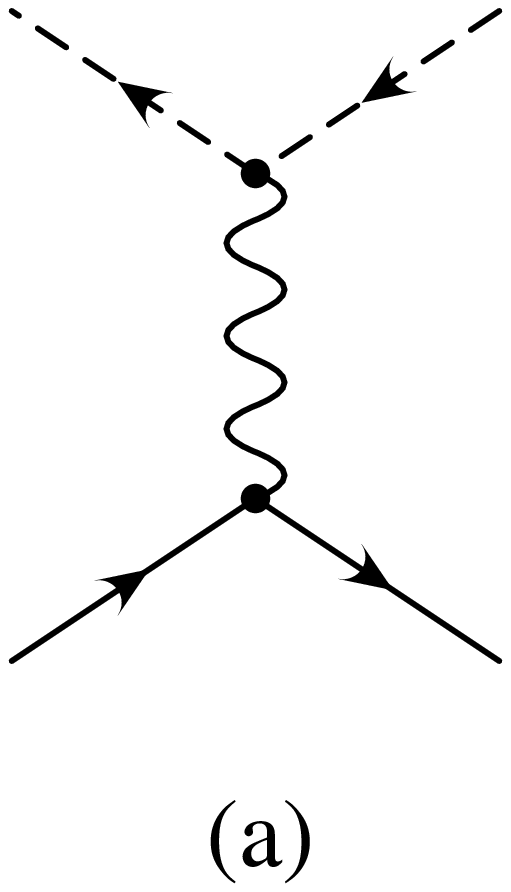}\hspace{2cm}
\includegraphics[width=2.5cm]{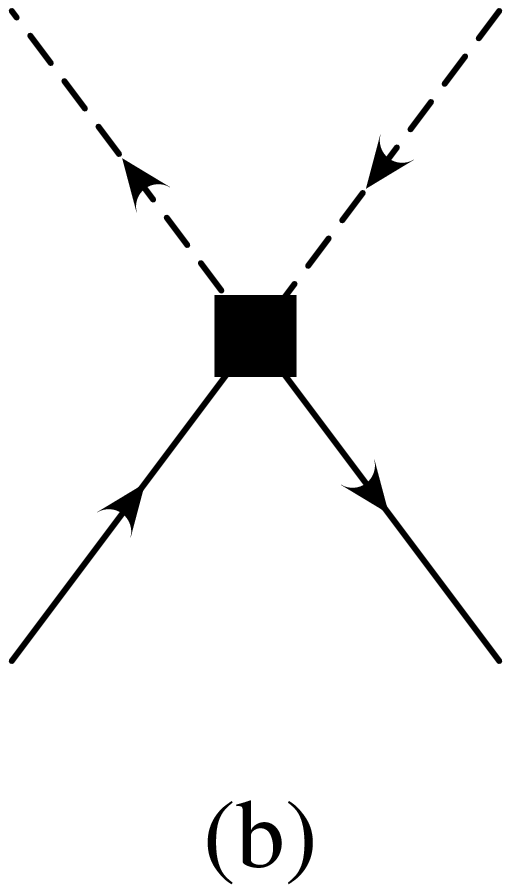} 
\end{center}
\caption{\label{FgE1} Tree level squark production in (a) the full theory  and (b) the effective theory.}
\end{figure}
As the final example, we consider heavy scalar (squark) pair production via $q(p_1) + \bar{q}(p_2) \to \tilde{t}(p_3) + \tilde{t}^*(p_4)$, where $\tilde t,\tilde t^*$ are the squark and anti-squark.  The squarks are taken to have mass $m \ll \sqrt s$.  
This example shows how one can use the scalar and scalar/fermion results in \ptwo\ to compute squark production. The discussion parallels that for heavy quark production in the previous section. The only difference is that since some of the particles are scalars, we need to use the $\bar{\psi} \phi$, $\phi^\dagger \psi $ and $\phi^\dagger \phi$ entries from the tables given in \ptwo. 

The first step is to match onto SCET at the scale $\mu \sim \sqrt{s}$.  The four-particle operators are 
\begin{eqnarray}
\label{EqE3}
{\mathcal{O}}_1 &=&   [ \Phi^\dagger_{n_4} W_{n_4}]  (i t^a \mathcal{D}_3+i\mathcal{D}_4 t^a)_\mu  [W_{n_3}^\dagger \Phi_{n_3}]  \nn
&&\times [\bar{\xi}_{n_2} W_{n_2}]  t^a \gamma^\mu [W_{n_1}^\dagger \xi_{n_1}]  \nn[5pt]
{\mathcal{O}}_2 &=& [\Phi^\dagger_{n_4} W_{n_4}](i\mathcal{D}_3+i\mathcal{D}_4)^\mu [W_{n_3}^\dagger \Phi_{n_3}]  \nn
&&\times [ \bar{\xi}^\dagger_{n_2} W_{n_2}]  \gamma^\mu [W_{n_1}^\dagger \xi_{n_1}] .
\end{eqnarray}
where $i\mathcal{D}_3 = \mathcal{P}+g( \bar n_3 \cdot A_{n_3 q})(n_3/2)$,  $i\mathcal{D}_4 = \mathcal{P}+g( \bar n_4 \cdot A_{n_4 q}) (n_4/2)$ are the label covariant derivatives on particles 3 and 4, respectively.

The tree-level coefficents are
\begin{eqnarray}
C^{(0)}_{1} &=& 4\pi\alpha/s\nn
C^{(0)}_{2} &=& 0
\end{eqnarray}
from the graph in Fig.~(\ref{FgE1}).
 
 The anomalous dimension in SCET below the scale $Q$ is given by using Eq.~(\ref{34}), and the values for the graphs in the region between $Q$ and $m$ given in \ptwo. For the anomalous dimension matrix, this means the replacements
\begin{eqnarray}
\mathcal{S}_{12}(-2p_1\cdot p_2) &\to& \gamma_{1\psi\psi}(-s)\nn
\mathcal{S}_{34}(-2p_3 \cdot p_4 &\to&\gamma_{1\phi\phi}(-s) \nn
\mathcal{S}_{13}(2p_1\cdot p_3) &\to& \gamma_{1\psi\phi}(-u)
\nn\mathcal{S}_{14}(2p_1\cdot p_4) &\to& \gamma_{1\psi\phi}(-t)\nn
\mathcal{S}_{23}(2p_2\cdot p_3) &\to& \gamma_{1\psi\phi}(-t)\nn
\mathcal{S}_{24}(2p_2\cdot p_4) &\to& \gamma_{1\psi\phi}(-u)\,.
\label{43d}
\end{eqnarray}
The anomalous dimensions are given in Table~I of \ptwo. The subscript $\psi \psi$, $\psi\phi$ and $\phi\phi$ means we use the anomalous dimension for bi-fermion operators, fermion-scalar, and bi-scalar operators respectively. The anomalous dimension is
\begin{eqnarray}
\gamma^{(1)} &=& \widetilde{\gamma}^{(1)}\, \openone +\gamma_S^{(1)}\nn
\widetilde{\gamma}^{(1)} &=& C_F\left(\gamma_{1\psi\psi}(-s)+\gamma_{1\phi\phi}(-s)\right)\nn
&=& 2C_F\left(4 \Ls  -7 \right)\nn
r_1 &\to& 2\gamma_{2\psi\phi}(-t)-2\gamma_{2\psi\phi}(-u)=8 \log \frac{t}{u}\nn
r_2 &\to&2\gamma_{2\psi\phi}(-t)+2\gamma_{2\psi\phi}(-u)-2\gamma_{1\psi\psi}(-s)-2\gamma_{1\phi\phi}(-s)\nn
&=& 8\log \frac{ut}{s^2}\nn
\gamma_S^{(1)} &=& \left[ \begin{array}{cc} 
2 C_d  \Ltu +2 C_A\Luts
& 8   \Ltu \\[5pt]
8 C_1    \Ltu & 0
\end{array}\right] .
\label{53}
\end{eqnarray}

After running the operators down to $\mu \sim m$ using Eq.~(\ref{53}), one matches to an effective theory in which the scalars are replaced by HQET fields. This is given by using Eq.~(\ref{42}), where the scalar values of  $R_{hh}$ and $R_{hl}$ are used. This means in Eq.~(\ref{43}), $R+T$ should be replaced by the bi-scalar value  on the second rows of Tables~III and IV, and for $\Gamma_{ij}$, $R$ should be replaced by $R_{\phi_2^\dagger \psi_1}$, the entry on the fourth row of Table~III for a bilinear with a heavy scalar and massless fermion:
\begin{eqnarray}
\mathcal{R}^{(1)} &=& \widetilde{\mathcal{R}}^{(1)}\, \openone +\mathcal{R}_S^{(1)}\nn
\widetilde{\mathcal{R}}^{(1)} &=& C_F \left(R_{\phi \phi}+T_{\phi\phi}\right)\nn
&=& C_F \left(\lm^2-2\lm+\frac{\pi^2}{6}+4\right)\nn
r_1 &=& 0\nn
r_2 &=& 4R_{\phi_2^\dagger \psi_1}-2R_{\phi \phi}-2T_{\phi\phi}=0\nn
\Rightarrow\mathcal{R}_S^{(1)} &=& 0\,.
\label{46s}
\end{eqnarray}

The running in the HQET/SCET theory below $m$, and the matching at $M$ is identical to Eq.~(\ref{48}) and Eq.~(\ref{50}) in the previous section, since  it does not matter whether the HQET field is a scalar or a fermion.

In summary, the computation proceeds as follows: (a) Match at $\mu \sim \sqrt{s}$ using Eq.~(\ref{EqE3}) (b) Run between $\sqrt{s}$ and $m$ using Eq.~(\ref{39})
(c) Matching at $m$ using Eq.~(\ref{46s}) (d) Run between $m$ and $M$ using Eq.~(\ref{48}) (e) Match at $M$ using Eq.~(\ref{50}).

If the squark mass is not much larger than $M$,  one can replace (c), (d) and (e) by a single step, ($c^\prime$) Integrate out the squark and gauge bosons simultaneously at $\mu \sim m \sim M$, as in Sec.~VIII~G,D of \ptwo. In this case, the matching is given by
\begin{eqnarray}
\mathcal{S}_{12}(-2p_1\cdot p_2) &\to& D_{\psi\psi}(-s)\nn
\mathcal{S}_{34}(-2p_3 \cdot p_4) &\to&D_{\phi\phi}(-s)+2 f_S(z)-h_S(z) \nn
\mathcal{S}_{13}(2p_1\cdot p_3) &\to& D_{\psi\phi}(-u)+ f_S(z)-h_S(z)/2 \nn
\mathcal{S}_{14}(2p_1\cdot p_4) &\to& D_{\psi\phi}(-t)+ f_S(z)-h_S(z)/2\nn
\mathcal{S}_{23}(2p_2\cdot p_3) &\to& D_{\psi\phi}(-t)+ f_S(z)-h_S(z)/2\nn
\mathcal{S}_{24}(2p_2\cdot p_4) &\to& D_{\psi\phi}(-u)+ f_S(z)-h_S(z)/2\nn
z &=& \frac{m^2}{M^2}
\label{43e}
\end{eqnarray}
where the functions $f_S$ and $h_S$ are given in Appendix~B of \ptwo. The matching matrix becomes
\begin{eqnarray}
R^{(1)} &=& \widetilde{R}^{(1)}\, \openone +R_S^{(1)}\nn
\widetilde{R}^{(1)} &=& C_F\left(D_{\psi\psi}(-s)+D_{\phi\phi}(-s)+2 f_S(z)-h_S(z)\right)\nn
&=& 2C_F\Bigl(-\lM^2+2\lM \Ls -\frac32\lM+4\nn
&&-\frac{5\pi^2}{6} +f_S(z)-\frac12h_S(z)\Bigr)\nn
r_1 &\to& 2D_{\phi\phi}(-t)-2D_{\phi\phi}(-u)=4\lM \log \frac{t}{u}\nn
r_2 &\to&2D_{\phi\phi}(-t)+2D_{\phi\phi}(-u)-4D_{\phi\phi}(-s)\nn
&=&4 \lM \log \frac{ut}{s^2}\nn
R_S^{(1)} &=& \lM \left[ \begin{array}{cc} 
C_d \Ltu +  C_A \Luts
& 4 \Ltu  \\[5pt]
4 C_1 \Ltu  & 0
\end{array}\right] \,.
\label{51b}
\end{eqnarray}

\subsection{Extension to more particles}
\label{sec:pairs}

In the previous examples, we saw that the four-particle $S$-matrix elements could be obtained by summing the two-particle $S$-matrix elements over all pairs of particles. This result can be generalized to \emph{gauge singlet} operators with an arbitrary of particles.

The SCET graphs do not depend on the Lorentz structure of the operators, the non-trivial dependence is on the gauge structure of the operators. We write the operators with all incoming fields. An outgoing particle can be represented as an incoming field in the complex conjugate representation. The incoming fields are combined into a net gauge singlet, and we have
\begin{equation}
\label{EqC1}
\left( \sum_{\alpha} T^a_\alpha \right) \mathcal{O}_i = 0
\end{equation}
where $T^a_\alpha$ acts on the indices $\mathcal{O}_i$ associated with field $\alpha$.
To make the notation clear: Assume $\psi$ and $\chi$ transform in the fundamental and anti-fundamental of $SU(N)$, and  $\mathcal{O}=\chi^i \psi_i$. The action of $T^a_\psi$ and $T^a_\chi$ on $\mathcal{O}$ are:
\begin{eqnarray}
\label{EqC3}
T^a_\psi \mathcal{O} &=& \chi^i \left(T^a\right)_{i}{}^j \psi_j\nn
T^a_\chi \mathcal{O} &=& \left(T^a\right)^i{}_j\chi^j \psi_i\,.
\end{eqnarray}
Here $\left(T^a\right)_{i}{}^j$ and $\left(T^a\right)^i{}_j$ are the representation matrices in the fundamental and anti-fundamental representations, so that
\begin{eqnarray}
 \left(T^a\right)^i{}_j = - \left(T^a\right)_{j}{}^i
\end{eqnarray}
from which it follows that
\begin{eqnarray}
\left(T^a_\psi +T^a_\chi \right)\mathcal{O} &=& 0.
\end{eqnarray}

The sum of graphs with gauge boson exchange between particles $\alpha$ and $\beta$, without any gauge factors, will be denoted by $\Gamma_{\alpha \beta}(2p_\alpha \cdot p_\beta)$, as in the preceeding section. The graph is computed with momentum $p_\alpha$ incoming, and $p_\beta$ outgoing. Treating all particles as incoming for both color and momentum flow means that the graph including color factors is $-\Gamma_{\alpha \beta}(-2 p_\alpha \cdot p_\beta ) T_\alpha^a T_\beta^a$. The minus sign of the argument takes care of the change in momentum labeling for $\beta$, and the overall minus sign is the charge conjugation minus sign from reversing the color flow of $\beta$.

The sum of  graphs including gauge factors is then
\begin{eqnarray}
\sum_{\vev{\alpha \beta}} - \Gamma_{\alpha \beta}(-2 p_\alpha \cdot p_\beta) 
\vev{T_\alpha^a T_\beta^a \mathcal{O}_i}^{(0)}
\end{eqnarray}
where we sum over all pairs ${\vev{\alpha \beta}}$, and $\vev{T_\alpha^a T_\beta^a \mathcal{O}_i}^{(0)}$ is the tree-level matrix element of the operator after the action of the gauge operators.

 The one loop contribution to the on-shell matrix element is
\begin{eqnarray}
\label{EqC4}
\langle \mathcal{O}_j\rangle^{(0)} R_{ji }&=& \sum_{\vev{\alpha \beta}} -\Gamma_{\alpha \beta}(-2  p_\alpha \cdot p_\beta) 
\vev{T_\alpha^a T_\beta^a \mathcal{O}_i}^{(0)}\nn
&&-\frac12 \sum_\alpha \delta Z^{-1}_\alpha \vev{T_\alpha^a T_\alpha^a \mathcal{O}_i}^{(0)}
\end{eqnarray}
including the wavefunction corrections for each external leg. This can be rewritten as
\begin{eqnarray}
&&\langle \mathcal{O}_j\rangle^{(0)} R_{ji }\nn
&=& \sum_{\vev{\alpha \beta}} - \left[\Gamma_{\alpha \beta}(-2  p_\alpha \cdot p_\beta)-\frac12\delta Z^{-1}_\alpha-\frac12 \delta Z^{-1}_\beta\right]
\vev{T_\alpha^a T_\beta^a \mathcal{O}_i}^{(0)}\nn
&&-\sum_{\vev{\alpha \beta}} \left[\frac12\delta Z^{-1}_\alpha+\frac12 \delta Z^{-1}_\beta\right]
\vev{T_\alpha^a T_\beta^a \mathcal{O}_i}^{(0)}\nn
&&-\frac12 \sum_\alpha \delta Z^{-1}_\alpha \vev{T_\alpha^a T_\alpha^a \mathcal{O}_i}^{(0)}\,.
\label{71h}
\end{eqnarray}
The first term in square brackets is the on-shell Sudakov form-factor for the two-particle case, including the wavefunction correction,
\begin{eqnarray}
\Gamma_{\alpha \beta}(-2  p_\alpha \cdot p_\beta)-\frac12\delta Z^{-1}_\alpha-\frac12 \delta Z^{-1}_\beta=\mathcal{S}_{\alpha \beta}(-2p_\alpha \cdot p_\beta)\,. \nn
\end{eqnarray}
We can simplify the remaining terms using
\begin{eqnarray}
0&=&\left( \sum_{\alpha} \delta Z^{-1}_\alpha T^a_\alpha \right)\left( \sum_{\beta} T^a_\beta \right) \mathcal{O}_i \nn
&=&  \sum_{\alpha,\beta} \delta Z^{-1}_\alpha T^a_\alpha T^a_\beta  \mathcal{O}_i \nn
&=&   \sum_{\vev{\alpha\beta}} \left[\delta Z^{-1}_\alpha+\delta Z^{-1}_\beta\right] T^a_\alpha T^a_\beta  \mathcal{O}_i
+  \sum_{\alpha } \delta Z^{-1}_\alpha T^a_\alpha T^a_\alpha  \mathcal{O}_i \nn
\label{79}
\end{eqnarray}
which follows from Eq.~(\ref{EqC1}), and reduces Eq.~(\ref{71h}) to
\begin{eqnarray}
\langle \mathcal{O}_j\rangle^{(0)} R_{ji }
&=& \sum_{\vev{\alpha \beta}} -\mathcal{S}_{\alpha \beta}(-2  p_\alpha \cdot p_\beta)
\vev{T_\alpha^a T_\beta^a \mathcal{O}_i}^{(0)}\,.\nn
\label{63}
\end{eqnarray}
The final answer can be written directly in terms of the on-shell two-particle matrix elements, as we found in the previous section for the four-particle case. Equation~(\ref{63}) is valid even without a summation on the gauge index $a$, and this will be useful in breaking up the electroweak corrections into the $W$, $Z$ and $\gamma$ contributions.

It is conventional to take the multiparticle scattering amplitude and divide it by the Sudakov form factors,
\begin{eqnarray}
A &\equiv& \sqrt{\prod_\alpha F_\alpha(Q^2)}\ A_S
\end{eqnarray} 
where $A$ is the scattering amplitude, and $F_r(Q^2)$ is the Sudakov form factor for particle $r$ at some reference momentum, e.g.\ $Q^2=-s$. $A_S$ is called the soft amplitude in the literature. With this definition, the soft amplitude has the form at one-loop
\begin{eqnarray}
A_S &=&  \sum_{\alpha} -\frac12\mathcal{S}_{\alpha \alpha}(Q^2)
\vev{T_\alpha^a T_\alpha^a \mathcal{O}_i}^{(0)} \nn
&&+\sum_{\vev{\alpha \beta}} -\mathcal{S}_{\alpha \beta}(-2  p_\alpha \cdot p_\beta)
\vev{T_\alpha^a T_\beta^a \mathcal{O}_i}^{(0)}\,,
\end{eqnarray}
since the one-loop Sudakov form factor for particle $\alpha$ is $T^a_\alpha T^a_\alpha \mathcal{S}_{\alpha \alpha}(Q^2)$. 

The Sudakov form factor has the form at one-loop (see the next section)
\begin{eqnarray}
\mathcal{S}_{\alpha \beta}(-2p_\alpha \cdot p_\beta ) &=& A \log
\frac{-2p_\alpha \cdot p_\beta}{\mu^2} + B_\alpha + B_\beta,
\label{s83}
\end{eqnarray}
where $A$ is a universal coefficient independent of particle type proportional to the cusp anomalous dimension which is known to be universal~\cite{korchemsky}, plus one-particle terms $B_\alpha$ which depend on the particle type, but are independent of $p_\alpha\cdot p_\beta$.

Using Eq.~(\ref{79}) with $\delta Z_\alpha^{-1} \to B_\alpha$, and with $\delta Z^{-1}_\alpha \to 1$ shows that the soft amplitude is given by a sum of the cusp part of the Sudkaov form factors, with coefficients which add up to zero, i.e.\ it can be written as differences of $A$-terms. The $B$ terms all cancel. We have seen this explicitly in Eq.~(\ref{34}). Thus the soft amplitude is universal, proportional to the cusp anomalous dimension, and formally has no large log terms since the differences of two $A$ terms gives a logarithm whose argument is order unity in the power counting, e.g.\ $\log t/u=\log (-t)-\log(-u)$.
This also implies that the soft anomalous dimension is proportional to the cusp anomalous dimension. While the above argument is at one-loop, we believe the general structure persists at higher loops. This property has been seen explicitly at two-loops in a very interesting recent computation~\cite{aybat}.

\section{Factorization}
\label{sec:fact}

There are strong constraints on the form of the scattering amplitude in SCET. We will discuss these in the context of the analytic regulator. The results hold for the $S$-matrix elements, and so are independent of any specific regulator. We have obtained the same results using a different regulator~\cite{hoang2}. We study the case where there is only a single amplitude to avoid problems with matrix ordering. This is the case, for example for scattering in a $U(1)$ gauge theory. If there are several gauge structures which can contribute, then the amplitude $A$ is a matrix, and one has to worry about matrix ordering. For example in $SU(N)$ gauge theory, there are two gauge invariant four-particle operators, $T^a \otimes T^a$ and $\mathbf{1} \otimes \mathbf{1}$, so $A$ is a $2 \times 2$ matrix. We briefly comment on the matrix ordering problem at the end of this section.

The $r$-particle scattering amplitudes are given by $n_i$-collinear sectors, $i=1,\ldots,r$ and the ultrasoft graphs. With the analytic regulator, the on-shell ultrasoft graphs vanish, and we only have to consider the collinear sectors.\footnote{With other regulators, the ultrasoft graphs can be non-zero.}

The $n_i$-collinear graphs have the form Fig.~\ref{fig:n1scet}, where particle $i$ is given by the SCET field, and all the other particles are Wilson lines.
The on-shell graph depends on the particle masses $\{m_k\}$, the renormalization scale $\mu$, and the analytic regulator parameters. The particle masses are the masses of any particles given by $n_i$-collinear SCET fields, such as the gauge boson masses, and the mass of particle $i$. They can also include the masses of other particles which couple to particle $i$. For example, in the standard model, a graph with a final $n_1$-collinear $t$-quark can depend on $m_t$ and $m_b$, since $n$-collinear $W$ bosons couple $t$ to $b$. The analytic regulator parameters for an $n_i$-collinear graph are $\nu_i^2$ from the SCET field, and $\nu_i^{(j)},\ j\not=i$ from the Wilson lines. Boost invariance requires $\nu_i^{(j)}$ to occur in the combination $\nu_i^{(j)}/(\bar n_i \cdot p_i)=\nu_j^2/(2 p_i \cdot p_j)$, as noted in Sec.~\ref{sec:toy theory}. The analytic regulator parameters only occur in logarithms, and we use the abbreviations
\begin{eqnarray}
\mathsf{L}_i &=& \log \nu_i^2,\nn
\mathsf{P}_{ij} &=& \log (2 p_i \cdot p_j)\qquad (i \not = j)\nn
\mathsf{P}_{ii} &\equiv& 0 \nn
\mathsf{P}_{ij} &\equiv& \mathsf{P}_{ji}\,.
\end{eqnarray}
The total $n_i$-collinear amplitude has the form
\begin{eqnarray}
\exp A_i(\left\{\mathsf{L}_j-\mathsf{P}_{ij}\right\},\{m_k\},\mu,\{\delta_k\})
\end{eqnarray}
$A_i$ depends on the momenta of all the particles in the process through its dependence on $2p_i \cdot p_j$ in $\mathsf{L}_{ij}$. The total connected amplitude $\exp A$ is given by the product of the different collinear sectors, so that
\begin{eqnarray}
A(\{p_k\},\{m_k\},\mu) &=& \sum_{i=1}^r A_i(\left\{\mathsf{L}_j-\mathsf{P}_{ij}\right\},\{m_k\},\mu,\{\delta_k\}).\nn
\label{f1}
\end{eqnarray}
The amplitudes $A_i$ and $A$ begin at order $\alpha$. The tree-level amplitude is the $1$ in the expansion of $\exp A$.

The individual terms $A_i$ depend on the regulator parameters $\{\nu_k\}$ and $\{\delta_k\}$ and are singular as $\{\delta_k\} \to 0$, as can be seen explicitly in the one-loop results given in I and \ptwo. However, the sum $A$ is finite as $\{\delta_k\} \to 0$ and independent of the analytic regulator parameters $\{\nu_k\}$. It can only depend on the particle masses (including internal particles), the external momenta, and $\mu$, as written in Eq.~(\ref{f1}).

The cancellation of the $\mathsf{L}_i$ dependence is a powerful constraint on the form of the SCET amplitudes. We showed in \ptwo\ how it implied that the low-scale matching $D$ when the massive gauge bosons were integrated out had to be at most linear in $\log Q^2$ to all orders in perturbation theory. In Eq.~(\ref{f1}), the right hand side depends on momenta only through the terms $\mathsf{P}_{ij}$, which occur only in the  combination $\mathsf{L}_j-\mathsf{P}_{ij}$ in $A_i$. The $\mathsf{L}_i$ cancellation implies that  $A_i$ and $A$ can be at most linear in $\mathsf{P}_{ij}$, to all orders in perturbation theory. The proof follows from a straightforward but tedious application of the principle of separation of variables used in partial differential equations --- if $f(x)+g(y)$ is a constant, and $x$ and $y$ are independent variables, then $f(x)$ and $g(y)$ must both be constant.

The linearity of $A$ in $\mathcal{P}_{ij}$ implies that the anomalous dimension and low-scale matching conditions $D$ are linear in $\mathcal{P}_{ij}$, since they are determined by the infinite and finite parts of $A$, respectively. The only multi-particle dependence of $A$ is through the $\mathcal{P}_{ij}$ dependence in the analytic regulator. Since $A$ is  linear in $\mathcal{P}_{ij}$, this leads to a two-particle dependence, plus one-particle terms, i.e. $A$ has the form Eq.~(\ref{s83}) to all orders. The $A$ term is universal; it cannot depend on the properties of the particles such as masses, because it is generated from Wilson line vertices which are independent of $m$. The $m$-dependence must be in one-particle contributions. 

If $A$ is a matrix, then the analysis becomes more complicated, but the general features discussed above continue to hold. The SCET anomalous dimension still contains only a single logarithm to all orders in perturbation theory~\cite{dis,Bauer:2003pi}. The amplitude Eq.~(\ref{13a}) is now a matrix equation, and the anomalous dimension integration is path-ordered in $\mu$. The $\lQ$ terms in the anomalous dimension $\gamma$ are proportional to the unit matrix $\mathbf{1}$, and can be pulled out as an overall multiplicative factor that commutes with the non-Abelian exponentiation of the integral of the rest of $\gamma$.

The high-scale matching need not be a square matrix, and one should replace $\exp C(Q) \to c(Q)$ in Eq.~(\ref{13a}). There are no large logs in either $C(Q)$ or $c(Q)$. The low-scale matching $D$ is also not a square matrix. It has the form
\begin{eqnarray}
d_0(\alpha(M)) e^{D_1(\alpha(M)) \log Q^2/M^2}
\label{mat:d}
\end{eqnarray}
where $d_0$ is a matrix, and $D_1$ is a number, i.e.\ it is proportional to the unit matrix. 
Thus our result that the low scale matching has the form $\exp D$, where $D$ has a single log to all orders in perturbation theory still holds in the matrix case, in the form Eq.~(\ref{mat:d}).

The structure of the amplitudes discussed in this and the previous section are a very powerful constraint. They follow from renormalization invariance of the effective theory and the universality of the cusp anomalous dimension. More extensive comments will be given elsewhere~\cite{hoang2}.

\section{Application to the Standard Model}
\label{sec:SM}

In this section, we apply the methods developed so far to compute radiative corrections in the standard model. There are several major differences between the toy theory and the standard model.  The standard model is a chiral theory and the couplings of the matter fields to the gauge fields are more complicated, with matter fields in several different representations of the gauge group. The gauge group is not simple and, we have to treat several different gauge interactions. After electroweak symmetry breaking,  there is electroweak gauge boson mixing between the $W_3$ and $B$, which gives $W$ and $Z$ bosons with different masses, and a massless photon. Finally, there are also Higgs exchange corrections proportional to the fermion mass, which are relevant for the top quark. It is straightforward to obtain the results for the standard model, following the same procedure used for the toy model. We have already shown in \ptwo\ how to obtain the Sudakov form-factor for the standard model including all these effects. In this section, we use the methods demonstrated in the previous section to calculate the radiative corrections to dijet, dilepton, top quark  and squark production in the standard model. These calculation are a non-trival example of the techniques developed in the previous sections and in  \ptwo. There are eighty independent amplitudes we need to compute, not including those related by crossing or flavor symmetry.

The left handed quarks and right handed quarks are in different representations of the unbroken gauge group of the standard model.  The left-handed quark doublets will be denoted by $Q^{(i)}_L$, where $i=u,c,t$ is a flavor index, the right handed charge $2/3$ quarks by $U^{(i)}_R$, the right-handed charge $-1/3$ quarks by $D^{(i)}_R$, the left-handed lepton doublets by  $L^{(i)}_L$ and the right-handed lepton singlets by  $E^{(i)}_R$.

Written in terms of $SU(2)$ components, $Q^{(i)}$ is
\begin{eqnarray}
Q^{(i)} = \left( \begin{array}{cc} U^{(i)}_L \\[5pt] D_L^{\prime(i)} \end{array}\right)=
\left( \begin{array}{cc} U^{(i)}_L \\[5pt] V_{ij} D_L^{(j)} \end{array}\right)
\end{eqnarray}
and  $L^{(i)}$ is
\begin{eqnarray}
L^{(i)}= \left( \begin{array}{cc} \nu^{(i)}_L \\[5pt] E_L^{(i)} \end{array}\right)
\end{eqnarray}
where the neutrinos are weak-eigenstates.

All the lepton and down-type quark masses can be neglected in our calculation, so we can work in the weak eigenstate basis, the CKM matrix $V$ does not enter the SCET computation, and generation number is conserved. The only place where $V$ enters is in the matrix element of SCET operators in the proton state, i.e.\ in computing the cross-section from the amplitude using the parton distribution functions, since these are given in terms of mass-eigenstate quarks.

\subsection{Matching at $Q$}

For scattering processes, we need to consider four-particle operators in \sceth\ generated at the scale $Q$ by the graphs in Fig.~\ref{FgO2} and Fig.~\ref{FgA1}. The SCET fields corresponding to a standard model field will be given by the replacement $Q^{(i)}
\to \xi_r(Q^{(i)})$, etc.\ We will drop the $n,p$ labels on $\xi$, and instead use the subscript $r=1,\ldots,4$ to denote the particle label in the scattering process. Thus $\xi_1(u)$ described collinear $u$ quarks with momentum $p_1$, etc.\

At the scale $Q$, the effective Lagrangian is the sum of terms representing the scattering of the various particles. If the initial and final particles are both quark doublets, then the Lagrangian is
\begin{eqnarray}
\mathcal{L}_{QQ} &=&
C_{QQ11,fi} \left[\bar \xi_4(Q^{(f)})W_{4} T^At^b \gamma_\mu W^\dagger_{3}\xi_3(Q^{(f)})\right]_L\, \nn
&&\times \left[\bar \xi_2(Q^{(i)})W_{2}T^A t^b\gamma^\mu W^\dagger_{1}\xi_1(Q^{(i)})\right]_L\nn
&& + C_{QQ12,fi}\left[\bar \xi_4(Q^{(f)})W_{4}t^a \gamma_\mu  W^\dagger_{3}\xi_3(Q^{(f)})\right]_L\,\nn
&&\times \left[\bar \xi_2(Q^{(i)})W_{2}t^a \gamma^\mu W^\dagger_{1}\xi_1(Q^{(i)})\right]_L \nn
&&+ C_{QQ21,fi}\left[\bar \xi_4(Q^{(f)})W_{4} T^A \gamma_\mu W^\dagger_{3}\xi_3(Q^{(f)})\right]_L\,\nn
&& \left[\bar \xi_2(Q^{(i)})W_{2}T^A \gamma^\mu W^\dagger_{1} \xi_1(Q^{(i)})\right]_L \nn
&&+C_{QQ22,fi}\left[\bar \xi_4(Q^{(f)})W_{4}\gamma_\mu  W^\dagger_{3}\xi_3(Q^{(f)})\right]_L\,\nn
&& \left[\bar \xi_2(Q^{(i)})W_{2} \gamma^\mu W^\dagger_{1}\xi_1(Q^{(i)})\right]_L \,.
\label{72}
\end{eqnarray}
We will write this in the abbreviated form
\begin{eqnarray}
\mathcal{L}_{QQ} &=& C_{QQ11,fi} [T^A t^a]_L \otimes [T^A t^a]_L +C_{QQ12,fi} [t^a]_L \otimes [t^a]_L\nn[5pt]
&& +C_{QQ21,fi} [T^A]_L \otimes [T^A]_L+C_{QQ22,fi} [\mathbf{1}]_L \otimes [\mathbf{1}]_L\,.\nn
\label{73}
\end{eqnarray}
The flavor quantum numbers are encoded in the subscripts on $C$. Recall that $T^A$ are the $SU(3)$ generators, and $t^a$ are the $SU(2)$ generators. The subscript $1$ 
is used for $T^A \otimes T^A$ or $t^a \otimes t^a$, and the subscript 2 for $\mathbf{1} \otimes \mathbf{1}$. Similarly, one has the other terms
\begin{eqnarray}
\mathcal{L}_{QU} &=& C_{QU1,fi} [T^A]_L \otimes [T^A]_R +C_{QU2,fi} [\mathbf{1}]_L \otimes [\mathbf{1}]_R\nn[5pt]
\mathcal{L}_{QD} &=& C_{QD1,fi} [T^A]_L \otimes [T^A]_R +C_{QD2,fi} [\mathbf{1}]_L \otimes [\mathbf{1}]_R\nn[5pt]
\mathcal{L}_{QL} &=& C_{QL1,fi} [t^a]_L \otimes [t^a]_L +C_{QL2,fi} [\mathbf{1}]_L \otimes [\mathbf{1}]_L\nn[5pt]
\mathcal{L}_{QE} &=& C_{QE,fi} [\mathbf{1}]_L \otimes [\mathbf{1}]_R\nn[5pt]
\mathcal{L}_{UQ} &=& C_{UQ1,fi} [T^A]_R \otimes [T^A]_L +C_{UQ2,fi} [\mathbf{1}]_R \otimes [\mathbf{1}]_L\nn[5pt]
\mathcal{L}_{UU} &=& C_{UU1,fi} [T^A]_R \otimes [T^A]_R +C_{UU2,fi} [\mathbf{1}]_R \otimes [\mathbf{1}]_R\nn[5pt]
\mathcal{L}_{UD} &=& C_{UD1,fi} [T^A]_R \otimes [T^A]_R +C_{UD2,fi} [\mathbf{1}]_R \otimes [\mathbf{1}]_R\nn[5pt]
\mathcal{L}_{UL} &=& C_{UL,fi} [\mathbf{1}]_R \otimes [\mathbf{1}]_L\nn[5pt]
\mathcal{L}_{UE} &=& C_{UE,fi} [\mathbf{1}]_R \otimes [\mathbf{1}]_R\nn[5pt]
\mathcal{L}_{DQ} &=& C_{DQ1,fi} [T^A]_R \otimes [T^A]_L +C_{DQ2,fi} [\mathbf{1}]_R \otimes [\mathbf{1}]_L\nn[5pt]
\mathcal{L}_{DU} &=& C_{DU1,fi} [T^A]_R \otimes [T^A]_R +C_{DU2,fi} [\mathbf{1}]_R \otimes [\mathbf{1}]_R\nn[5pt]
\mathcal{L}_{DD} &=& C_{DD1,fi} [T^A]_R \otimes [T^A]_R +C_{DD2,fi} [\mathbf{1}]_R \otimes [\mathbf{1}]_R\nn[5pt]
\mathcal{L}_{DL} &=& C_{DL,fi} [\mathbf{1}]_R \otimes [\mathbf{1}]_L\nn[5pt]
\mathcal{L}_{DE} &=& C_{DE,fi} [\mathbf{1}]_R \otimes [\mathbf{1}]_R\nn[5pt]
\mathcal{L}_{LQ} &=& C_{LQ1,fi} [t^a]_L \otimes [t^a]_L+C_{LQ2,fi} [\mathbf{1}]_L \otimes [\mathbf{1}]_L\nn[5pt]
\mathcal{L}_{LU} &=& C_{LU,fi} [\mathbf{1}]_L \otimes [\mathbf{1}]_R\nn[5pt]
\mathcal{L}_{LD} &=& C_{LD,fi} [\mathbf{1}]_L \otimes [\mathbf{1}]_R\nn[5pt]
\mathcal{L}_{LL} &=& C_{LL1,fi} [t^a]_L \otimes [t^a]_L+C_{LL2,fi} [\mathbf{1}]_L \otimes [\mathbf{1}]_L\nn[5pt]
\mathcal{L}_{LE} &=& C_{LE,fi} [\mathbf{1}]_L \otimes [\mathbf{1}]_R\nn[5pt]
\mathcal{L}_{EQ} &=& C_{EQ,fi} [\mathbf{1}]_R \otimes [\mathbf{1}]_L\nn[5pt]
\mathcal{L}_{EU} &=& C_{EU,fi} [\mathbf{1}]_R \otimes [\mathbf{1}]_R\nn[5pt]
\mathcal{L}_{ED} &=& C_{EU,fi} [\mathbf{1}]_R \otimes [\mathbf{1}]_R\nn[5pt]
\mathcal{L}_{EL} &=& C_{EL,fi} [\mathbf{1}]_R \otimes [\mathbf{1}]_R\nn[5pt]
\mathcal{L}_{EE} &=& C_{EE,fi} [\mathbf{1}]_R \otimes [\mathbf{1}]_R\,.
\label{74}
\end{eqnarray}

The tree-level matching coefficients from the graph in Fig.~\ref{FgO2} are ($f\not=i$)
\begin{eqnarray}
&&C^{(0)}_{QQ11, fi}  =   0   \nn 
&&s C^{(0)}_{QQ12, fi} =  {4\pi \alpha_2} \nn
&&sC^{(0)}_{QQ21, fi} =  {4\pi \alpha_3}  \nn
&&sC^{(0)}_{QQ22, fi}  =   {4\pi \alpha_1}  Y_{Q}^2  \nn
&&sC^{(0)}_{QU1,fi} =  {4\pi \alpha_3}\nn
&&sC^{(0)}_{QU2,fi}  =   {4\pi \alpha_1}Y_{Q} Y_U \nn
&&sC^{(0)}_{QL1,fi} =  {4\pi \alpha_2} \nn
&&sC^{(0)}_{QL2,fi} =  {4\pi \alpha_1}Y_Q Y_L \nn 
&&sC^{(0)}_{QE,fi} =    {4\pi \alpha_1}Y_Q Y_E \nn
&&sC^{(0)}_{UD1,fi} =  {4\pi \alpha_3} \nn
&&sC^{(0)}_{UD2,fi} =  {4\pi \alpha_1}Y_U Y_D \nn
&&sC^{(0)}_{UL,fi } =   {4\pi \alpha_1}Y_U Y_L \nn
&&sC^{(0)}_{UE,fi}  =   {4\pi \alpha_1}Y_U Y_E \nn
&&sC_{LL1, fi} =   {4\pi \alpha_2}\nn
&&sC^{(0)}_{LL2, fi} =   {4\pi \alpha_1}Y_L^2 \nn
&&sC^{(0)}_{LE,fi} =  {4\pi \alpha_1}Y_L Y_E \nn
&&sC^{(0)}_{EE, fi} =  {4\pi \alpha_1} Y_E^2
\end{eqnarray}
and the one loop matching coefficients are
\begin{widetext}
\begin{eqnarray}
&&sC^{(1)}_{QQ11,fi}  =   - 2\alpha_2\alpha_3 \tilde f(s,t)   \nn 
&&sC^{(1)}_{QQ12,fi} =  \alpha_2^2 \biggl[ X_2(s,t) -\frac{(C_{d_2}+C_{A_2})}{4}  \tilde f(s,t) \biggr] +  2\biggl[ \alpha_1 \alpha_2Y_{Q}^2  + \alpha_2 \alpha_3  C_{F_3} \biggr]  W  - 2\alpha_1 \alpha_2 Y_{Q}^2 \tilde f(s,t) \nn
&&sC^{(1)}_{QQ21,fi} =  \alpha_3^2 \biggl[ X_3(s,t) -\frac{(C_{d_3}+C_{A_3})}{4}  \tilde f(s,t) \biggr] + 2\biggl[ \alpha_1 \alpha_3 Y_{Q}^2  + \alpha_2 \alpha_3  C_{F_2} \biggr] W -2 \alpha_1 \alpha_3 Y_{Q}^2  \tilde f(s,t) \nn
&&sC^{(1)}_{QQ22,fi}  =   - \biggl[ \alpha_3^2 C_{1_3}  +  \alpha_2^2 C_{1_2} + \alpha_1^2 Y_Q^4 \biggr] \tilde f(s,t) + \alpha_1^2 Y^2_{Q} \Pi_1 + 2\biggl[ \alpha_1\alpha_3 Y_Q^2 C_{F_3}  +\alpha_1\alpha_2   Y_Q^2 C_{F_2} + \alpha_1^2Y^4_{Q}  \biggr]  W\nn
&&sC^{(1)}_{QU1,fi} =  \alpha_3^2 \biggl[ X_3(s,u) +\frac{(C_{d_3}-C_{A_3})}{4}  \tilde f(s,u) \biggr]  +  \biggl[ \alpha_1 \alpha_3 (Y_{Q}^2 +Y_{U}^2 ) + \alpha_2 \alpha_3  C_{F_2} \biggr] W + 2\alpha_1 \alpha_3 Y_{Q} Y_{U}  \tilde f(s,u) \nn
&&sC^{(1)}_{QU2,fi}  =    \biggl[   \alpha_3^2 C_{1_3}  +\alpha_1^2Y_{Q}^2 Y^2_{U}   \biggr]\tilde f(s,u)   + \alpha_1^2  Y_{U} Y_{Q} \Pi_1   +\biggl[ \alpha_1 \alpha_2  Y_{U} Y_{Q} C_{F_2}  + 2\alpha_1 \alpha_3 Y_{U} Y_{Q} C_{F_3} +\alpha_1^2 ( Y_{Q}^3 Y_{U}+ Y_{U}^3 Y_{Q} )   \biggr] W  \nn
&&sC^{(1)}_{QL1,fi} =  \alpha_2^2 \biggl[ X_2(s,t) -\frac{(C_{d_2}+C_{A_2})}{4}  \tilde f(s,t) \biggr]  + \biggl[ \alpha_2\alpha_3 C_{F_3} + \alpha_1\alpha_2(Y_Q^2 + Y_L^2) \biggl] W - 2\alpha_1\alpha_2 Y_L Y_Q  \tilde f(s,t) \nn
&&sC^{(1)}_{QL2,fi} =  - \biggl[ \alpha_2^2 C_{1_2}  +\alpha_1^2 Y_L^2Y_Q^2\biggr] \tilde f(s,t) + \alpha_1^2 Y_{L} Y_{Q}\Pi_1   +\biggl[ \alpha_1\alpha_3  Y_{L}Y_{Q} C_{F_3} +2\alpha_1\alpha_2 Y_{L}Y_{Q}C_{F_2}+ \alpha_1^2 ( Y_{L}^3 Y_{Q}+ Y_{Q}^3 Y_{L} )   \biggr]  W \nn
&&sC^{(1)}_{QE,fi} =    \alpha_1^2 Y_E^2Y_Q^2  \tilde f(s,u) + \alpha_1^2  Y_{E} Y_{Q} \Pi_1 + \biggl[ \alpha_1\alpha_3 Y_{E}Y_{Q}C_{F_3}  +\alpha_1\alpha_2 Y_{E}Y_{Q}C_{F_2} + \alpha_1^2 ( Y_{E}^3 Y_{Q}+ Y_{Q}^3 Y_{E} )   \biggr]  W\nn
&&sC^{(1)}_{UD1,fi} =  \alpha_3^2 \biggl[ X_3(s,t) -\frac{(C_{d_3}+C_{A_3})}{4}  \tilde f(s,t) \biggr]  - 2\alpha_1 \alpha_3 Y_{D} Y_{U} \tilde f(s,t) +  \alpha_1 \alpha_3( Y_{D}^2 +Y_{U}^2 )W \nn
&&sC^{(1)}_{UD2,fi} =   - \biggl[  \alpha_3^2 C_{1_3} +\alpha_1^2 Y^2_U Y^2_D  \biggr] \tilde f(s,t) + \alpha_1^2  Y_{U} Y_{D} \Pi_1   + \biggl[  2\alpha_1 \alpha_3 Y_{U} Y_{D}C_{F_3}  +\alpha_1^2 ( Y_{D}^3 Y_{U}+ Y_{U}^3 Y_{D} )   \biggr] W \nn
&&sC^{(1)}_{UL,fi } =    \alpha_1^2 Y_L^2Y_U^2 \tilde f(s,u) + \alpha_1^2 Y_{L} Y_{U} \Pi_1  + \biggl[ \alpha_1\alpha_3  Y_{L}Y_{U} C_{F_3}+\alpha_1\alpha_2 Y_{L}Y_{U} C_{F_2}+ \alpha_1^2 ( Y_{L}^3 Y_{U}+ Y_{U}^3 Y_{L} )   \biggr] W\nn
&&sC^{(1)}_{UE,fi} =   - \alpha_1^2 Y_E^2Y_U^2 \tilde f(s,t) + \alpha_1^2  Y_{E} Y_{U} \Pi_1  + \biggl[ \alpha_1\alpha_3 Y_{E}Y_{U} C_{F_3} + \alpha_1^2 ( Y_{E}^3 Y_{U}+ Y_{U}^3 Y_{E} )   \biggr] W \nn
&&sC_{LL1} =  \alpha_2^2 \biggl[ X_2(s,t) -\frac{(C_{d_2}+C_{A_2})}{4}  \tilde f(s,t) \biggr]   +   2\alpha_1 \alpha_2 Y_{L}^2  W - 2 \alpha_1 \alpha_2 Y_{L}^2  \tilde f(s,t) \nn
&&sC^{(1)}_{LL2} =  -\biggl[ \alpha_2^2 C_{1_2} + \alpha_1^2 Y_L^4  \biggr]\tilde f(s,t)  + \alpha_1^2  Y^2_{L} \Pi_1  + 2\biggl[ \alpha_1\alpha_2 Y_L^2 C_{F_2}  + \alpha_1^2 Y^4_{L}  \biggr] W \nn
&&sC^{(1)}_{LE,fi} =    \alpha_1^2 Y_E^2 Y_L^2  \tilde f(s,u)     +  \alpha_1^2  Y_{E}Y_{L}  \Pi_1  + \biggl[ \alpha_1\alpha_2  Y_L Y_E  C_{F_2}  + \alpha_1^2( Y^3_{L}Y_E + Y^3_E Y_L ) \biggr] W \nn
&&sC^{(1)}_{EE,fi} =  -  \alpha_1^2 Y_E^4 \tilde f(s,t) + \alpha_1^2  Y_{E}^2 \Pi_1 + 2\alpha_1^2 Y^4_{E}   W
\end{eqnarray}
where
\begin{eqnarray}
X_N(s,t) &=& 2C_{F_N} W + C_{A_N}  \biggl(2\Ls^2-2\LL_{-s-t} \Ls-\frac{11}{3} \Ls+\pi ^2+\frac{85}{9} \biggr) + \left(\frac23\Ls-\frac{10}{9}\right)T_{F_N} n_{F_N} + \left(\frac13\Ls-\frac{8}{9}\right)T_{F_N} n_{S_N}  \nn
W &=& -\Ls^2+3\Ls+\frac{\pi^2}{6}-8\nn
\Pi_1 &=&\frac{41}{6}\Ls - \frac{104}{9}\nn
\tilde f(s, t) &=&  -\frac{2s}{s+t} \LL_{t/s} + \frac{s(s+2t)}{(s+t)^2}\biggl(\LL^2_{t/s} + \pi^2 \biggr)+4 \Ls \LL_{t/(-s-t)} \,,
\end{eqnarray} 
\end{widetext}
for $N = 2, 3$ for $SU(2)$ and $SU(3)$, respectively.   $n_{F_N} (n_{S_N})$ denotes the number of Weyl fermions and complex scalars in the fundamental representation of $SU(N)$. The matching is symmetric between initial and final fermions, so that $C_{UQ,1}=C_{QU,1}$, etc. The coefficients $C_{QD,j}$ are given by $C_{QU,j}$ with $Y_U \to Y_D$, $C_{UU,j}$ by $C_{UD,j}$ with $Y_D \to Y_U$, $C_{DD,j}$ by $C_{UD,j}$ with $Y_U \to Y_D$, $C_{DL,j}$ by $C_{UL,j}$ with $Y_U \to Y_D$, and$C_{DE,j}$ by $C_{UE,j}$ with $Y_U \to Y_D$, and so have not been listed above. For identical particles (i.e.\ $C_{QQ,f=i}$,
$C_{LL,f=i}$, etc.) there is also the crossed-channel contribution as discussed in Appendix~\ref{app:box}.

The above matching coefficients do not include Higgs exchange contributions. The Yukawa couplings are proportional to the fermion masses, and the only Yukawa coupling large enough to be relevant is the top quark Yukawa coupling. Higgs corrections only arise at one-loop for LHC processes, since the initial state is $pp$, and contains no $t$-quarks.\footnote{One can always treat the proton as a hadron in QCD with all heavy flavors integrated out. Heavy quark distribution functions in the proton are calculable in terms of light-quark distribution functions; see e.g. Ref.~\cite{kaplan}.}  The Higgs contributions to the matching for operators containing $Q^{(t)}$ in the final state are
 \begin{eqnarray}
\delta C_{QQ12,ti}   &=&   \frac{g_t^2 \alpha_2}{4\pi s}  \biggl[ \frac{3}{2} - \frac{1}{2}\Ls \biggr] \nn
\delta C_{QQ21,ti}   &=&   \frac{g_t^2 \alpha_3}{4\pi s}    \biggl[\frac{1}{2}-\frac{1}{2}\Ls \biggr]   \nn
\delta C_{QQ22,ti}   &=&   \frac{g_t^2 \alpha_1}{4\pi s}  Y_{Q_u}  \biggl[ -\frac{5}{12}-\frac{1}{12}\Ls   \biggr]  \nn
\delta C_{QU1,ti}   &=&   \frac{g_t^2 \alpha_3}{4\pi s}    \biggl[\frac{1}{2}-\frac{1}{2}\Ls \biggr]   \nn
\delta C_{QU2,ti}   &=&   \frac{g_t^2 \alpha_1}{4\pi s}  Y_{U}  \biggl[ -\frac{5}{12}-\frac{1}{12}\Ls   \biggr]  \nn
\delta C_{QD1,ti}   &=&   \frac{g_t^2 \alpha_3}{4\pi s}    \biggl[\frac{1}{2}-\frac{1}{2}\Ls \biggr]   \nn
\delta C_{QD2,ti}   &=&   \frac{g_t^2 \alpha_1}{4\pi s}  Y_{D}  \biggl[ -\frac{5}{12}-\frac{1}{12}\Ls   \biggr]  \nn
\delta C_{QL1,ti}   &=&   \frac{g_t^2 \alpha_2}{4\pi s}    \biggl[\frac{3}{2}-\frac{1}{2}\Ls \biggr]   \nn
\delta C_{QL2,ti}   &=&   \frac{g_t^2 \alpha_1}{4\pi s}  Y_{L}  \biggl[ -\frac{5}{12}-\frac{1}{12}\Ls   \biggr]  \nn
\delta C_{QE,ti}   &=&   \frac{g_t^2 \alpha_1}{4\pi s}  Y_{E}  \biggl[ -\frac{5}{12}-\frac{1}{12}\Ls   \biggr],
\end{eqnarray}
whereas the contribution matching for operators containing $U^{(t)}$ in the final state are
 \begin{eqnarray}
 \delta C_{UQ1,ti}   &=&   \frac{g_t^2 \alpha_3}{4\pi s}    \biggl[1-\Ls \biggr]   \nn
\delta C_{UQ2,ti}   &=&   \frac{g_t^2 \alpha_1}{4\pi s}  Y_{Q}  \biggl[ \frac{4}{3}-\frac{1}{3}\Ls   \biggr]  \nn
\delta C_{UU1,ti}   &=&   \frac{g_t^2 \alpha_3}{4\pi s}    \biggl[1-\Ls \biggr]   \nn
\delta C_{UU2,ti} &=&   \frac{g_t^2 \alpha_1}{4\pi s}  Y_{U}  \biggl[ -\frac{4}{3}-\frac{1}{3}\Ls   \biggr]  \nn
\delta C_{UD1,ti}   &=&   \frac{g_t^2 \alpha_3}{4\pi s}    \biggl[1- \Ls \biggr]   \nn
\delta C_{UD2,ti}   &=&   \frac{g_t^2 \alpha_1}{4\pi s}  Y_{D}  \biggl[ \frac{4}{3}-\frac{1}{3}\Ls   \biggr]  \nn
\delta C_{UL,ti}   &=&   \frac{g_t^2 \alpha_1}{4\pi s}  Y_{L}  \biggl[ \frac{4}{3}-\frac{1}{3}\Ls   \biggr]  \nn
\delta C_{UE,ti}   &=&   \frac{g_t^2 \alpha_1}{4\pi s}  Y_{E}  \biggl[ \frac{4}{3}-\frac{1}{3}\Ls   \biggr],
\end{eqnarray}
with $i=u,c$ for $Q$ or $U$, and $i=d,s,b$ for $D$.  The logarithmic terms for $\delta C_{QE,ti}$ and $\delta C_{UE,ti}$ agree with Ref.~\cite{Melles:2000ia}.

Once we match onto SCET, Higgs vertex corrections are power suppressed, as shown in Ref.~\cite{cgkm2}, and the only Higgs contributions in SCET are wavefunction renormalization corrections.

\subsection{Anomalous Dimension below $Q$}

The anomalous dimensions in SCET between $Q$ and $m$ are obtained using the results of Sec.~\ref{ssec:qq_scattering}. The anomalous dimension due to gluon exchange depends on the color quantum numbers of the initial and final fermions. If both are color triplets, then the operators have the the color structure $C_1\, T^A \otimes T^A+C_2\, \mathbf{1} \otimes \mathbf{1}$. The anomalous dimension is given by Eq.~(\ref{39}) with group invariants replaced by their values for $N=3$, and with the $\alpha/(4\pi)$ prefactor for QCD,
\begin{eqnarray}
\gamma_{SU(3)} &=& \widetilde{\gamma}_{SU(3)}\, \openone +\gamma_{S,SU(3)}\nn[5pt]
\widetilde{\gamma}_{SU(3)} &=& \frac{8}{3}\frac{\alpha_3}{4\pi} \left(4 \Ls -6\right)\nn[5pt]
\gamma_{S,SU(3)} &=& \frac{\alpha_3}{4\pi} \left[ \begin{array}{cc} 
\frac{10}{3} \Ltu +6 \Luts
& 8 \Ltu \\[5pt]
 \frac{16}{9}\Ltu & 0
\end{array}\right]\,.
\label{77}
\end{eqnarray}
This $2\times 2$ anomalous dimension matrix acts on operators with color structure $T^A \otimes T^A$ and $\mathbf{1} \otimes \mathbf{1}$, and does not mix different flavors, chiralities or $SU(2)$ quantum numbers. Thus the renormalization group equation has the form
\begin{eqnarray}
\mu \frac{\rd}{\rd \mu}\left[ \begin{array}{cc} C_1 \\ C_2 \end{array}\right] &=& 
\gamma\left[ \begin{array}{cc} C_1 \\ C_2 \end{array}\right]
\label{78}
\end{eqnarray}
where $(C_1, C_2)$ are the pairs $(C_{QQ11,fi}, C_{QQ12,fi})$, $(C_{QQ21,fi}, C_{QQ22,fi})$, $(C_{QU1,fi}, C_{QU2,fi})$, $(C_{QD1,fi}, C_{QD2,fi})$, $(C_{UQ1,fi}, C_{UQ2,fi})$, $(C_{UU1,fi}, C_{UU2,fi})$, $(C_{UD1,fi}, C_{UD2,fi})$, $(C_{DQ1,fi}, C_{DQ2,fi})$,  $(C_{DU1,fi}, C_{DU2,fi})$, and $(C_{DD1,fi},  C_{DD2,fi})$.

If one of the fermions is a color triplet and the other is a color singlet, the operator has the color structure $C\, \mathbf{1} \otimes \mathbf{1}$. The QCD anomalous dimension 
for $C$ is identical to the Sudakov form-factor case,
\begin{eqnarray}
\gamma_{SU(3)} &=&\frac{4}{3}\frac{\alpha_3}{4\pi} \left(4 \Ls -6\right)\,.
\label{80}
\end{eqnarray}
If both fermions are color singlets, then
\begin{eqnarray}
\gamma_{SU(3)} &=&0\,.
\label{82}
\end{eqnarray}

The anomalous dimension due to $SU(2)$ gauge boson exchange is obtained similalrly. If both fermions are doublets, the operator has the form $C_1\, t^A \otimes t^A+C_2\, \mathbf{1} \otimes \mathbf{1}$, and the anomalous dimension matrix is
\begin{eqnarray}
\gamma_{SU(2)} &=& \widetilde{\gamma}_{SU(2)}\, \openone +\gamma_{S,SU(2)}\nn[5pt]
\widetilde{\gamma}_{SU(2)} &=& \frac{3}{2}\frac{\alpha_2}{4\pi} \left(4 \Ls -6\right)\nn[5pt]
\gamma_{S,SU(2)} &=&\frac{\alpha_2}{4\pi} \left[ \begin{array}{cc} 
4 \Luts
& 8 \Ltu \\[5pt]
\frac{3}{2} \Ltu & 0
\end{array}\right] 
\label{822}
\end{eqnarray}
where the $(C_1, C_2)$ pairs are $(C_{QQ11,fi}, C_{QQ21,fi})$, $(C_{QQ12,fi}, C_{QQ22,fi})$, $(C_{QL1,fi}, C_{QL2,fi})$, $(C_{LQ1,fi}, C_{LQ2,fi})$, and $(C_{LL1,fi}, C_{LL2,fi})$.

If one of the fermions is a weak doublet and the other is a weak singlet, the operator has the structure $C\, \mathbf{1} \otimes \mathbf{1}$, and the $SU(2)$ anomalous dimension  for $C$ is identical to the Sudakov form-factor case,
\begin{eqnarray}
\gamma_{SU(2)} &=&\frac{3}{4}\frac{\alpha_2}{4\pi} \left(4 \Ls -6\right)\,.
\end{eqnarray}
If both fermions are weak singlets, then
\begin{eqnarray}
\gamma_{SU(2)} &=&0\,.
\end{eqnarray}

$B$ exchange gives the diagonal contribution
\begin{eqnarray}
\label{EqF5}
\gamma_{U(1)} &=& \frac{\alpha_1}{4\pi}   \biggl[\left(Y_i^2+Y_f^2\right) \left(4\Ls - 6\right) +8 Y_i Y_f \Ltu \biggr] \nn
\end{eqnarray}
where $Y_i$ and $Y_f$ are the hypercharges of the initial and final representations. Note that $Y(U_R)=2/3$, $Y(D_R)=-1/3$ and $Y(E_R)=-1$.

The Higgs wavefunction graphs give the diagonal contribution
\begin{eqnarray}
\label{EqF8}
\gamma_{H}(Q^{(t)}) &=& \frac{g_t^2}{16\pi^2} \frac12.
\end{eqnarray}
to an operator for each $Q^{(t)}$ field, and
\begin{eqnarray}
\label{EqF8b}
\gamma_{H}(t_R) &=& \frac{g_t^2}{16\pi^2} 
\end{eqnarray}
for each $t_R=U_R^{(t)}$ field. This term breaks the flavor symmetry in the anomalous dimension.
The total anomalous dimension is the sum,
\begin{equation}
\label{EqF8a}
\gamma = \gamma_{H}+ \gamma_{U(1)} + \gamma_{SU(2)} + \gamma_{SU(3)},
\end{equation}
and is used to run the operators from $\mu = s$ to $\mu \sim M_Z$. 

\subsection{Matching at the low scale to \scetl}

At a low scale $\mu$ of order $M_Z$ (or $m_t$) one matches from \sceth\ with dynamical gluons and electroweak bosons onto \scetl\ with dynamical gluons and photons, by integrating out the $W$ and $Z$ bosons. The electroweak symmetry is broken in \scetl\ so the operators in Eq.~(\ref{74}) must now be decomposed into separate $SU(2)$ component fields.

We start by considering the case where all particles have mass much smaller than $m_t$, i.e.\ for all particles except the $t$-quark. This includes all the operators in Eq.~(\ref{74}) except those that contain $Q^{(t)}_L$ and $U^{(t)}_R$. The photon and gluon graphs are the same in \sceth\ and \scetl\ and do not contribute to the matching condition. The $W$ contribution depends on whether the particles invovled are $SU(2)$ doublets or singlets. For the case of two doublets, consider the operators
\begin{eqnarray}
C_{QQ12,fi} [t^a]_L \otimes [t^a]_L+C_{QQ22,fi} [\mathbf{1}]_L \otimes [\mathbf{1}]_L
\label{85}
\end{eqnarray}
with $i\not=t, f\not=t$, which are two of the terms in Eq.~(\ref{73}). For definiteness, let $f=c$ and $i=u$. These operators match onto a linear combination of
\begin{eqnarray}
\widehat{\mathcal{O}}_{12} &=& [\bar{c}_{L4}  \gamma_\mu c_{L3}] [\bar{u}_{L2}  \gamma^\mu u_{L1}]  \nn
\widehat{\mathcal{O}}_{22} &=&  [\bar{c}_{L4}   \gamma_\mu c_{L3}][\bar{d}_{L2}^\prime  \gamma^\mu d_{L1}^\prime] \nn
\widehat{\mathcal{O}}_{32} &=& [\bar{s}_{L4}^\prime   \gamma_\mu s_{L3}^\prime] [\bar{u}_{L2} \gamma^\mu  u_{L1}]  \nn
\widehat{\mathcal{O}}_{42} &=& [\bar{s}_{L4}^\prime   \gamma_\mu s_{L3}^\prime] [\bar{d}_{L2}^\prime \gamma^\mu  d_{L1}^\prime] \nn
\widehat{\mathcal{O}}_{52} &=&   [\bar{s}_{L4}^\prime   \gamma_\mu c_{L3}][\bar{u}_{L2} \gamma^\mu  d_{L1}^\prime]\nn
\widehat{\mathcal{O}}_{62} &=&  [\bar{c}_{L4}   \gamma_\mu s_{L3}^\prime][\bar{d}_{L2}^\prime\gamma^\mu   u_{L1}] 
\label{81}
\end{eqnarray}
where the flavor label represents the \scetl\ fields $W^\dagger \xi$.

The matching from Eq.~(\ref{85}) in \sceth\ onto Eq.~(\ref{81}) in \scetl\ is computed as in Sec.~\ref{ssec:pair_production}. As shown in Sec.~\ref{sec:pairs}, the matching can be written as the sum of the Sudakov form-factor $S$-matrix elements, even though we are considering $W^{\pm}$ exchange, $Z$ exchange and $\gamma$ exchange separately, and not  summing over all the $SU(2)$ gauge bosons. The matching matrix is
\begin{eqnarray}
\left[ \begin{array}{c}
\widehat C_{12} \\
\widehat C_{22} \\
\widehat C_{32} \\
\widehat C_{42} \\
\widehat C_{52} \\
\widehat C_{62} \\
\end{array}
\right] &=& R \left[ \begin{array}{c}
\widehat C_{QQ12,cu} \\[5pt]
\widehat C_{QQ22,cu} \end{array}\right]
\label{87}
\end{eqnarray}
where $\widehat C_{i2}$ are the coefficients of $\widehat{\mathcal{O}}_{i2}$. At tree-level
$R$ is
\begin{eqnarray}
R^{(0)} &=& \left[\begin{array}{rc} \frac14 &1 \\[5pt]
-\frac14 & 1\\[5pt]
 -\frac14 & 1 \\[5pt]
 \frac14 & 1 \\[5pt]
\frac12 & 0  \\[5pt]
\frac12 & 0  
\end{array}\right]\,.
\label{88}
\end{eqnarray}
At one-loop
\begin{eqnarray}
R^{(1)}_W &=& \frac{\alpha_{\text{em}}}{4\pi \sin^2\theta_W}\frac12\Biggl\{ 2 F_g(-s,M_W^2,\mu^2) R^{(0)}\nn
&&+ 2\log\frac{M_W^2}{\mu^2} \left[\begin{array}{cc} 
\log\frac{t}{s} & 0\\[5pt]
-\log\frac{u}{s} & 0 \\[5pt]
-\log\frac{u}{s} & 0 \\[5pt]
\log\frac{t}{s} & 0\\[5pt]
\frac12\log\frac{ut}{s^2} & 2\log\frac{t}{u} \\[5pt]
\frac12\log\frac{ut}{s^2} & 2\log\frac{t}{u} \\[5pt]
\end{array}\right]\Biggr\}
\label{89}
\end{eqnarray}
where
\begin{eqnarray}
F_g(Q^2,M^2,\mu^2) &=&  - \log^2 \frac{M^2}{\mu^2}+2  \log \frac{M^2}{\mu^2} \log \frac{Q^2}{\mu^2}\nn
&&- 3 \log \frac{M^2}{\mu^2} + \frac92-\frac{5\pi^2}{6}\,.
\label{90}
\end{eqnarray}

The $Z$ exchange contribution is
\begin{eqnarray}
R^{(1)}_Z &=&\frac{\alpha_{\text{em}}}{4\pi \sin^2\theta_W\cos^2\theta_W} \left[ \begin{array}{rc}
\frac14 r_1 & r_1 \\[5pt]
-\frac14 r_2 & r_2 \\[5pt]
-\frac14 r_3 & r_3 \\[5pt]
\frac14 r_4 & r_4 \\[5pt]
\frac12r_5 & 0 \\[5pt]
\frac12 r_5 & 0
\end{array}\right]
\label{95}
\end{eqnarray}

\begin{eqnarray}
r_1  &=& (g_{Lc}^2+g_{Lu}^2)  F_g(-s,M_Z^2,\mu^2)  + 4 g_{Lc}g_{Lu}\log\frac{M_Z^2}{\mu^2}\log\frac{t}{u} \nn
r_2 &=& (g_{Lc}^2+g_{Ld}^2)  F_g(-s,M_Z^2,\mu^2) + 4 g_{Lc} g_{Ld} \log\frac{M_Z^2}{\mu^2}\log\frac{t}{u} \nn
r_3 &=& (g_{Lu}^2+g_{Ls}^2) F_g(-s,M_Z^2,\mu^2) +4 g_{Lu}g_{Ls}\log\frac{M_Z^2}{\mu^2}\log\frac{t}{u} \nn
r_4  &=& (g_{Ld}^2+g_{Ls}^2) F_g(-s,M_Z^2,\mu^2) +4 g_{Ld}g_{Ls}\log\frac{M_Z^2}{\mu^2}\log\frac{t}{u} \nn
r_5 &=&\frac12\left(g_{Lc}^2+g_{Ls}^2+g_{Lu}^2+g_{Ld}^2\right) F_g(-s,M_Z^2,\mu^2)\nn
&&+2\left(g_{Lu}g_{Lc}+ g_{Ld}g_{Ls}\right) \log\frac{M_Z^2}{\mu^2}\log\frac{t}{s}\nn
&&-2\left(g_{Lu}g_{Ls}+ g_{Ld}g_{Lc}\right) \log\frac{M_Z^2}{\mu^2}\log\frac{u}{s}  
\label{96}
\end{eqnarray}
where
\begin{eqnarray}
g_{Lc}&=&g_{Lu}=\frac12-\frac23 \sin^2\theta_W \nn
g_{Ls}&=&g_{Ld}=-\frac12+\frac13 \sin^2\theta_W 
\end{eqnarray}
are the couplings to the $Z$. The total one-loop matching is $R_W^{(1)}+R^{(1)}_Z$.

The remaining two operators in $\mathcal{L}_{QQ}$,
\begin{eqnarray}
\mathcal{L}_{QQ} &=& C_{QQ11,fi} [T^A t^a]_L \otimes [T^A t^a]_L \nn
&& +C_{QQ21,fi} [T^A]_L \otimes [T^A]_L
\end{eqnarray}
match onto
\begin{eqnarray}
\widehat{\mathcal{O}}_{11} &=& [\bar{c}_{L4} T^A  \gamma_\mu c_{L3}] [\bar{u}_{L2} T^A \gamma^\mu u_{L1}]  \nn
\widehat{\mathcal{O}}_{21} &=&  [\bar{c}_{L4}T^A   \gamma_\mu c_{L3}][\bar{d}_{L2}^\prime T^A \gamma^\mu d_{L1}^\prime] \nn
\widehat{\mathcal{O}}_{31} &=& [\bar{s}_{L4}^\prime T^A \gamma_\mu s_{L3}^\prime] [\bar{u}_{L2}  T^A\gamma^\mu  u_{L1}]  \nn
\widehat{\mathcal{O}}_{41} &=& [\bar{s}_{L4}^\prime T^A \gamma_\mu s_{L3}^\prime] [\bar{d}_{L2}^\prime  T^A\gamma^\mu  d_{L1}^\prime] \nn
\widehat{\mathcal{O}}_{51} &=&   [\bar{s}_{L4}^\prime T^A \gamma_\mu  c_{L3}][\bar{u}_{L2}T^A\gamma^\mu   d_{L1}^\prime]\nn
\widehat{\mathcal{O}}_{61} &=&  [\bar{c}_{L4}  T^A \gamma_\mu s_{L3}^\prime][\bar{d}_{L2}^\prime  T^A\gamma^\mu  u_{L1}] \,.
\label{81c}
\end{eqnarray}
Since $W$ exchange leaves the color indices unaffected, the matching matrix is identical to Eq.~(\ref{88}), (\ref{89}), (\ref{95}), and the matching relation is given by Eq.~(\ref{87}) with the replacement $\widehat C_{i2} \to \widehat C_{i1}$, $\widehat C_{QQi2,cu} \to \widehat C_{QQi1,cu}$. The results Eq.~(\ref{88}), (\ref{89}) hold for all cases where both fermions are doublets. If the final quark doublet is replaced by a lepton doublet, the coupling constants in Eq.~(\ref{96}) have the obvious replacement $g_{Lc} \to g_{L\nu}$, $g_{Ls} \to g_{Le}$ with
\begin{eqnarray}
g_{L\nu}&=&\frac12\nn
g_{L e}&=&-\frac12+ \sin^2\theta_W 
\end{eqnarray}
and similarly if the initial doublet is a lepton doublet, or both doublets are lepton doublets.

The second case is where one fermion is a doublet and the other is a singlet. As an example, consider $C_{QU2,fi} [\mathbf{1}]_L \otimes [\mathbf{1}]_R$ with $f=c$ and $i=u$. This matches onto a linear combination of
\begin{eqnarray}
\widehat{\mathcal{O}}_{12} &=& [\bar{c}_{L4}   \gamma_\mu c_{L3}] [\bar{u}_{R2} \gamma^\mu u_{R1}]  \nn
\widehat{\mathcal{O}}_{22} &=& [\bar{s}_{L4}^\prime  \gamma_\mu s_{L3}^\prime] [\bar{u}_{R2} \gamma^\mu  u_{R1}]  \,.
\label{97}
\end{eqnarray}
The matching matrix is
\begin{eqnarray}
\left[ \begin{array}{c}
\widehat C_{12} \\
\widehat C_{22} \\
\end{array}
\right] &=& R \, \widehat C_{QU2,cu} 
\label{98}
\end{eqnarray}
where $\widehat C_{i2}$ are the coefficients of $\widehat{\mathcal{O}}_{i2}$. At tree-level
$R$ is
\begin{eqnarray}
R^{(0)} &=& \left[\begin{array}{cc} 
 1\\
1 \\
\end{array}\right]\,.
\label{99}
\end{eqnarray}
At one-loop
\begin{eqnarray}
R^{(1)}_W &=& \frac{\alpha_{\text{em}}}{4\pi \sin^2\theta_W} \frac12 F_g(-s,M_W^2,\mu^2) R^{(0)}\nn
R^{(1)}_Z &=&\frac{\alpha_{\text{em}}}{4\pi \sin^2\theta_W\cos^2\theta_W}
\Biggl\{\nn
&&
F_g(-s,M_Z^2,\mu^2)\left[\begin{array}{cc} 
 g_{Lc}^2+g_{Ru}^2 \\[5pt]
g_{Ls}^2+g_{Ru}^2 
\end{array}\right]\nn
&& +4 \log\frac{M_Z^2}{\mu^2}\log\frac{t}{u} \left[\begin{array}{cc} 
 g_{Lc}g_{Ru} \\[5pt]
g_{Ls} g_{Ru}
\end{array}\right]\Biggr\} 
\label{100}
\end{eqnarray}
where $F_g$ is given in Eq.~(\ref{90}). The singlet fermion $Z$ couplings are
\begin{eqnarray}
g_{Ru}&=& -\frac23 \sin^2\theta_W \nn
g_{Rd}&=&\frac13\sin^2\theta_W \nn
g_{Re}&=& \sin^2\theta_W\,.
\end{eqnarray}

Equations~(\ref{98}), (\ref{99}), (\ref{100}) apply to all cases where one fermion is weak doublet, and the other is a weak singlet, with the obvious replacement of the $Z$ charges for lepton doublets. Since electroweak exchange does not affect the color indices, the same matching matrix applies, for example, to the transition from $C_{QU1,fi} [T^A]_L \otimes [T^A]_R$ to
\begin{eqnarray}
\widehat{\mathcal{O}}_{11} &=& [\bar{c}_{L4}  T^A \gamma_\mu c_{L3}] [\bar{u}_{R2} T^A \gamma^\mu u_{R1}]  \nn
\widehat{\mathcal{O}}_{21} &=& [\bar{s}_{L4}^\prime T^A \gamma_\mu s_{L3}^\prime] [\bar{u}_{R2} T^A \gamma^\mu  u_{R1}]  \, .
\label{105}
\end{eqnarray}

The last case is if both fermions are weak singlets---take $C_{UU2,fi} [\mathbf{1}]_R \otimes [\mathbf{1}]_R$ as an example with $f=c$ and $i=u$. The operator matches to
\begin{eqnarray}
\widehat{\mathcal{O}} &=& [\bar{c}_{R4}  \gamma_\mu c_{R3}] [\bar{u}_{R2}  \gamma^\mu u_{R1}]  \,.
\label{106}
\end{eqnarray}
The one-loop matching condition is $\widehat C = (1+R_W^{(1)}+R_Z^{(1)} )C_{UU2,cu}$ with
\begin{eqnarray}
R_W^{(1)} &=& 0 \nn
R_Z^{(1)}  &=& \frac{\alpha_{\text{em}}}{4\pi \sin^2\theta_W\cos^2\theta_W}
\Bigl\{ (g_{Rc}^2+g_{Ru}^2)  F_g(-s,M_Z^2,\mu^2)\nn
&& + 4 g_{Rc}g_{Ru}\log\frac{M_Z^2}{\mu^2}\log\frac{t}{u}\Bigr\}\,.
\label{107}
\end{eqnarray}
Again, the same matching coefficient holds for the matching between $C_{UU1,fi} [T^A]_R \otimes [T^A]_R$ and $ [\bar{c}_{R4}T^A  \gamma_\mu c_{R3}] [\bar{u}_{R2}  T^A \gamma^\mu u_{R1}]  $, and the equations hold with an obvious substitution of $Z$ charges if the quarks are replaced by leptons.

\centerline{\it Anomalous dimensions in \scetl}
\medskip

Finally, one computes the anomalous dimension of the operator in \scetl\ between $\mu\sim M_Z$ and some low scale $\mu_0$, at which point one takes operator matrix elements to compute the desired observables. The matrix elements of the initial state SCET fields in the proton are the usual parton distribution functions, and the final state fields are used to construct jet observables. The scale $\mu_0$ is chosen to minimize logarithms in the matrix element computation. For LHC jet observables, it is of order the typical invariant mass of a single jet.

The QCD anomalous dimensions given in Eq.~(\ref{77}), (\ref{80}), (\ref{81}) continue to hold for the case of two quarks, one quark and one lepton, or two leptons, respectively. The QED anomalous dimension is
\begin{eqnarray}
\gamma_{\text{em}} &=& \frac{ \alpha_{\text{em}}}{4 \pi}\Bigl[\frac12\left(q_1^2+q_2^2+q_3^2+q_4^2\right)\left(4 \log \frac{-s}{\mu^2}-6\right)\nn
&&+4\left(q_1 q_4+q_2q_3\right) \log \frac{t}{s}-4\left(q_1 q_3+q_2q_4\right) \log \frac{u}{s}\Bigr]\nn
\end{eqnarray}
where $q_{1-4}$ are the charges of the four fields, and $q_1+q_3=q_2+q_4$. The initial particle charges are $q_1$ and $-q_2$, and the final particle charges are $-q_3$ and $q_4$.

\subsection{$t$-quark Production}

In processes involving the $t$-quark, $m_t^2/M_{W,Z}^2$ terms must be included in the loop graphs, as discussed in \ptwo. At a scale $\mu$, we transition to an effective theory in which the $t$-quark is represented by an HQET field, whereas the light quarks are still represented by SCET fields. Since $m_t$ is not much larger than $M_W$ and $M_Z$, it is convenient to make this transition at the same time that the $W$, $Z$ and Higgs bosons are integrated out of the theory in the transition from \sceth\ to \scetl. This method was used in \ptwo\ for the Sudakov form-factor of the $t$-quark, and allows one to include the complete $m_t^2/M_{W,Z}^2$ dependence in the matching computation. Here we apply the same procedure to the operators relevant for $t$-quark production---the operators in Eqs.~(\ref{73},\ref{74}) which contain either $Q^{(t)}_L$ or $t_R$ fields. The initial state in proton-proton collisions contains only light quarks, so we will only look at operators with top-quarks in the final state and light quarks in the initial state.

The matching at $Q$ and the anomalous dimension below $Q$ are mass independent, and identical to those for light quarks. The $m_t$ dependent terms give an additive correction to the low-scale matching matrices $R_{W,Z}^{(1)}$ of the previous section. There are also contributions  $R_{\gamma,g}^{(1)}$ to low-scale matching from the photon and gluon, because of the transition to an HQET field for the $t$-quark. The graphs in \sceth\ use a SCET field for the $t$-quark, and those in \scetl\ use a HQET field for the $t$-quark, so there is a matching correction even for massless gauge bosons, as computed in \ptwo.

The Higgs only contributes through wavefunction renormalization in SCET. The matching contribution from the Higgs is
\begin{eqnarray}
H(t_L) &=& -\frac12\frac{y_t^2}{16\pi^2}\bigl[\frac12 F_h(M_H^2,m_t^2)+\frac12 F_h(M_Z^2,m_t^2)\nn
&&+\frac12 \tilde a(h_t,h_t)+\frac12 \tilde a(z_t,z_t) + \tilde c(h_t,h_t)+\tilde c(z_t,z_t)\nn
&&  + \tilde c(w_t,0)-\tilde b(h_t,h_t)+\tilde b(z_t,z_t)\bigr] \nn
H(t_R) &=& H(t_L)-\frac12 \frac{y_t^2}{16\pi^2} \left( F_h(M_W^2,m_t^2)+ \tilde a(w_t,0) \right)\nn
H(b^\prime_L) &=& -\frac12\frac{y_t^2}{16\pi^2} \bigl[ F_h(M_W^2,m_t^2) + \tilde a(0,w_t)\bigr] ,\nn
F_h(M^2,\mu^2)&=&\frac14-\frac12\lM \nn
h_t &=& \frac{m_t^2}{M_H^2},\ w_t = \frac{m_t^2}{M_W^2},\ z_t = \frac{m_t^2}{M_Z^2}
\label{110}
\end{eqnarray}
where the functions are tabulated in Appendix~\ref{app:integrals}. For each $t_L$, $t_R$ or $b_L$ field, one adds $H(t_L)$, $H(t_R)$ or $H(b^\prime_L)$ to the matching matrix. For example the operator $\bar t_L \gamma^\mu t_L \bar b_L^\prime \gamma_\mu b_L^\prime$ gets the Higgs matching contribution $2 H(t_L) + 2 H(b^\prime_L)$.

The matching for operators containing $Q^{(t)}$ and a light quark doublet, Eq.~(\ref{85}) with $f=t$ and $i=u,c$, is given by Eq.~(\ref{89}), with the additional additive contribution
\begin{eqnarray}
\delta R_H^{(1)} &=& 
\left[\begin{array}{cc}
\frac 12 H(t_L) & 2 H(t_L) \\[5pt]
-\frac12 H(t_L) & 2 H(t_L) \\[5pt]
-\frac12 H(b^\prime_L) & 2 H(b^\prime_L) \\[5pt]
\frac12 H(b^\prime_L) & 2 H(b^\prime_L) \\[5pt]
\frac12H(t_L) + \frac12 H(b^\prime_L) & 0\\[5pt]
\frac12 H(t_L) + \frac12 H(b^\prime_L) & 0
\end{array}\right]
\end{eqnarray}
from the Higgs,
\begin{eqnarray}
\delta R_W^{(1)} &=& \frac{\alpha_{\text{em}}}{4\pi \sin^2\theta_W} \frac12
\left[\begin{array}{cc}
\frac 12 W_1 & 2 W_1 \\[5pt]
-\frac12 W_1 & 2 W_1 \\[5pt]
-\frac12 W_2 & 2 W_2\\[5pt]
\frac12 W_2 & 2 W_2\\[5pt]
\frac12 W_1 + \frac12 W_2 & 0\\[5pt]
\frac12 W_1 + \frac12 W_2 & 0
\end{array}\right]
\end{eqnarray}
from the $W$,
\begin{eqnarray}
\delta R^{(1)}_Z &=&\frac{\alpha_{\text{em}}}{4\pi \sin^2\theta_W\cos^2\theta_W} U_1 \left[ \begin{array}{rc}
\frac12  & 2  \\[5pt]
-\frac12 & 2  \\[5pt]
0 & 0 \\[5pt]
0 & 0 \\[5pt]
\frac12 & 0 \\[5pt]
\frac12  & 0
\end{array}\right]
\end{eqnarray}
from the $Z$, 
\begin{eqnarray}
\delta R^{(1)}_\gamma &=&\frac{\alpha_{\text{em}}}{4\pi } q_t^2
\left(\frac{\pi^2}{12}+2 \right)
\left[ \begin{array}{rc}
\frac12  & 2 \\[5pt]
-\frac12 & 2  \\[5pt]
0 & 0 \\[5pt]
0 & 0 \\[5pt]
\frac12& 0 \\[5pt]
\frac12  & 0
\end{array}\right]
\end{eqnarray}
from the photon, where $q_t=2/3$ is the $t$-quark charge, and
\begin{eqnarray}
\delta R^{(1)}_g &=&\frac{\alpha_s}{4\pi} \frac{4}{3} \left(\frac{\pi^2}{12}+2 \right)\left[ \begin{array}{rc}
\frac12 & 2  \\[5pt]
-\frac12 & 2 \\[5pt]
0 & 0 \\[5pt]
0 & 0 \\[5pt]
\frac12 & 0 \\[5pt]
\frac12  & 0
\end{array}\right]
\end{eqnarray}
from the gluon
where
\begin{eqnarray}
W_1 &=& f_F(w_t,0)-\frac12 a(w_t,0)-\frac12c(w_t,0)\nn
W_2 &=& f_F(0,w_t)-\frac12 a(0,w_t)\nn
U_1 &=& g_{Lt}^2 f_F(z_t,z_t)-\frac12g_{Lt}^2 a(z_t,z_t)\nn
&&-\frac12 \left(g_{Lt}^2+g_{Rt}^2\right) c(z_t,z_t)+g_{Lt} g_{Rt} b(z_t,z_t)\nn
U_2 &=& g_{Rt}^2 f_F(z_t,z_t)-\frac12g_{Rt}^2 a(z_t,z_t)\nn
&&-\frac12 \left(g_{Lt}^2+g_{Rt}^2\right) c(z_t,z_t)+g_{Lt} g_{Rt} b(z_t,z_t)\nn
\label{108}
\end{eqnarray}
and the functions $f_F$, $a$, $b$ and $c$ are tabulated in Appendix~\ref{app:integrals}. The matching matrix multiplied by $(C_{QQ1l,tq}, C_{QQ2l,tq})$ gives the coefficients ${\widehat C}_{kl}$, $k=1,\ldots,6$, $l=1,2$ of the operators in \scetl\ listed in Eq.~(\ref{81}), (\ref{81c}) with $c \to t$ and $s \to b^\prime$ for the final state quarks, and the initial state flavors replaced by the two members of the light quark doublet $q$, $(u,d^\prime)\to (c,s^\prime)$, or $(u,d^\prime)\to(u,d^\prime)$.
Note that $\mathcal{O}_{51},\mathcal{O}_{52},\mathcal{O}_{61},\mathcal{O}_{62}$ are relevant for single-top production.

The gluon matching $\delta R_g^{(1)}$ is diagonal in color space, and does not mix the $\mathbf{1}\otimes \mathbf{1}$ and $T^A \otimes T^A$ operators. This follows from Eq.~(\ref{34}) and the additive nature of the mass corrections to the amplitudes.

The matching for operators containing $Q^{(t)}$ and a light quark singlet are given by Eq.~(\ref{100}) with the additional terms
\begin{eqnarray}
\delta R_H^{(1)} &=& 
\left[\begin{array}{cc}
2 H(t_L) \\[5pt]
2 H(b^\prime_L) 
\end{array}\right]\nn
\delta R_W^{(1)} &=& \frac{\alpha_{\text{em}}}{4\pi \sin^2\theta_W} \frac12
\left[\begin{array}{cc}
2 W_1 \\[5pt]
2 W_2 
\end{array}\right]\nn
\delta R_Z^{(1)} &=& \frac{\alpha_{\text{em}}}{4\pi \sin^2\theta_W\cos^2\theta_W} 
\left[\begin{array}{cc}
2 U_1 \\[5pt]
0 
\end{array}\right]\nn
\delta R_\gamma^{(1)} &=& \frac{\alpha_{\text{em}}}{4\pi} q_t^2\left(\frac{\pi^2}{6}+4\right)
\left[\begin{array}{cc}
1\\[5pt]
0 
\end{array}\right]\nn
\delta R_g^{(1)} &=& \frac{\alpha_s}{4\pi } 
\frac43\left(\frac{\pi^2}{6}+4\right) \left[\begin{array}{cc}
1 \\[5pt]
0 
\end{array}\right]\,.
\end{eqnarray}
with $X_{1,2}$ and $U_1$ given in Eq.~(\ref{108}).

The matching for operators containg $t_R$ and a light quark doublet are given by Eq.~(\ref{100}) with the additional terms
\begin{eqnarray}
\delta R_H^{(1)} &=& 
\left[\begin{array}{cc}
2 H(t_R) \\[5pt]
2 H(t_R) 
\end{array}\right]\nn
\delta R_W^{(1)} &=& \frac{\alpha_{\text{em}}}{4\pi \sin^2\theta_W} \frac12
\left[\begin{array}{cc}
-c(w_t,0) \\[5pt]
-c(w_t,0) 
\end{array}\right]\nn
\delta R_Z^{(1)} &=& \frac{\alpha_{\text{em}}}{4\pi \sin^2\theta_W\cos^2\theta_W} 
\left[\begin{array}{cc}
2 U_2 \\[5pt]
2U_2 
\end{array}\right]\nn
\delta R_\gamma^{(1)} &=& \frac{\alpha_{\text{em}}}{4\pi} q_t^2\left(\frac{\pi^2}{6}+4\right)
\left[\begin{array}{cc}
1\\[5pt]
1 
\end{array}\right]\nn
\delta R_g^{(1)} &=& \frac{\alpha_s}{4\pi } 
\frac43\left(\frac{\pi^2}{6}+4\right) \left[\begin{array}{cc}
1 \\[5pt]
1 
\end{array}\right]\,.
\end{eqnarray}

The matching for operators containing $t_R$ and singlet light quarks is given by Eq.~(\ref{107}) with the additional terms
\begin{eqnarray}
\delta R_H^{(1)} &=& 2H(t_R)\nn
\delta R_W^{(1)} &=& \frac{\alpha_{\text{em}}}{4\pi \sin^2\theta_W} \frac12
\left( -c(w_t,0)\right)\nn
\delta R_Z^{(1)} &=&  \frac{\alpha_{\text{em}}}{4\pi \sin^2\theta_W\cos^2\theta_W} 2U_2\nn
\delta R_\gamma^{(1)} &=& \frac{\alpha_{\text{em}}}{4\pi} q_t^2\left(\frac{\pi^2}{6}+4\right) \nn
\delta R_g^{(1)} &=& \frac{\alpha_s}{4\pi } 
\frac43\left(\frac{\pi^2}{6}+4\right) \,.
\end{eqnarray}

\centerline{\it Anomalous Dimension in \scetl}
\medskip

The anomalous dimension in \scetl\ after integrating out the electroweak gauge bosons and switching to HQET for the top quarks, is given by gluon and photon exchange. For $t$ pair production, particles $3$ and $4$ are HQET $t$-quarks, and $\gamma$ can be obtained from Eq.~(\ref{34a}), using the heavy-heavy anomalous dimension ($\gamma_1$ of \ptwo) for exchange between $(3,4)$, the heavy-light anomalous dimension  ($\gamma_2$ of \ptwo) for exchange between $(3,4)$ and $(1,2)$ and the light-light anomalous dimension  ($\gamma_3$ of \ptwo) for exchange between $(1,2)$.
This gives Eq.~(\ref{48a}) for the QCD part of the anomalous dimension, with $\alpha \to \alpha_s$, and group theory factors replaced by their $SU(3)$ values,
\begin{eqnarray}
\gamma^{(1)} &=& \widetilde{\gamma}^{(1)}\,\openone+\gamma_S^{(1)}\nn[5pt]
\widetilde{\gamma}^{(1)} &=& \frac{\alpha_s}{4 \pi}\frac43 \left(8 \Ls-4\LL_{m_t} -10\right)\nn[5pt]
\gamma_S^{(1)} &=&\frac{\alpha_s}{4 \pi}\left[ \begin{array}{cc} 
\frac{10}{3}\Ltu +6 \Luts
& 8\Ltu  \\[5pt]
\frac{16}{9}\Ltu  & 0
\end{array}\right] \,.
\label{120}
\end{eqnarray}
The QED anomalous dimension is
\begin{eqnarray}
\gamma &=& \frac{\alpha_{\text{em}}}{4\pi} \Bigl[ 
q_l^2 \left(4 \log \frac{-s}{\mu^2}-6\right) + q_t^2 \left( 4 \log \frac{-s}{m_t^2}-4\right)\nn
&& + 8 q_l q_t \log \frac{t}{u} \Bigr]
\label{121}
\end{eqnarray}
where $q_t=2/3$ and $q_l=2/3,-1/3$ is the charge of the light quark.

For single-top production from the operators ${\widehat O}_{51}$, ${\widehat O}_{61}$, ${\widehat O}_{52}$, ${\widehat O}_{62}$, there is only one heavy quark in the final state, and
\begin{eqnarray}
\gamma &=& \widetilde{\gamma}^{(1)}\,\openone+\gamma_S^{(1)}\nn[5pt]
\widetilde{\gamma}^{(1)} 
&=& \frac{\alpha_s}{4 \pi}\frac43 \left(8 \Ls -11 -2 \log\frac{m_t^2}{\mu^2}\right)\nn[5pt]
\gamma_S^{(1)} &=&\frac{\alpha_s}{4 \pi} \left[ \begin{array}{cc} 
\frac{10}{3}\Ltu +6\Luts
& 8\Ltu  \\[5pt]
\frac{16}{9}\Ltu & 0
\end{array}\right]
\end{eqnarray}
and the QED anomalous dimension
\begin{eqnarray}
\gamma &=& \frac{\alpha_{\text{em}}}{4\pi} \Bigl[ 
\left(q_u^2+q_d^2\right)\left(8 \log \frac{-s}{\mu^2} -11 -2 \log\frac{m_t^2}{\mu^2}\right)\nn
&&+ 8 q_u q_d \log \frac{s}{u} \Bigr]
\end{eqnarray}
where $q_u=2/3$ and $q_d=-1/3$ are the charges of the up-type and down-type quarks, respectively.

This completes the computation of radiative corrections in the standard model. The formul\ae\ derived in this section will be used for the numerical computations in the next section. The only case we have not treated is when both initial and final particles are top quarks. This can be obtained from the case we have analyzed, with a heavy quark in the final state, by also adding heavy quark corrections terms for the initial quark.

\subsection{$ gg \to q \bar q$, $gq \to gq$ and $g \bar q \to g \bar q$}
\label{ssec:glue}

\begin{figure}
\begin{center}
\includegraphics[width=3cm]{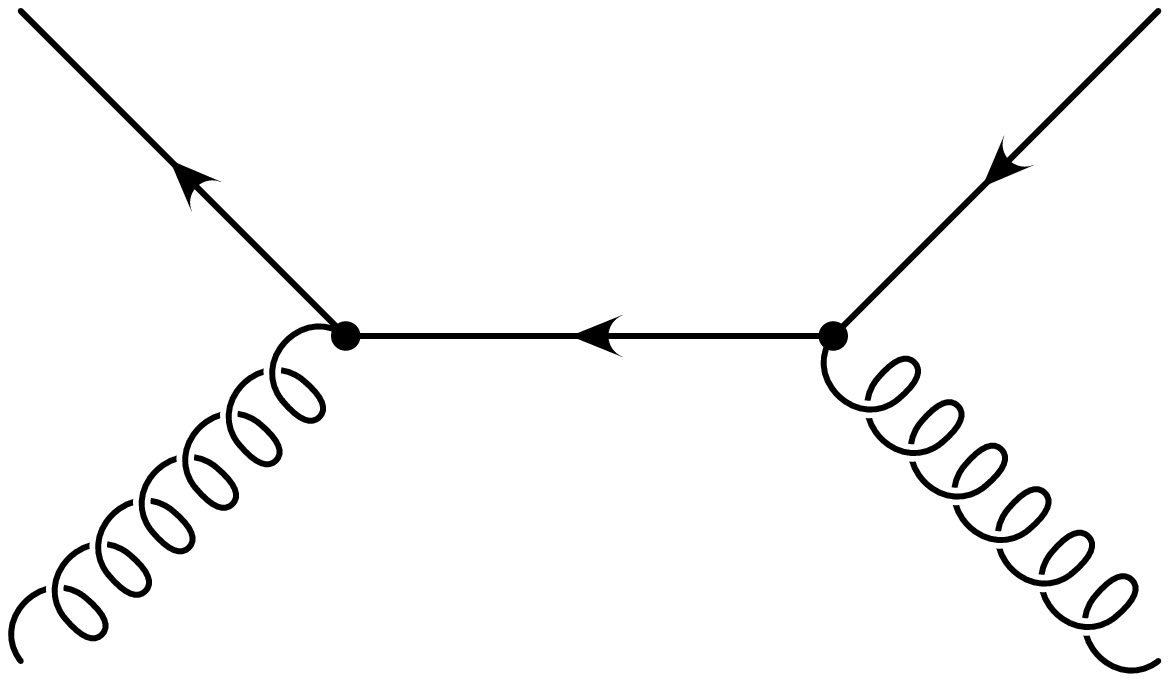}\hspace{1cm}
\includegraphics[width=3cm]{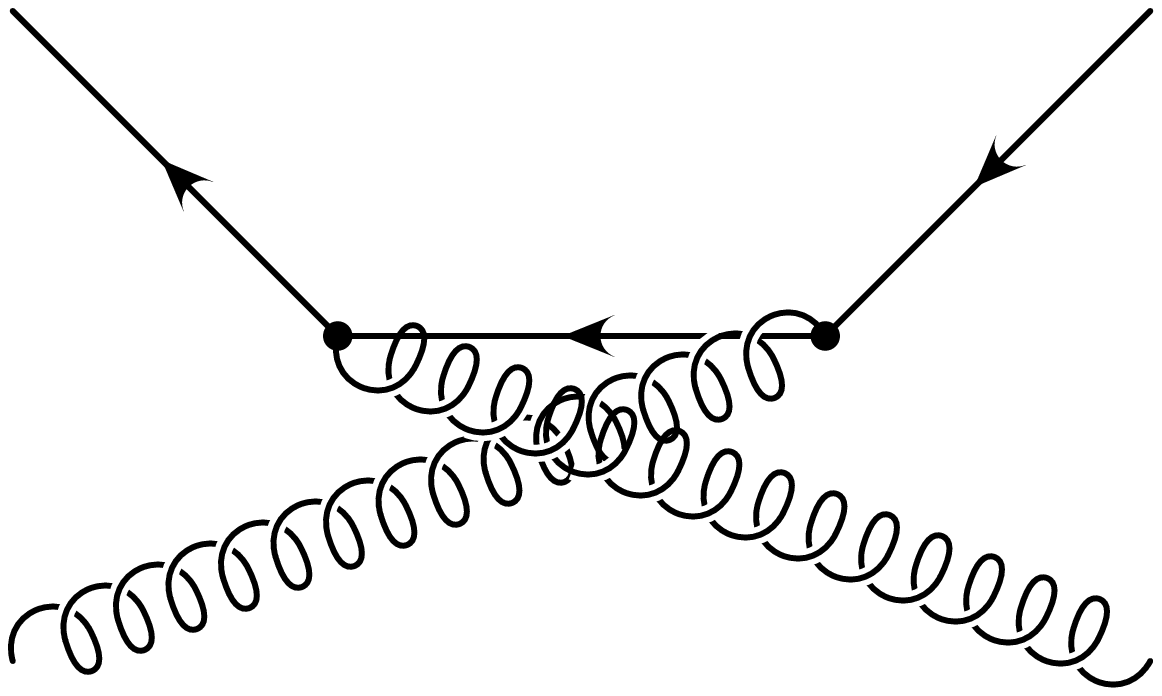}\\
\vspace{1.0cm}
\includegraphics[width=2cm]{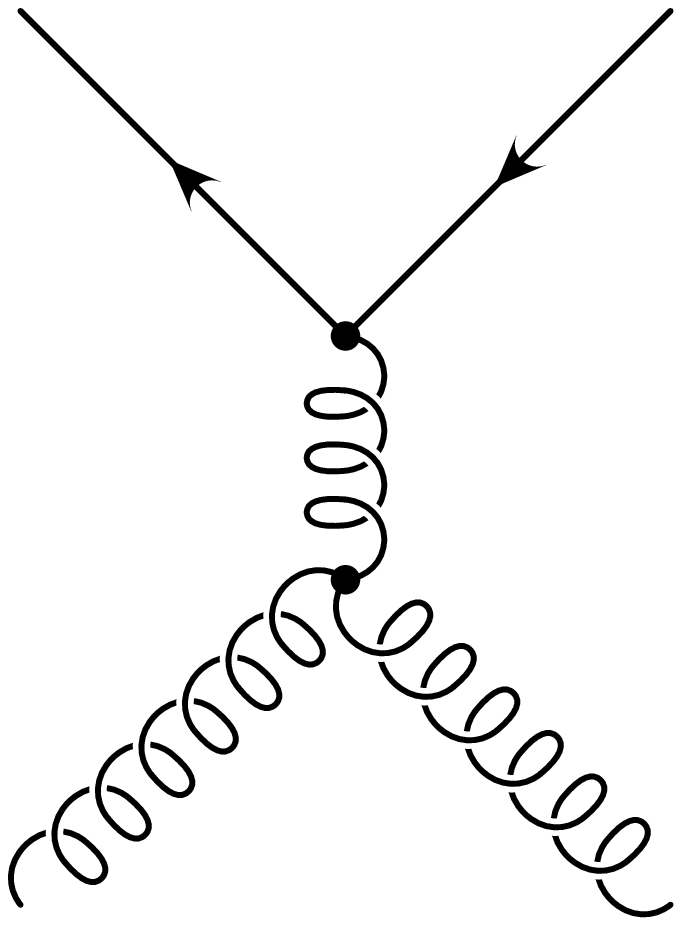}
\end{center}
\caption{\label{fig:ggqq} Graphs contributing to $gg \to q \bar q$ in the full theory.}
\end{figure}

The computations in this paper have been restricted to those involving external matter fields. In top-quark production and in jet production, processes involving external gluons are also important. Consider, for definiteness, the case $gg \to q \bar q$. At the high-scale $Q$, the tree-graphs which contribute to $gg \to q \bar q$ are shown in Fig.~\ref{fig:ggqq}.
\begin{figure}
\begin{center}
\includegraphics[width=3cm]{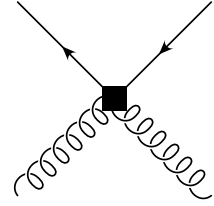}
\end{center}
\caption{\label{fig:ggqqEFT} Operator contributing to $gg \to q \bar q$ in the EFT.}
\end{figure}
In the EFT, one generates a local operator which involves the fields $q$, $\bar q$, and two collinear-gluon field strength tensors, shown graphically in Fig.~\ref{fig:ggqqEFT}. The QCD corrections  involve studying operators with gauge field strength tensors, and will be discussed elsewhere. The QCD corrections are known from existing fixed-order computations~\cite{kidonakis}. The new feature discussed in this article is the electroweak correction. If we restrict ourselves to the electroweak corrections alone, then we can compute these using the results in \ptwo. The gluon field strength tensor is an electroweak singlet, and so the $gg q \bar q$ operator in the EFT is equivalent to the electroweak singlet currents $\bar q \gamma^\mu P_{L,R} q$ studied in \ptwo\, and the running and matching corrections in the effective theory are identical, with the identification $-Q^2 \to s$. Thus the total radiative corrections are given by combining the known QCD corrections, with the electroweak corrections for the current given in \pone, \ptwo. The other important parton subprocesses which contributes to dijet production are $g q \to g q$, $g \bar q \to g \bar q$, and $gg \to gg$. For gluon-quark or gluon-antiquark scattering, the EFT operator is a $gg q \bar q$ operator, and the electroweak corrections are the same as those for the Sudakov form-factor, with $Q^2 \to -t$. For $gg$ scattering, there are no electroweak corrections to the order we are working, since the gluons do not couple to the electroweak gauge bosons, and the radiative corrections can be computed using the known QCD corrections.

\subsection{Squark Production}
\label{ssec:squark}

The techniques developed in the previous sections can be used to calculate the radiative corrections in a theory involving scalar particles in the final state such as SUSY. 
To perform a high precision computation requires specifying a particular supersymmetric theory, and computing the matching conditions and radiative corrections using the given SUSY particle spectrum. This is beyond the scope of the present work.  

To estimate the size of electroweak Sudakov corrections in squark production, we will compute the $SU(2)$ corrections in the toy model of Sec.~\ref{sec:squark_production}, assuming the squark is a doublet, and $\alpha \to \alpha_2$, the weak interaction coupling constant. This gives the expected size of electroweak Sudakov corrections in squark production.

\section{Numerics}
\label{sec:numerics}

The formul\ae\ for the EFT computation of standard model scattering processes have been given in Sec.~\ref{sec:SM}. As discussed in Sec.~\ref{sec:exp}, the anomalous dimensions are integrated using the two-loop $\beta$-functions, and we also include the known two-loop QCD anomalous dimensions~\cite{aybat} in addition to the one-loop results of Sec.~\ref{sec:SM}.  The corrections have a very small dependence on the Higgs mass (much less than 1\%). In the numerics, we use a Higgs mass of 200~GeV. The EFT coefficients should be run down to a scale $\mu_0$ of order a typical jet invariant mass. We have chosen to use $\mu_0=30$~GeV. The electroweak corrections are insenstive to this scale, because the only electroweak correction below $M_Z$ is due to photon exchange.

The matching corrections at the high scale $Q$ are about 2\%, and dominated by the QCD contribution. The low-scale matching due to integrating out the $W$ and $Z$ is about 2\%. Both matching corrections are not very strongly dependent on $Q$. The largest corrections are from the anomalous dimension running. These corrections grow rapidly with energy. The one-loop QCD corrections are very large, and reduce the rate by factors of $3$--$30$ in the range $\sqrt{\hat s}$ between 1 and 5~TeV. The two-loop QCD cusp anomalous dimension reduces the rate by about 10\% at $\sqrt{\hat s}=5$~TeV. This is smaller than the electroweak corrections, but not negligible. The two-loop non-cusp QCD anomalous dimension (the $B$ term in Eq.~(\ref{13a})) increases the rate by about 2\%. We have included the QCD two-loop cusp and non-cusp terms in the numerical results.  The two loop cusp anomalous dimension has been shown to be proportional to the one-loop result~\cite{aybat}, and we use their $K$ factor to determine the two-loop cusp anomalous dimension (the $A$ term in Eq.~(\ref{13a})).  The two-loop non-cusp anomalous dimension was determined in \ptwo\ by comparing the EFT result with the two-loop results of Jantzen and Smirnov~\cite{js}. The two-loop cusp anomalous dimension also determines the two-loop contribution to the soft anomalous dimension matrix $\gamma_S$. The non-cusp contribution vanishes, since $\gamma_S$ depends on differences of anomalous dimensions. The three-loop QCD cusp anomalous dimension contribution~\cite{MVV} is less than $0.1$\%, and can be omitted.  The one-loop electroweak anomalous dimension corrections are significant, ranging from 5\% at 1~TeV to around 30\% at 5~TeV. Higher order electroweak corrections, such as the two-loop electroweak cusp anomalous dimension are smaller than $0.1$\%. The numerical results are accurate at the one-percent level, so that the error in LHC cross-sections is dominated by other uncertainties, such as in the parton distribution functions.

The EFT analysis neglected power corrections of the form $M^2/\hat s$, $M^2/\hat t$ and $M^2/\hat u$. The dominant power corrections arise from one-loop QCD graphs, so we use the estimate $(\alpha_s M^2/\pi) \times 1/( \hat s, \hat t, \hat u)$ since the graphs have a color factor of (roughly) $4 C_F \alpha_s/(4\pi)$. To keep the power corrections below 1\% requires $\sqrt{{\hat s, \abs{\hat t}, \abs{\hat u}}}$ to be larger than about 200~GeV for light-quark processes, where the largest $M$ is $M_Z$, and larger than about 350~GeV for processes involving the top-quark. Note that we have included all power corrections that depend on ratios such as $m_Z/m_t$ or $M_Z/M_H$, and not expanded in these ratios. There are tree-level power corrections due to gauge boson mass effects in the propagators, e.g the $s$-channel propagator $\hat s - M_Z^2$ is approximated as $\hat s$. These trivial effects cancel in our results, because we normalize all amplitudes to their tree-level values.

The LHC cross-sections are given by using the coefficients computed earlier to compute the parton scattering cross-sections, and then convoluting them with parton distribution functions. For processes involving four-quark operators, the effective interaction at the low scale is a linear combination of two color structures,
\begin{eqnarray}
\mathcal{O} &=& C_1 \left(T^a \otimes T^a\right) + C_2 \left(\mathbf{1} \otimes \mathbf{1} \right).
\end{eqnarray}
Color-averaging over initial particles and color summing over final particles lead to a contribution to the cross-section which is proportional to an effective coefficient $C$, with
\begin{eqnarray}
\abs{C}^2 &=& \frac29 \abs{C_1}^2 + \abs{C_2}^2.
\end{eqnarray}

For $q \bar q \to q^\prime \bar q^\prime$, e.g. $ u \bar u \to b \bar b$, the parton scattering cross-section is
\begin{eqnarray}
\frac{\rd \hat \sigma}{\rd \hat t} &=& \frac{1}{16 \pi \hat s^2}
 \Bigr[\left(\abs{C_{LL}}_{s,t}^2+\abs{C_{RR}}_{s,t}^2\right) \hat u^2\nn
 &&+\left(\abs{C_{LR}}_{s,t}^2+\abs{C_{RL}}_{s,t}^2\right) \hat t^2\Bigr]
 \label{144}
 \end{eqnarray}
where $C_{LL}$, etc.\ are the coefficients of the $LL$, etc.\ operators. The $\,\hat{}\,$ denote partonic variables. The subscript $s,t$ is a reminder that one uses the coefficients as computed in Sec.~\ref{sec:SM} with annihilation channel kinematics.
From this, one can compute hadronic cross-sections. For example, the dijet invariant mass distribution from the partonic subprocess $u \bar u \to d \bar d$ is given by
\begin{eqnarray}
M^2 \frac{\rd^2 \sigma}{\rd M^2 \rd E_T} &=& 2E_T
\sqrt{\frac{\hat s}{\hat s -4 E_T^2}} \left.\frac{\rd \hat \sigma}{\rd \hat t}\right|_{\hat s=\tau s}\ \tau \frac{\rd L_{u\bar u}}{\rd \tau}\nn
\label{146}
 \end{eqnarray}
where $M^2$ is the dijet invariant mass, $E_T$ is the transverse energy of the jet, $\tau=M^2/s$, $\sqrt{s}=14$~TeV is the LHC center of mass energy, and the parton luminosity function $L_{ij}$ is defined by
\begin{eqnarray}
\frac{\rd L_{ij}}{\rd \tau} 
&=&\frac{1}{1+\delta_{ij}}\nn
&&\times \int_\tau^1 \frac{\rd x}{x}\,  \left[ f_i^{(1)}(x) f_j^{(2)}(\tau/x)+ f_j^{(1)}(x) f_i^{(2)}(\tau/x)\right] \nn
\end{eqnarray}
where $f_i^{(1,2)}$ are the distribution functions for parton $i$ in beams 1 and 2.\footnote{$2E_T=\sqrt{\hat s} \sin \theta$ and $\hat t = - \hat s \sin^2 \theta/2$, where $\theta$ is the center of mass scattering angle. Thus in Eq.~(\ref{146}), a given $E_T$ values gets contributions from two values of $\hat t$, or equivalently, one should symmetrize $\rd \hat \sigma /\rd \hat t$ under $\hat t \leftrightarrow \hat u$. We will plot $\rd \hat \sigma /\rd \hat t$ before symmetrizing.} For the LHC, both are proton distribution functions. The $1+\delta_{ij}$ is the symmetry factor for identical partons in the initial state. For the case $q=q^\prime$, e.g. $u \bar u \to u \bar u$, Eq.~(\ref{144}) still holds, and the coefficients $C_{LL,RR}$ get contributions from both the direct and crossed graphs.

For identical particles, e.g. $u \bar u \to u \bar u$, the partonic cross-section has the schematic form
\begin{eqnarray}
\frac{ {\rm d}\hat \sigma}{{\rm d}\hat t} &=& \frac{1}{16 \pi \hat s^2}
 \Bigl[\left(\abs{C_{LL}+\widetilde C_{LL}}_{s,t}^2+\abs{C_{RR}+\widetilde C_{LL}}_{s,t}^2\right) \hat u^2\nn
 &&+\left(\abs{C_{LR}}_{s,t}^2+\abs{C_{RL}}_{s,t}^2\right) \hat t^2\nn
 &&+\left(\abs{C_{LR}}_{t,s}^2+\abs{C_{RL}}_{t,s}^2\right) \hat s^2\Bigr]\,.
 \label{22a}
\end{eqnarray}
There is the direct channel as well as the crossed-channel with $s \leftrightarrow t$. For $LL$ and $RR$, the crossed-channel amplitudes have the same fermion chiralities as the direct channel, and are included as  $\widetilde C$, which includes both $s \leftrightarrow t$ and the crossing matrix. One has to add the amplitudes in the two channels before squaring. For $LR$ and $RL$, the crossed-diagrams do not interfere because the initial and final chiralities do not match, and one adds the probabilities.

For $qq^\prime$ scattering processes not involving identical particles the cross-section is
\begin{eqnarray}
\frac{ {\rm d}\hat \sigma}{{\rm d}\hat t} &=& \frac{1}{16 \pi \hat s^2}
 \Bigl[\left(\abs{C_{LL}}_{t,u}^2+\abs{C_{RR}}_{t,u}^2\right) \hat s^2\nn
 &&+\left(\abs{C_{LR}}_{t,u}^2+\abs{C_{RL}}_{t,u}^2\right) \hat u^2 \Bigr]
 \end{eqnarray}
and for $q \bar q^\prime$ scattering
\begin{eqnarray}
\frac{ {\rm d}\hat \sigma}{{\rm d}\hat t} &=& \frac{1}{16 \pi \hat s^2}
 \Bigl[\left(\abs{C_{LL}}_{t,s}^2+\abs{C_{RR}}_{t,s}^2\right) \hat u^2\nn
 &&+\left(\abs{C_{LR}}_{t,s}^2+\abs{C_{RL}}_{t,s}^2\right) \hat s^2 \Bigr]
\end{eqnarray}
The subscripts $t,u$, etc.\ are a reminder the one has to use the amplitudes of Sec.~\ref{sec:SM} with the replacements
$s \to t$, $t \to u$, etc.

For identical quark scattering, $qq \to qq$, e.g. $uu \to uu$, the cross-section is
\begin{eqnarray}
\frac{ {\rm d}\hat \sigma}{{\rm d}\hat t} &=& \frac{1}{16 \pi \hat s^2}
 \Bigl[\frac12\left(\abs{C_{LL}+\widetilde C_{LL}}_{t,u}^2+\abs{C_{RR}+\widetilde C_{LL}}_{t,u}^2\right) \hat s^2\nn
 &&+2 \abs{C_{LR}}_{t,u}^2 \hat u^2+2\abs{C_{LR}}_{u,t}^2 \hat t^2\Bigr]\,.
 \end{eqnarray}
The $1/2$ is from final state phase space for identical particles. The initial state $1/2$ is included in the parton luminosity function.

There are 72 four-fermion amplitudes that have been computed in Sec.~\ref{sec:SM} in the $s$-channel, not including those which are identical by flavor symmetry, and another 72 amplitudes in the $t$-channel, and we cannot plot them all here. We will choose some representative examples to illustrate the size of the radiative corrections in high energy LHC processes. Rather than plot the hadronic cross-sections, which involve convolutions over rapidly falling parton luminosities, we have chosen to plot the ratio of the partonic cross-sections $\rd \hat \sigma/\rd \hat t$ to their tree-level values. From Eq.~(\ref{146}), it follows that this also gives the ratio of the hadronic cross-section to its tree-level value.
We will also neglect the CKM matrix in the plots, since the flavor dependence of the electroweak corrections is small, and the CKM factors enter as off-diagonal CKM matrix elements multiplied by the difference of electroweak corrections between $d^\prime$ and $s^\prime$, and $d^\prime$ and $b^\prime$.

\subsection{Plots}

\begin{figure}
\begin{center}
\includegraphics[width=8cm]{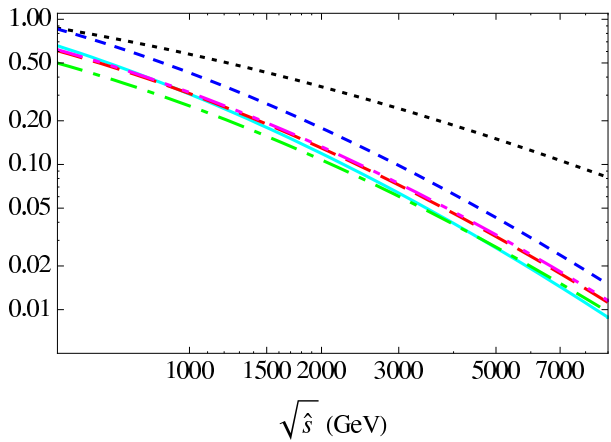}
\end{center}
\caption{\label{fig:num1} Rates for $u \bar u \to \mu^+ \mu^-$ (dotted black), $u\bar u \to u \bar u $ (solid cyan), $u\bar u \to c \bar c$ (long-dashed red), $u \bar u \to t \bar t $ (short-dashed blue), $u\bar u \to d \bar d$ (dot-dashed green) and $u \bar u \to b \bar b$  (double-dot-dashed magenta)  as a function of $\sqrt{\hat s}$ in GeV at $\theta=90^\circ$, normalized to their tree-level values without any electroweak corrections. Note the logarithmic scale.}
\end{figure}
Figure~\ref{fig:num1} show the ratio  $\rd \hat \sigma/\rd \hat t$ to its tree-level value for
$u \bar u \to \mu^+ \mu^-$, $ u \bar u$, $c \bar c$, $t \bar t $ and $b \bar b$ as a function of $\sqrt{\hat s}$ for $90^\circ$ scattering, $\hat t = - \hat s/2$, including QCD and electroweak corrections. The radiative corrections are enormous, and reduce the cross-sections by $1.15$--$2$ at $\sqrt{\hat s}=500$~GeV to a factor of $7$--$38$ at $\sqrt{\hat s}=5$~TeV compared to the tree-level value, depending on the process.\footnote{Note that the parton luminosity is falling by about four orders of magnitude over the same range.} The bulk of the correction is due to QCD effects. Some of the QCD corrections are included in parton shower Monte-Carlos, because gluon radiation from tree-level branching is related to the LL Sudakov series. However, the electroweak corrections, and some of the QCD corrections are not included in the shower algorithms, so the Monte-Carlo results can have substantial ($\sim 50$\%) corrections. 

\begin{figure}
\begin{center}
\includegraphics[width=8cm]{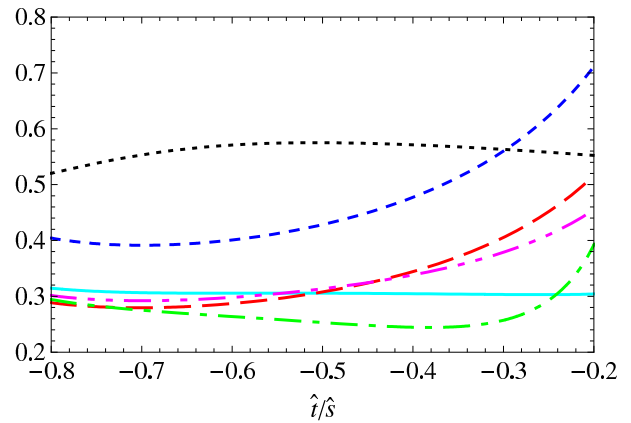}
\end{center}
\caption{\label{fig:angnum1} Rates for $u \bar u \to \mu^+ \mu^-$ (dotted black), $u\bar u \to u \bar u $ (solid cyan), $u\bar u \to c \bar c$ (long-dashed red), $u \bar u \to t \bar t $ (short-dashed blue), $u\bar u \to d \bar d$ (dot-dashed green)  and $u \bar u \to b \bar b$ (double-dot-dashed magenta) as a function of $\hat t/\hat s$  for $\sqrt{\hat s}=1$~TeV, normalized to their tree-level values without any electroweak corrections.}
\end{figure}

The $u \bar u$ rate differs from $c \bar c$ because of the crossed-channel graph for identical particles. The difference between $c \bar c$ and $t \bar t$, and between $s \bar s$ (not shown) and $b \bar b$ is due to top-quark mass effects.  The $u \bar u \to \mu^+ \mu^-$ rate has smaller QCD corrections, since the final state is a color singlet. The  anomalous dimension $\tilde \gamma$ is proportional to $4 \Ls - 6$. At large values of $\hat s$, the $4 \Ls$ term dominates, and produces the large Sudakov (double-log) suppression. At smaller values of $\hat s$, the $-6$ can compensate the $4\Ls$ term, leading to an enhancement of the cross-section. This leads to a flattening of the curves at the smallest values of $\hat s$. The cross-sections will decrease slightly if we continue the plot to even smaller values of $\hat s$. This effect can also be seen in the plots of Ref.~\cite{jkps4}. Figure~\ref{fig:angnum1} show the radiative corrections to the angular distribution for $u \bar u \to \mu^+ \mu^-$, $ u \bar u$, $c \bar c$, $t \bar t $ and $b \bar b$ at $\sqrt{\hat s}=1$~TeV.
There is about a factor of two variation in the radiative correction over the range $-0.8 \le \hat t/\hat s \le -0.2$.

\begin{figure}
\begin{center}
\includegraphics[width=8cm]{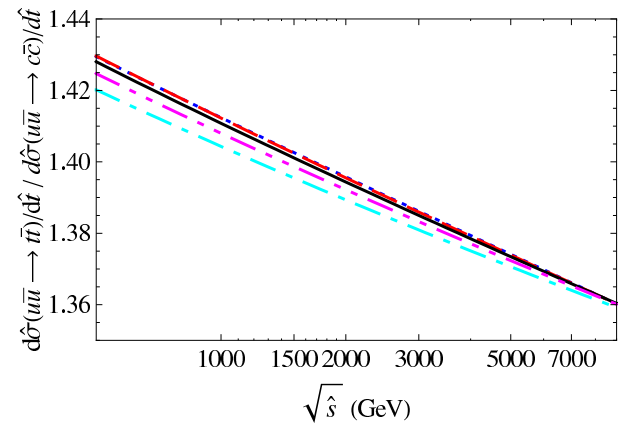}
\end{center}
\caption{\label{fig:uuttcc} The ratio $(u \bar u \to t \bar t)/(u \bar u \to c \bar c)$ at $\hat t = -0.2 \hat s$, (dotted blue), $\hat t = -0.35 \hat s$ (long-dashed red), $\hat t = -0.5 \hat s$ (solid black), $\hat t = -0.65 \hat s$ (double-dot-dashed magenta) and $\hat t = -0.8 \hat s$ (dot-dashed cyan) as a function of $\sqrt{\hat s}$ in GeV.}
\end{figure}

\begin{figure}
\begin{center}
\includegraphics[width=8cm]{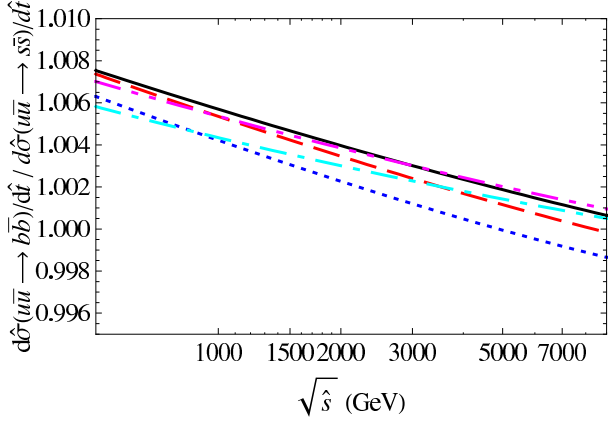}
\end{center}
\caption{\label{fig:uubbss} The ratio $(u \bar u \to b \bar b)/(u \bar u \to s \bar s)$ at $\hat t = -0.2 \hat s$, (dotted blue), $\hat t = -0.35 \hat s$ (long-dashed red), $\hat t = -0.5 \hat s$ (solid black), $\hat t = -0.65 \hat s$ (double-dot-dashed magenta) and $\hat t = -0.8 \hat s$ (dot-dashed cyan) as a function of $\sqrt{\hat s}$ in GeV.}
\end{figure}

\begin{figure}
\begin{center}
\includegraphics[width=8cm]{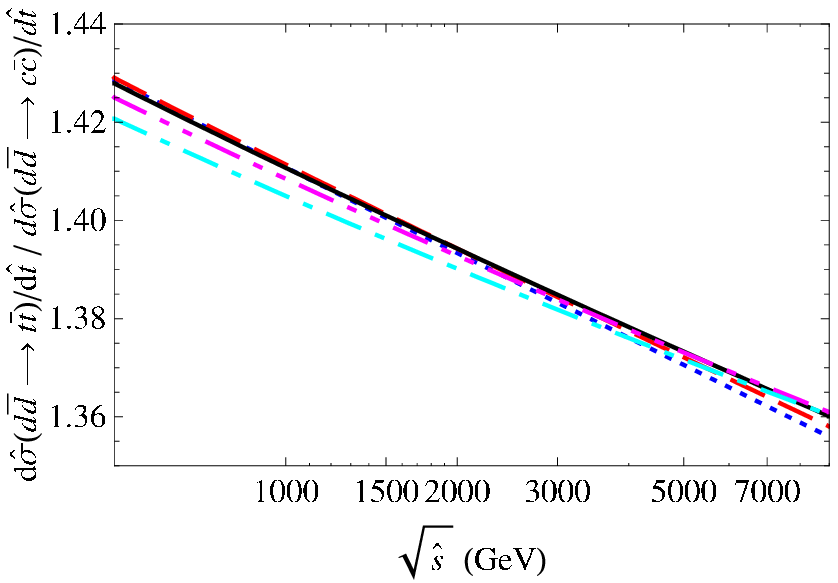}
\end{center}
\caption{\label{fig:ddttcc} The ratio $(d \bar d \to t \bar t)/(d \bar d \to c \bar c)$ at $\hat t = -0.2 \hat s$, (dotted blue), $\hat t = -0.35 \hat s$ (long-dashed red), $\hat t = -0.5 \hat s$ (solid black), $\hat t = -0.65 \hat s$ (double-dot-dashed magenta) and $\hat t = -0.8 \hat s$ (dot-dashed cyan) as a function of $\sqrt{\hat s}$ in GeV.}
\end{figure}

\begin{figure}
\begin{center}
\includegraphics[width=8cm]{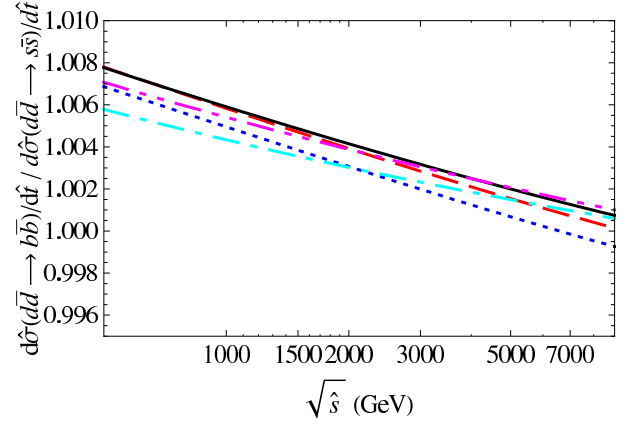}
\end{center}
\caption{\label{fig:ddbbss} The ratio of $(d \bar d \to b \bar b)/(d \bar d \to s \bar s)$ at $\hat t = -0.2 \hat s$, (dotted blue), $\hat t = -0.35 \hat s$ (long-dashed red), $\hat t = -0.5 \hat s$ (solid black), $\hat t = -0.65 \hat s$ (double-dot-dashed magenta) and $\hat t = -0.8 \hat s$ (dot-dashed cyan) as a function of $\sqrt{\hat s}$ in GeV.}
\end{figure}

Figure~\ref{fig:uuttcc} and Fig.~\ref{fig:uubbss}
show the ratios $\rd \hat \sigma(u \bar u \to t \bar t)/\rd \hat t/\rd \hat \sigma(u \bar u \to c \bar c)/\rd \hat t$ and $\rd \hat \sigma(u \bar u \to b \bar b)/\rd \hat t/\rd \hat \sigma(u \bar u \to s \bar s)/\rd \hat t$ as a function of $\sqrt{\hat s}$ for different values of $\hat t $ (i.e. the scattering angle $\theta$), including QCD and electroweak corrections. These ratios are unity in the absence of top-quark mass effects. There is a $\sim\!40$\% increase in the $t \bar t$ rate due to the top-quark mass. About $-4$\% is from the Higgs contribution, $-2$\% from mass effects in the low-scale electroweak matching, and the rest from mass effects in the QCD matching at $m_t$ and running below $m_t$. There is a much smaller enhancement of $\rd \hat \sigma(u \bar u \to b \bar b)/\rd \hat t/\rd \hat \sigma(u \bar u \to s \bar s)/\rd \hat t$ due to virtual top-quark effects in the $b$-sector.\footnote{Even though our individual radiative corrections have corrections under 1\%, ratios such as $\rd \hat \sigma(u \bar u \to b \bar b)/\rd \hat t/\rd \hat \sigma(u \bar u \to s \bar s)/\rd \hat t$ have much smaller errors, so that the deviation from unity in Fig.~\ref{fig:uubbss} is a real effect.} Figures.~\ref{fig:ddttcc} and \ref{fig:ddbbss} show the corresponding results for $d \bar d \to t \bar t, b \bar b$. 

\begin{figure}
\begin{center}
\includegraphics[width=8cm]{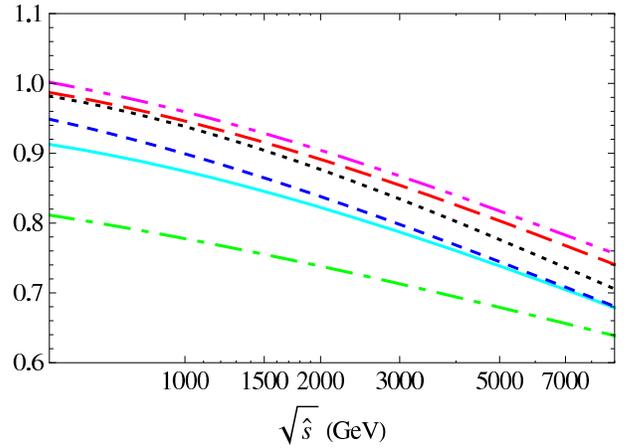}
\end{center}
\caption{\label{fig:num1EW} Electroweak corrections to $u \bar u \to \mu^+ \mu^-$ (dotted black), $u\bar u \to u \bar u $ (solid cyan), $u\bar u \to c \bar c$ (long-dashed red), $u \bar u \to t \bar t $ (short-dashed blue), $u\bar u \to d \bar d$ (dot-dashed green) and $u \bar u \to b \bar b$ (double-dot-dashed magenta) as a function of $\sqrt{\hat s}$ in GeV at $\theta=90^\circ$. The large corrections for $u\bar u \to d \bar d$ arise from the $t$-channel $W$ exchange graph. }
\end{figure}

The plots discussed above include QCD and electroweak corrections. To show the importance of electroweak corrections, we show in Fig.~\ref{fig:num1EW}, the same processes as in Fig.~\ref{fig:num1}, but instead of plotting the ratio of the partonic cross-section to the tree-level value,
\begin{figure}
\begin{center}
\includegraphics[width=8cm]{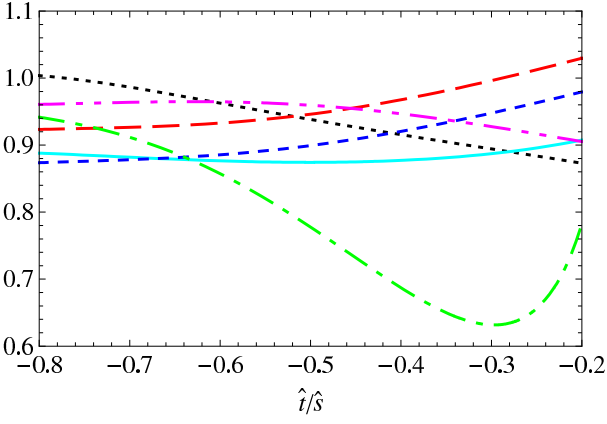}
\end{center}
\caption{\label{fig:angnum1EW} Electroweak corrections to $u \bar u \to \mu^+ \mu^-$ (dotted black), $u\bar u \to u \bar u $ (solid cyan), $u\bar u \to c \bar c$ (long-dashed red), $u \bar u \to t \bar t $ (short-dashed blue), $u\bar u \to d \bar d$ (dot-dashed green) and $u \bar u \to b \bar b$ (double-dot-dashed magenta)  as a function of $\hat t/\hat s$  for $\sqrt{\hat s}=1$~TeV. The large corrections for $u\bar u \to d \bar d$ arise from the $t$-channel $W$ exchange graph. }
\end{figure}
we plot the ratio of the cross-section to the value including only QCD corrections, i.e.\ with $\alpha_{1,2} \to 0$.\footnote{In $q \bar q \to \mu^+ \mu^-$, we include tree-level electroweak exchange, and keep $\alpha_s \alpha_{1,2}$ terms in the one-loop matching, but drop order $\alpha_{1,2}^2$ terms.} This ratio shows the additional effect of electroweak corrections beyond the QCD corrections, which have been computed previously. The electroweak corrections are significant, increasing from $(-4)$--$(-22)$\% at 1~TeV to $(-18)$--$(-32)$\% at 5~TeV, depending on the process. The electroweak corrections to the angular distribution are shown in Fig.~\ref{fig:angnum1EW}. There are 10--30\% variations in the corrections in the range $-0.8 \le \hat t / \hat s \le -0.2$ for $\sqrt{s}=1$~TeV.

\begin{figure}
\begin{center}
\includegraphics[width=8cm]{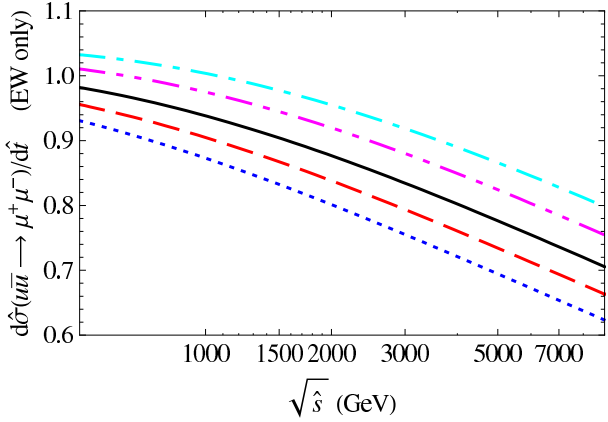}
\end{center}
\caption{\label{fig:uuee} Electroweak corrections to $u \bar u \to \mu^+ \mu^-$  at $\hat t = -0.2 \hat s$, (dotted blue), $\hat t = -0.35 \hat s$ (long-dashed red), $\hat t = -0.5 \hat s$ (solid black), $\hat t = -0.65 \hat s$ (double-dot-dashed magenta) and $\hat t = -0.8 \hat s$ (dot-dashed cyan) as a function of $\sqrt{\hat s}$ in GeV.}
\end{figure}

\begin{figure}
\begin{center}
\includegraphics[width=8cm]{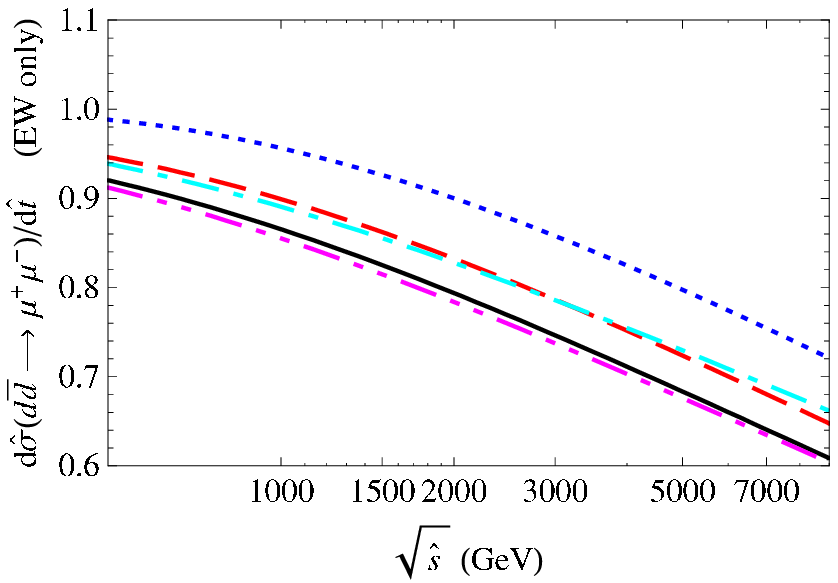}
\end{center}
\caption{\label{fig:ddee} Electroweak corrections to $d \bar d \to \mu^+ \mu^-$  at $\hat t = -0.2 \hat s$, (dotted blue), $\hat t = -0.35 \hat s$ (long-dashed red), $\hat t = -0.5 \hat s$ (solid black), $\hat t = -0.65 \hat s$ (double-dot-dashed magenta) and $\hat t = -0.8 \hat s$ (dot-dashed cyan) as a function of $\sqrt{\hat s}$ in GeV.}
\end{figure}

The electroweak corrections alone (defined as just discussed) for lepton pair production from $u$-quark and $d$-quark annihilation are shown in Fig.~\ref{fig:uuee} and Fig.~\ref{fig:ddee}
\begin{figure}
\begin{center}
\includegraphics[width=8cm]{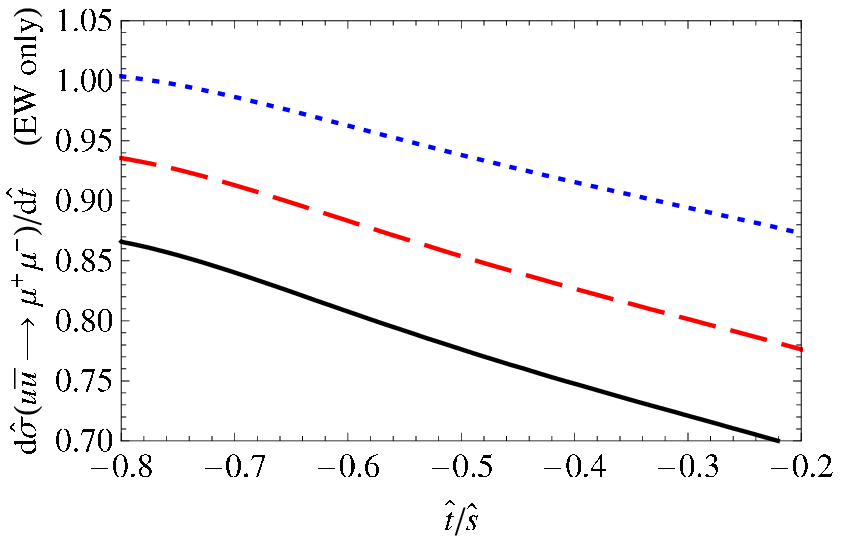}
\end{center}
\caption{\label{fig:anguuee}  Electroweak corrections to $u \bar u \to \mu^+ \mu^-$
at $\sqrt{\hat s} = 1$~TeV, (dotted blue), $\sqrt{\hat s} = 2.5$~TeV (long-dashed red) and$\sqrt{\hat s} = 5$~TeV (solid black) as a function of $\hat t/\hat s$.}
\end{figure}
\begin{figure}
\begin{center}
\includegraphics[width=8cm]{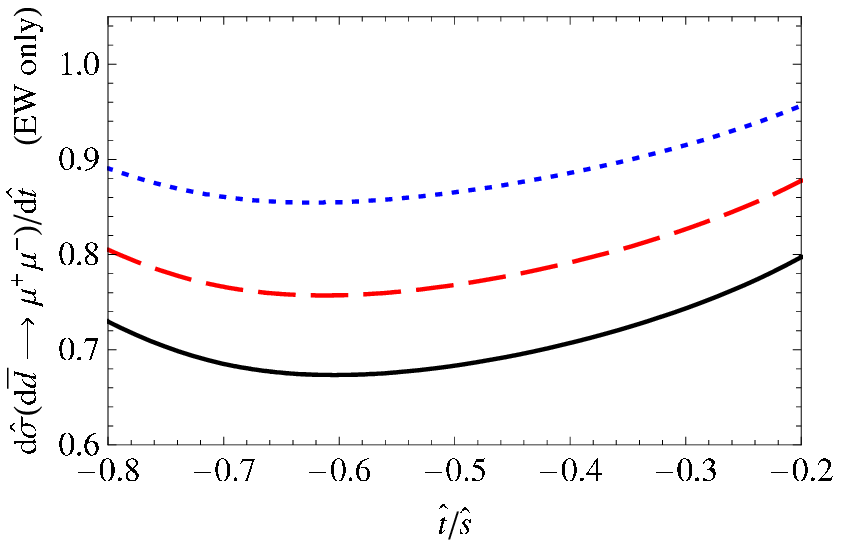}
\end{center}
\caption{\label{fig:angddee} Electroweak corrections to $d \bar d \to \mu^+ \mu^-$
at $\sqrt{\hat s} = 1$~TeV, (dotted blue), $\sqrt{\hat s} = 2.5$~TeV (long-dashed red) and$\sqrt{\hat s} = 5$~TeV (solid black) as a function of $\hat t/\hat s$.}
\end{figure}
for different values of $\hat t$. At $\sqrt{\hat s}=1$~TeV, the corrections range from $(0.4)$--$(-14)$\%, increasing to $(-13)$--$(-32)$\% at $\sqrt{\hat s}=5$~TeV. The electroweak corrections also change the angular distribution of the lepton pairs. Fig.~\ref{fig:anguuee} and Fig.~\ref{fig:angddee} show the $\hat t$ dependence of the cross-section for different values of $\hat s$. The angular dependence is approximately independent of $\hat s$. The reason is that the dominant $\hat t$ dependence arises from the soft anomalous dimension $\gamma_S$, which is a function only of the dimensionless ratio $\hat t/\hat s$. The angular dependence of the electroweak corrections differ for $u \bar u \to \mu^+ \mu^-$ and $d \bar d \to \mu^+ \mu^-$.

\begin{figure}
\begin{center}
\includegraphics[width=8cm]{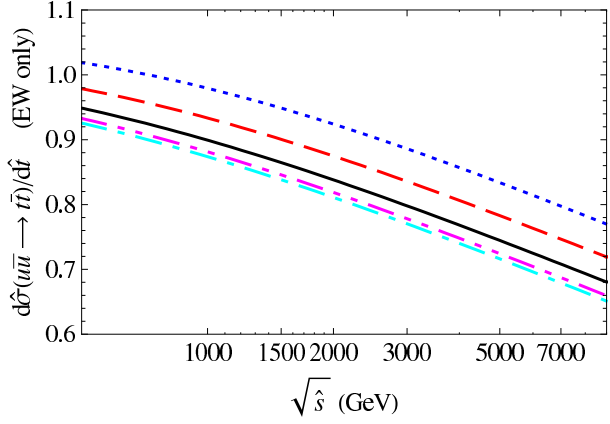}
\end{center}
\caption{\label{fig:uutt} Electroweak corrections to $u \bar u \to t \bar t $ at $\hat t = -0.2 \hat s$, (dotted blue), $\hat t = -0.35 \hat s$ (long-dashed red), $\hat t = -0.5 \hat s$ (solid black), $\hat t = -0.65 \hat s$ (double-dot-dashed magenta) and $\hat t = -0.8 \hat s$ (dot-dashed cyan) as a function of $\sqrt{\hat s}$ in GeV.}
\end{figure}

\begin{figure}
\begin{center}
\includegraphics[width=8cm]{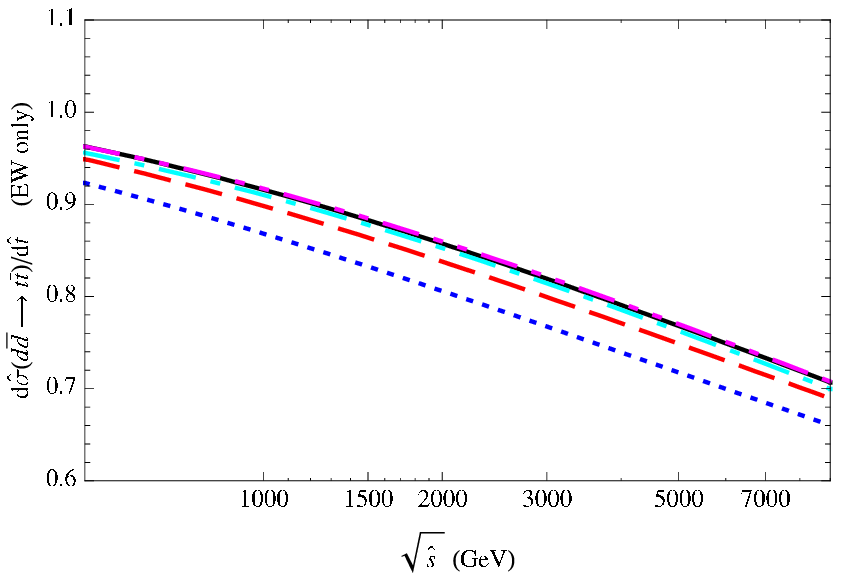}
\end{center}
\caption{\label{fig:ddtt} Electroweak corrections to $d \bar d \to t \bar t $ at $\hat t = -0.2 \hat s$, (dotted blue), $\hat t = -0.35 \hat s$ (long-dashed red), $\hat t = -0.5 \hat s$ (solid black), $\hat t = -0.65 \hat s$ (double-dot-dashed magenta) and $\hat t = -0.8 \hat s$ (dot-dashed cyan) as a function of $\sqrt{\hat s}$ in GeV.}
\end{figure}

\begin{figure}
\begin{center}
\includegraphics[width=8cm]{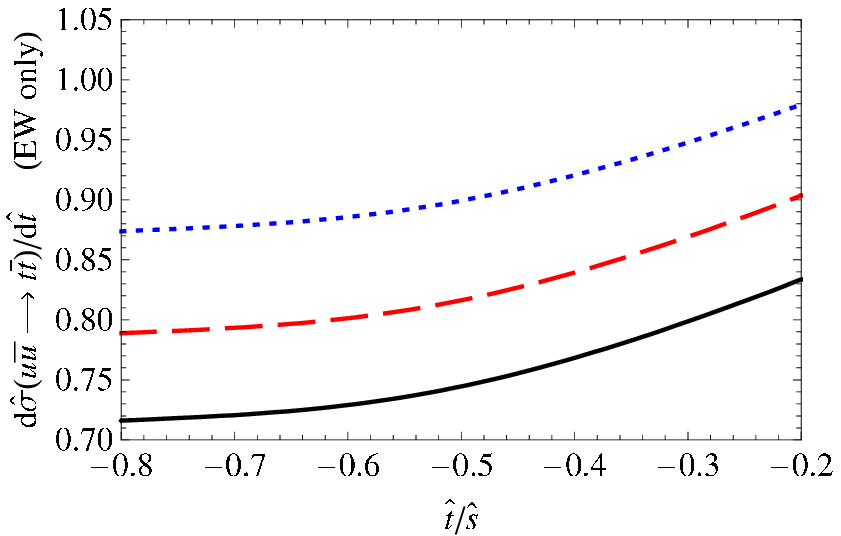}
\end{center}
\caption{\label{fig:anguutt} Electroweak corrections to $u \bar u \to t \bar t $ at $\sqrt{\hat s} = 1$~TeV, (dotted blue), $\sqrt{\hat s} = 2.5$~TeV (long-dashed red) and$\sqrt{\hat s} = 5$~TeV (solid black) as a function of $\hat t/\hat s$.}
\end{figure}

\begin{figure}
\begin{center}
\includegraphics[width=8cm]{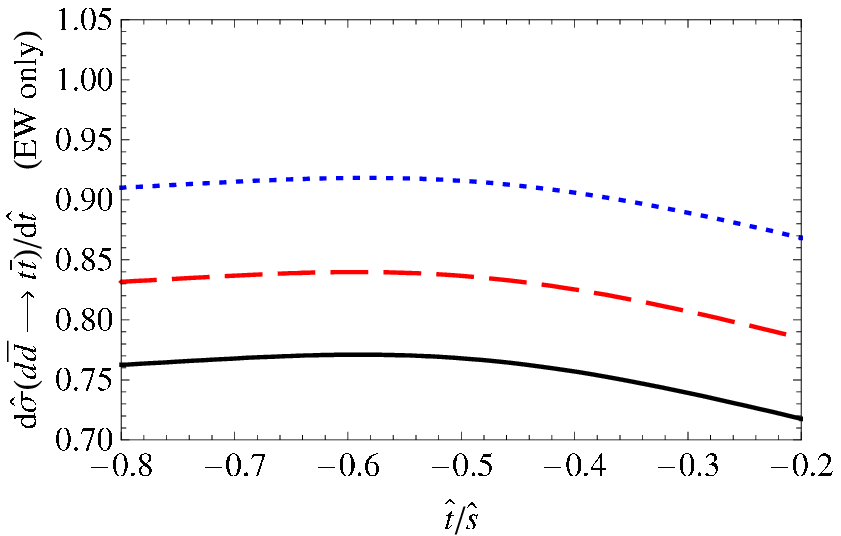}
\end{center}
\caption{\label{fig:angddtt} Electroweak corrections to $d \bar d \to t \bar t $  at $\sqrt{\hat s} = 1$~TeV, (dotted blue), $\sqrt{\hat s} = 2.5$~TeV (long-dashed red) and$\sqrt{\hat s} = 5$~TeV (solid black) as a function of $\hat t/\hat s$.}
\end{figure}

\begin{figure}
\begin{center}
\includegraphics[width=8cm]{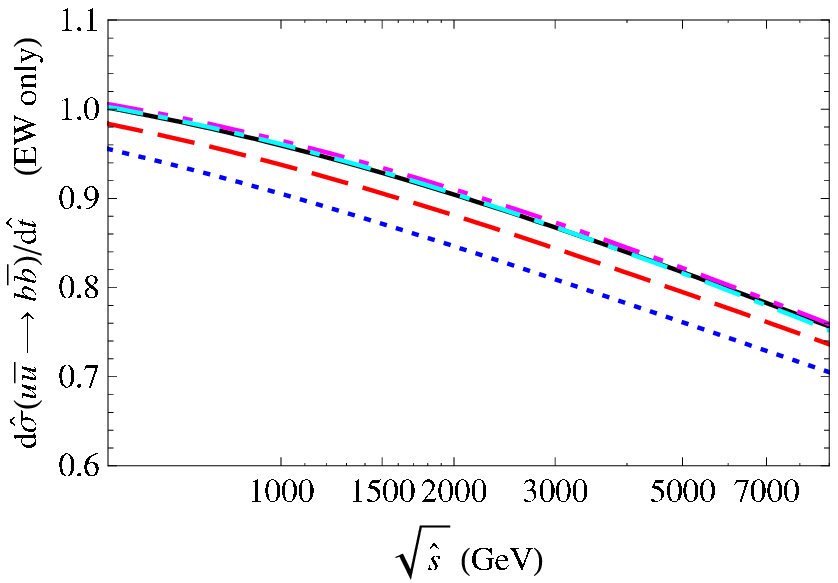}
\end{center}
\caption{\label{fig:uubb} Electroweak corrections to $u \bar u \to b \bar b $ at $\hat t = -0.2 \hat s$, (dotted blue), $\hat t = -0.35 \hat s$ (long-dashed red), $\hat t = -0.5 \hat s$ (solid black), $\hat t = -0.65 \hat s$ (double-dot-dashed magenta) and $\hat t = -0.8 \hat s$ (dot-dashed cyan) as a function of $\sqrt{\hat s}$ in GeV.}
\end{figure}

\begin{figure}
\begin{center}
\includegraphics[width=8cm]{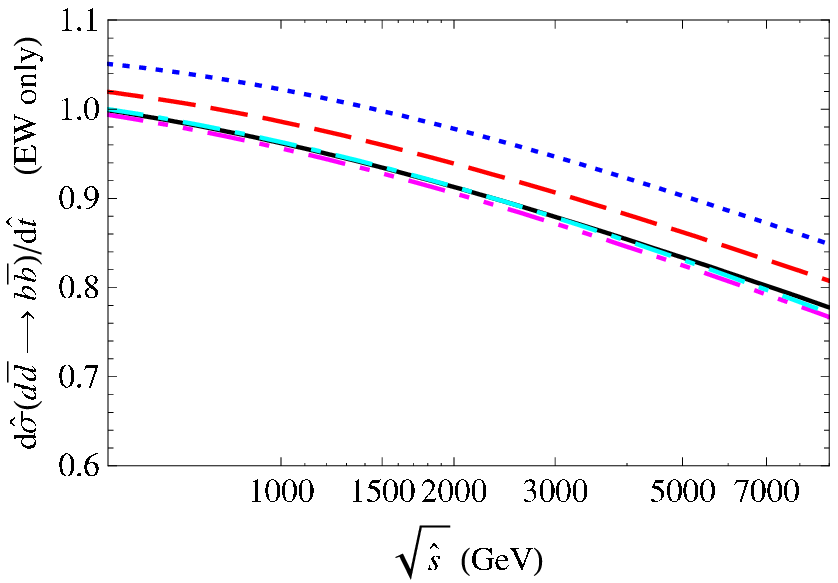}
\end{center}
\caption{\label{fig:ddbb} Electroweak corrections to $d \bar d \to b \bar b $  at $\hat t = -0.2 \hat s$, (dotted blue), $\hat t = -0.35 \hat s$ (long-dashed red), $\hat t = -0.5 \hat s$ (solid black), $\hat t = -0.65 \hat s$ (double-dot-dashed magenta) and $\hat t = -0.8 \hat s$ (dot-dashed cyan) as a function of $\sqrt{\hat s}$ in GeV.}
\end{figure}

\begin{figure}
\begin{center}
\includegraphics[width=8cm]{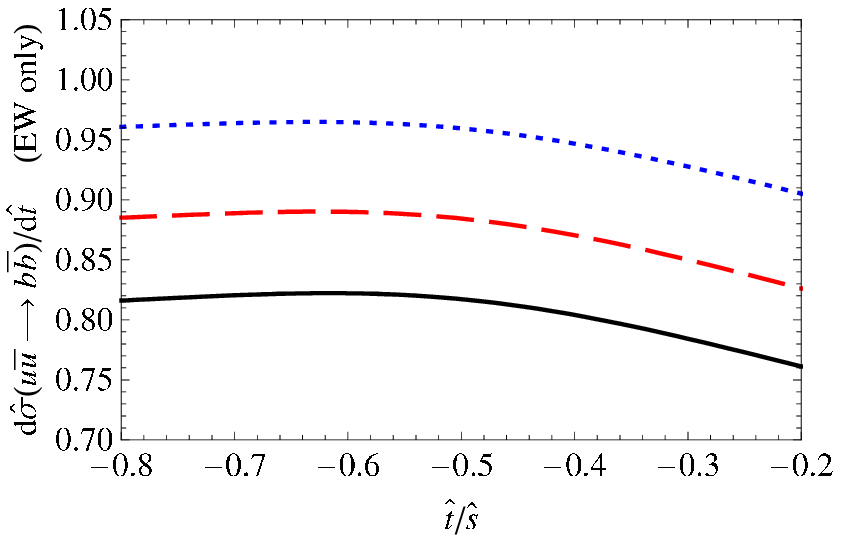}
\end{center}
\caption{\label{fig:anguubb} Electroweak corrections to $u \bar u \to b \bar b $ at $\sqrt{\hat s} = 1$~TeV, (dotted blue), $\sqrt{\hat s} = 2.5$~TeV (long-dashed red) and$\sqrt{\hat s} = 5$~TeV (solid black) as a function of $\hat t/\hat s$.}
\end{figure}

\begin{figure}
\begin{center}
\includegraphics[width=8cm]{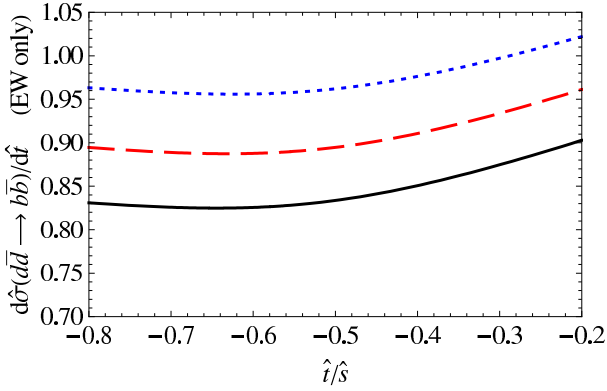}
\end{center}
\caption{\label{fig:angddbb} Electroweak corrections to $d \bar d \to b \bar b $ at $\sqrt{\hat s} = 1$~TeV, (dotted blue), $\sqrt{\hat s} = 2.5$~TeV (long-dashed red) and$\sqrt{\hat s} = 5$~TeV (solid black) as a function of $\hat t/\hat s$.}
\end{figure}

The electroweak corrections to heavy quark production via $u$ and $d$ quark annihilation are shown in Figs.~\ref{fig:uutt}, Fig.~\ref{fig:ddtt}, Fig.~\ref{fig:anguutt}, Fig.~\ref{fig:angddtt} for $t$-quark production, and Fig.~\ref{fig:uubb}, Fig.~\ref{fig:ddbb} Fig.~\ref{fig:anguubb}, Fig.~\ref{fig:angddbb} for $b$-quark production.
Electroweak corrections to heavy quark production have been computed previously~\cite{
Nason:1987xz,Beenakker:1988bq,Nason:1989zy,Beenakker:1990maa,Laenen:1991af, Kidonakis:1995wz, Berger:1997gz,Cacciari:2003fi,Kuhn:2006vh}. We find the same qualitative behavior---the electroweak corrections give a small ($\sim -6$\%) suppression, and the QCD corrections give a large ($\sim 50$\%) enhancement.

\begin{figure}
\begin{center}
\includegraphics[width=8cm]{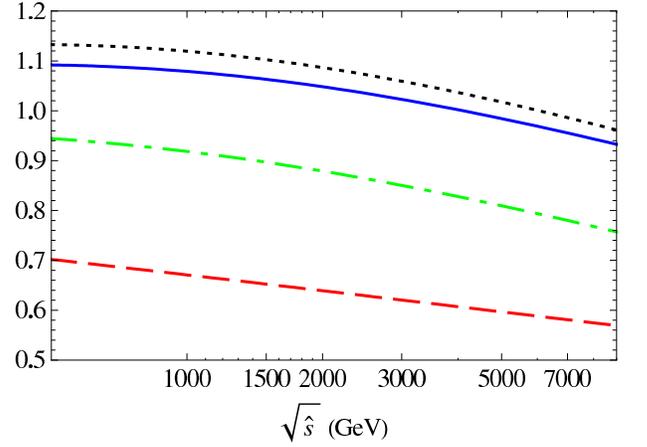}
\end{center}
\caption{\label{fig:scat} Electroweak corrections to $u u \to u u$ (dotted black), $u d \to u d$ (long-dashed red), $dd \to dd$ (solid blue) and $u \bar d \to u \bar d, d \bar u \to d \bar u$ (dot-dashed green) as a function of $\sqrt{\hat s}$ in GeV at $\theta=90^\circ$.}
\end{figure}

\begin{figure}
\begin{center}
\includegraphics[width=8cm]{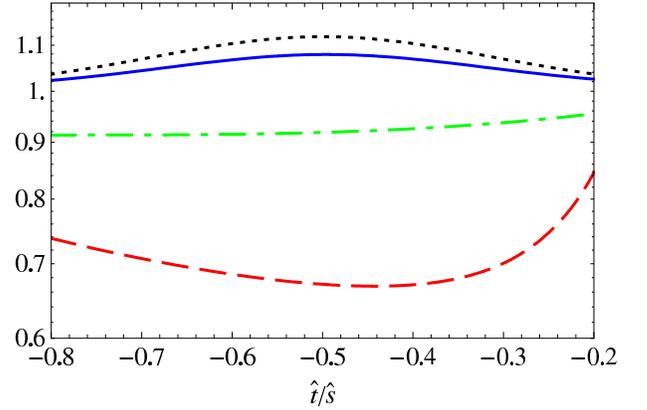}
\end{center}
\caption{\label{fig:angscat} Electroweak corrections to $u u \to u u$ (dotted black), $u d \to u d$ (long-dashed red), $dd \to dd$ (solid blue) and $u \bar d \to u \bar d, d \bar u \to d \bar u$ (dot-dashed green) as a function  $\hat t/\hat s$ at $\sqrt{\hat s}=1$~TeV.}
\end{figure}

The above plots have been for  $s$-channel proceeses. There are also $t$-channel parton subprocesses that contribute to dijet production. Rather than go through these in detail, we show two illustrative plots: Fig.~\ref{fig:scat} shows the electroweak corrections to $uu \to uu$, $u d \to ud$, $dd \to dd$ and $u \bar d \to u \bar d$ (which is equal to $d \bar u \to d \bar u$) as a function of $\sqrt{\hat s}$ for $90^\circ$ scattering, and Fig.~\ref{fig:angscat} shows the angular dependence of the electroweak corrections at $\sqrt{s}=1$~TeV.

\begin{figure}
\begin{center}
\includegraphics[width=8cm]{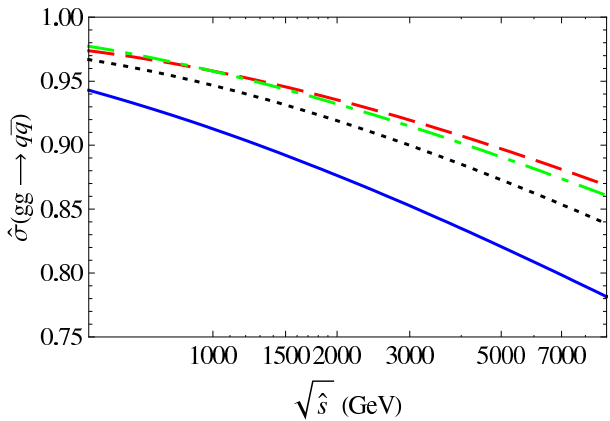}
\end{center}
\caption{\label{fig:qqgg} Electroweak corrections to $gg \to u \bar u,c \bar c$ (dotted black), $gg \to d \bar d, s \bar s$ (long-dashed red), $gg \to t \bar t$ (solid blue) and $gg \to b \bar b$ (dot-dashed green) as a function of $\sqrt{\hat s}$ in GeV. The electroweak corrections are independent of $\hat t$ to the order we are working. The same plot also gives the electroweak corrections to $g q \to gq$ and $g \bar q \to g \bar q$, as a function of $\sqrt{-\hat t}$.}
\end{figure}

There are also scattering processes involving external gluons. For $gg \to q \bar q$, $g q \to g q$, and $g \bar q \to g \bar q$, we have only computed the electroweak part of the correction, which is equal to that for the Sudakov form-factor. For the $s$-channel procees $gg \to q \bar q$, the electroweak correction only depends on $\sqrt{\hat s}$, and is shown in Fig.~\ref{fig:qqgg}.
The same plot also gives the electroweak correction to the $t$-channel scattering processes $g q \to g q$ and $g \bar q \to g \bar q$ as a function of $\sqrt{-\hat t}$, by crossing symmetry. The imaginary parts from the logarithmic branch cuts in the $s$-channel amplitude do not change the absolute value of the amplitude.

\begin{figure}
\begin{center}
\includegraphics[width=8cm]{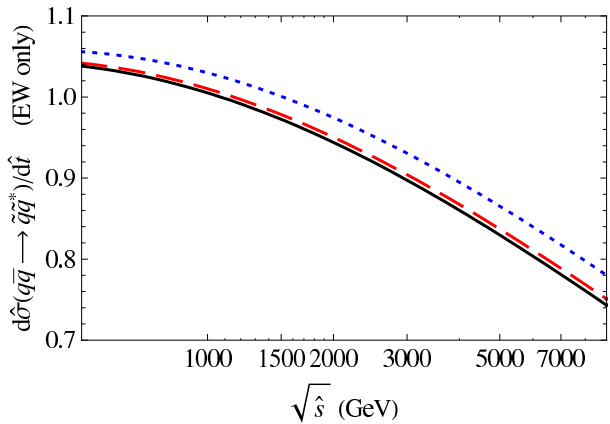}
\end{center}
\caption{\label{fig:squarks} Electroweak corrections to $ q \bar q \to \widetilde q \widetilde q^*$ in the toy theory at $\hat t = -0.2 \hat s$, (dotted blue), $\hat t = -0.35 \hat s$ (long-dashed red), $\hat t = -0.5 \hat s$ (solid black) as a function of $\sqrt{\hat s}$ in GeV. The rate is symmetric under $\theta \to 180^\circ-\theta$.}
\end{figure}

Finally, we show the electroweak corrections for squark production. As discussed earlier, we use the electroweak correction for squark production in the toy theory, with the gauge coupling constant set equal to $\alpha_2$ of the standard model. This gives an indication of the size of electroweak corrections to squark production in supersymmetric extensions of the standard model. A more precise computation depends on the specific scenario. The electroweak correction to squark production is shown in Fig.~\ref{fig:squarks} for
\begin{figure}
\begin{center}
\includegraphics[width=8cm]{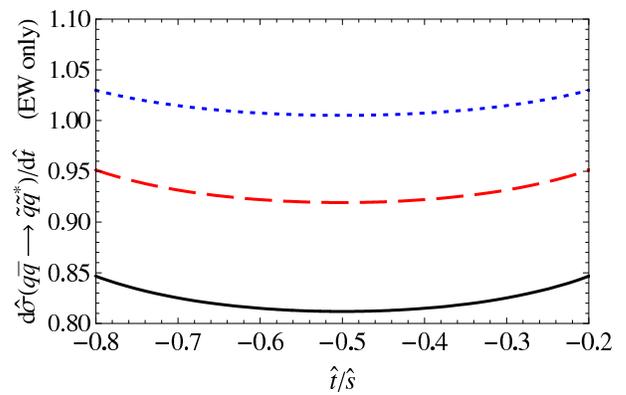}
\end{center}
\caption{\label{fig:angsquarks} Electroweak corrections to $ q \bar q \to \widetilde q \widetilde q^*$ in the toy theory at $\sqrt{\hat s} = 1$~TeV as a function of $-\hat t/\hat s$ at $\sqrt{\hat s}=1$~TeV (black), $\sqrt{\hat s}=2.5$~TeV (long-dashed red) and $\sqrt{\hat s}=5$~TeV (dotted blue).}
\end{figure}
a squark mass of 250~GeV. The radiative correction to the angular distribution is much smaller than for fermions. For discovering squarks, the only correction which matters is that at threshold, $\sqrt{\hat s}=2m_{\tilde q}$ since the parton luminosity falls steeply with $\hat s$. The electroweak corrections give a small (5\%) enhancement of the rate for $\hat s$ near threshold.

\section{Conclusions}

This paper extends the analysis of two previous publications~\cite{cgkm1,cgkm2}, and gives detailed numerical results for radiative corrections to high energy scattering processes in the standard model. The electroweak and QCD corrections have been computed using EFT methods, and the Sudakov logarithms have been summed using renormalization group methods.  The EFT also properly sums mixed higher order logarithms that depend on both $\alpha_s$ and $\alpha_{1,2}$, as well as those that depend on the top-quark Yukawa coupling. We have checked that our results agree with previous results when expanded in powers of $\alpha$.

The electroweak corrections can be important for  LHC processes, particularly in searches for new physics that look for deviations from the standard model. The corrections vary in size from about ($0.4$)--($-14$)\% at 1~TeV to about ($-13$)--($-32$)\% at 5~TeV, and need to be included to obtain LHC cross-sections with accuracies under 10\%.

We have also shown that the radiative corrections to four-quark operators are given in terms of those for two-quark operators by summing over pairs of particles. The relation between this and factorization, and with the two loop soft anomalous dimension of Aybat et al.~\cite{aybat} was discussed in Sec.~\ref{sec:fact}. Further work on this important topic is in progress.

We would like to thank F.~Golf for his contributions during an early stage of this work. We would also like to thank A.H.~Hoang, F.~W\"urthwein and A.~Yagil for helpful discussions. RK was supported by a fellowship under the LHC Theory Initiative of the National Science Foundation.

\begin{appendix}

\section{Matching at $Q$ including the case of  identical particles}
\label{app:box}

In this appendix, we summarize the matching computation at scale $Q$, including the case of identical particles. We start with the case of an $SU(N)$ gauge theory with  left-handed fermions in the fundamental representation.

\begin{figure}[tbh]
\begin{center}
\includegraphics[bb=206 285 406 519,width=2.5cm]{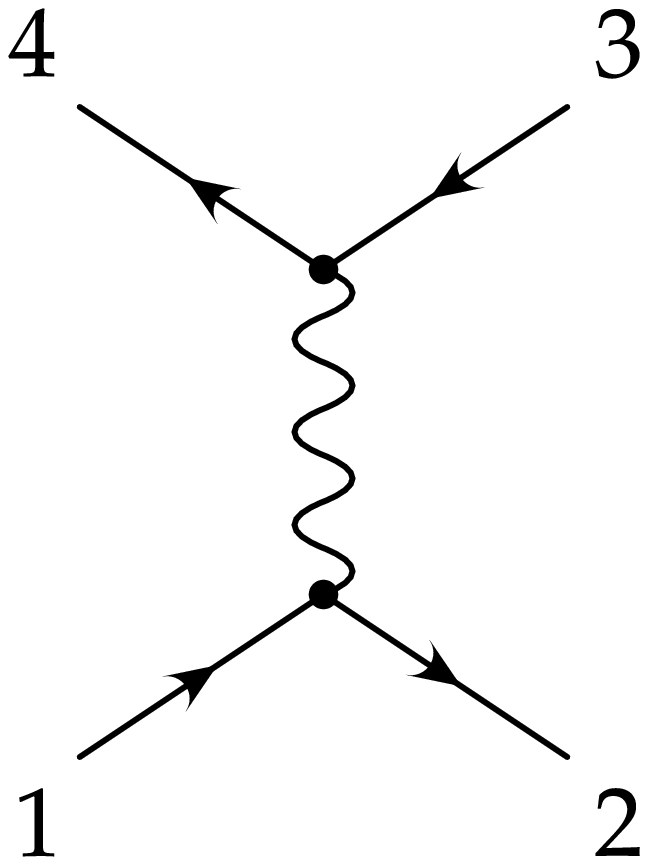}
\end{center}
\caption{\label{fig:app1}}
\end{figure}

The tree-graph in Fig.~\ref{fig:app1} gives
\begin{eqnarray}
A^{(s)} &=& \frac{4\pi \alpha}{s}\ \left[\bar u_4 \gamma^\mu T^a v_3\right]_L  \left[\bar v_2 \gamma^\mu T^a u_1\right]_L
\end{eqnarray}
which is written as
\begin{eqnarray}
A^{(s)} &=&\frac{4\pi \alpha}{s}\ A_{LL} \left(T^a \otimes T^a\right) \nn
A_{LL} &=&  \left[\bar u_4 \gamma^\mu v_3\right]_L  \left[\bar v_2 \gamma^\mu  u_1\right]_L
\label{A1}
\end{eqnarray}
factoring out the color structure from the spinor structure of the graph. The left $T^a$ is contracted with the color indices of particles 4 and 3, and the right $T^a$ with those of particles 2 and 1.

\begin{figure}[tbh]
\begin{center}
\includegraphics[bb=206 285 406 519,width=2.5cm]{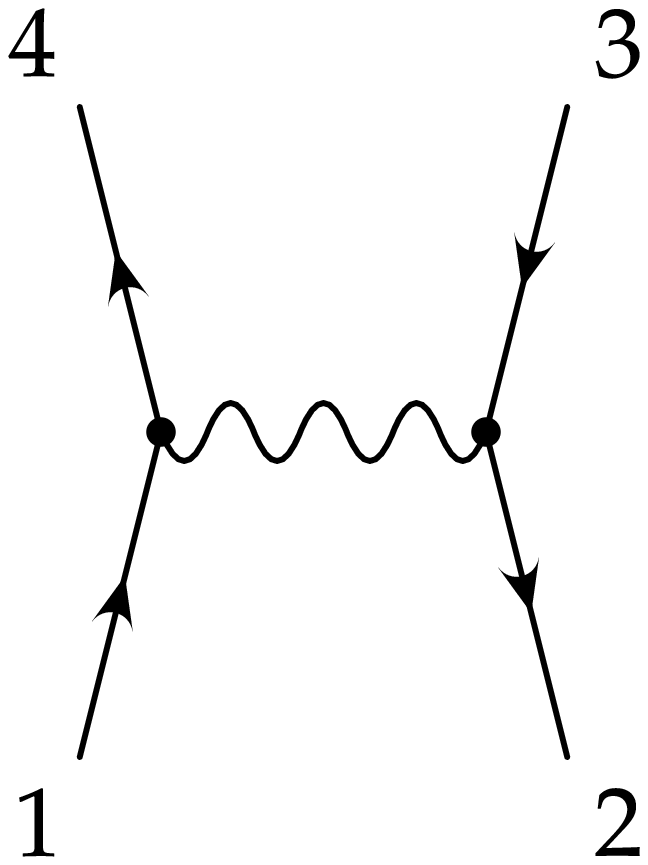}
\end{center}
\caption{\label{fig:app2}}
\end{figure}

If the initial and final particles are identical, there is also the $t$-channel graph in Fig.~\ref{fig:app2} which gives
\begin{eqnarray}
A^{(t)} &=&- \frac{4\pi \alpha}{t}\ \left[\bar u_4 \gamma^\mu T^a u_1\right]_L  \left[\bar v_2 \gamma^\mu T^a v_3\right]_L\nn
&=&- \frac{4\pi \alpha}{t}\ \left[\bar u_4 \gamma^\mu u_1\right]_L  \left[\bar v_2 \gamma^\mu  v_3\right]_L \left(T^a \otimes T^a\right)_c
\label{A2}
\end{eqnarray}
and the relative minus sign is from Wick's theorem. The subscript $c$ indicates that the color structure is in the crossed $t$-channel. The left $T^a$ is contracted with the color indices of particles 4 and 1, and the right $T^a$ with those of particles 2 and 3.

It is convenient to convert the $t$-channel graph to the standard basis used in the paper. The $t$-channel color structure can be converted to the $s$-channel using the $SU(N)$ color crossing matrix
\begin{eqnarray}
&& \left[ \begin{array}{cc} \left(T^a \otimes T^a\right)_c  & 
 \left(\mathbf{1} \otimes \mathbf{1}\right)_c \end{array}\right]= \left[ \begin{array}{cc} \left(T^a \otimes T^a\right) &
 \left(\mathbf{1} \otimes \mathbf{1}\right) \end{array}\right] M_N.\nn
 \label{A3}
\end{eqnarray}
The color Fierz identity
\begin{eqnarray}
(T^a)^i{}_j\, (T^a)^k{}_l &=& \frac 12 \delta^i_l \delta^k_j -\frac{1}{2N}\delta^i_j \delta^k_l
\label{A4}
\end{eqnarray}
can be written as
\begin{eqnarray}
\left(T^a \otimes T^a\right) &=& \frac12 \left(\mathbf{1} \otimes \mathbf{1}\right)_c- \frac1{2N} \left(\mathbf{1} \otimes \mathbf{1}\right)\,.
\end{eqnarray}
Using this and the same equation with direct and crossed channels exchanged, one finds
\begin{eqnarray}
M_N &=& \left[ \begin{array}{cc} 
-\frac1N & 2\\[5pt]
 \frac12-\frac1{2N^2}  & \frac{1}{N}
\end{array}\right]
\label{A5}
\end{eqnarray}
with $M_N^2=\mathbf{1}$.
There is no color crossing matrix required for a $U(1)$ gauge theory. For $SU(2)$ and $SU(3)$,
\begin{eqnarray}
M_2 = \left[ \begin{array}{cc} 
-\frac12 & 2\\[5pt]
 \frac38  & \frac{1}{2}
\end{array}\right] \qquad
M_3 = \left[ \begin{array}{cc} 
-\frac13 & 2\\[5pt]
 \frac49  & \frac{1}{3}
\end{array}\right]\,.
\label{A6}
\end{eqnarray}

The spinor Fierz is
\begin{eqnarray}
\left[\bar u_4 \gamma^\mu u_1\right]_L \left[ \bar v_2 \gamma^\mu  v_3\right]_L
&=& - \left[\bar u_4 \gamma^\mu v_3\right]_L  \left[\bar v_2 \gamma^\mu  u_1\right]_L
\label{A7}
\end{eqnarray}
so Eq.~(\ref{A2}) is
\begin{eqnarray}
A &=& \frac{4\pi \alpha}{t}\ A_{LL} \nn
&&\left[-\frac1 N T^a \otimes T^a+
\left(\frac12-\frac1{2N^2} \right)
\mathbf{1}\otimes \mathbf{1} \right]\,.
\label{A8}
\end{eqnarray}

Comparing with Eq.~(\ref{16}) in Sec.~\ref{sec:toy theory}, we see that the $s$-channel contribution to the matching coefficient is
\begin{eqnarray}
C_{1LL}^{(s)}(s,t) &=& \frac{4 \pi \alpha}{s}\nn
C_{2LL}^{(s)}(s,t) &=& 0
\end{eqnarray}
and the $t$-channel contribution is
\begin{eqnarray}
\left[ \begin{array}{c} 
C_{1LL}^{(t)}(s,t)\\[5pt]
C_{2LL}^{(t)}(s,t)
\end{array}\right]
&=&  M_N  \left[ \begin{array}{c} 
C_{1LL}^{(s)}(t,s)\\[5pt]
C_{2LL}^{(s)}(t,s)
\end{array}\right].
\label{A9}
\end{eqnarray}
The total contribution is
\begin{eqnarray}
C_{1LL} &=& C_{1LL}^{(s)}(s,t) +  C_{1LL}^{(t)}(s,t)\nn
&=& \frac{4\pi \alpha}{s} - \frac{4 \pi \alpha}{N t}\nn
C_{2LL} &=& C_{2LL}^{(s)}(s,t) + C_{2LL}^{(t)}(s,t)\nn
&=& \left(\frac12-\frac1{2N^2} \right) \frac{4\pi \alpha}{t}
\end{eqnarray}
where the $t$-channel pieces should only be included for identical particles.

This sets up the notation and procedure to be used for the one-loop matching computation. The full theory diagrams of Fig.~\ref{FgA1} were computed in order to match the full gauge theory onto SCET at $\mu = Q$.  Dimensional regularization was used to regulate both the infrared and ultra-violet divergences, which are distinguished by subscripts on $1/\epsilon$.  The diagrams are computed with all masses set to zero.
The logarithms are expressed using the short hand notation 
\begin{equation}
\LL_{x} = \log\frac{-x}{\mu^2},\qquad \LL_{x/y} = \log \frac{x}{y}
\end{equation}
for $x,y = s,t,u$. 

The first two vertex graphs of Fig.~\ref{FgA1} each give a contribution of 
 \begin{eqnarray}
V_{v}     &&=  \frac{\alpha^2}{s} A_{LL} \biggl( C_F - \frac{1}{2}C_A \biggr)  \biggl[ \frac{1}{\eUV}-\frac{2}{\eIR^2}-\frac{4}{\eIR}\nn
&&+\frac{2}{\eIR}\LL_{s}-\LL_{s}^2 + 3\LL_{s}-8+\frac{\pi^2}{6} \biggr] . 
\end{eqnarray}
The next two vertex graphs in Fig.~\ref{FgA1} each involve a triple gauge boson coupling, and give
\begin{eqnarray}
V_{g} & =&   \frac{\alpha^2}{s} A_{LL}  \frac{C_A}{2}   \left[  \frac{3}{\e_{\rm UV}}-\frac{4}{\e_{\rm IR}} - 2 + \LL_{s} \right]  .
\end{eqnarray}

\begin{figure}[htb]
\begin{center}
\includegraphics[bb=138 234 474 570,width=4cm]{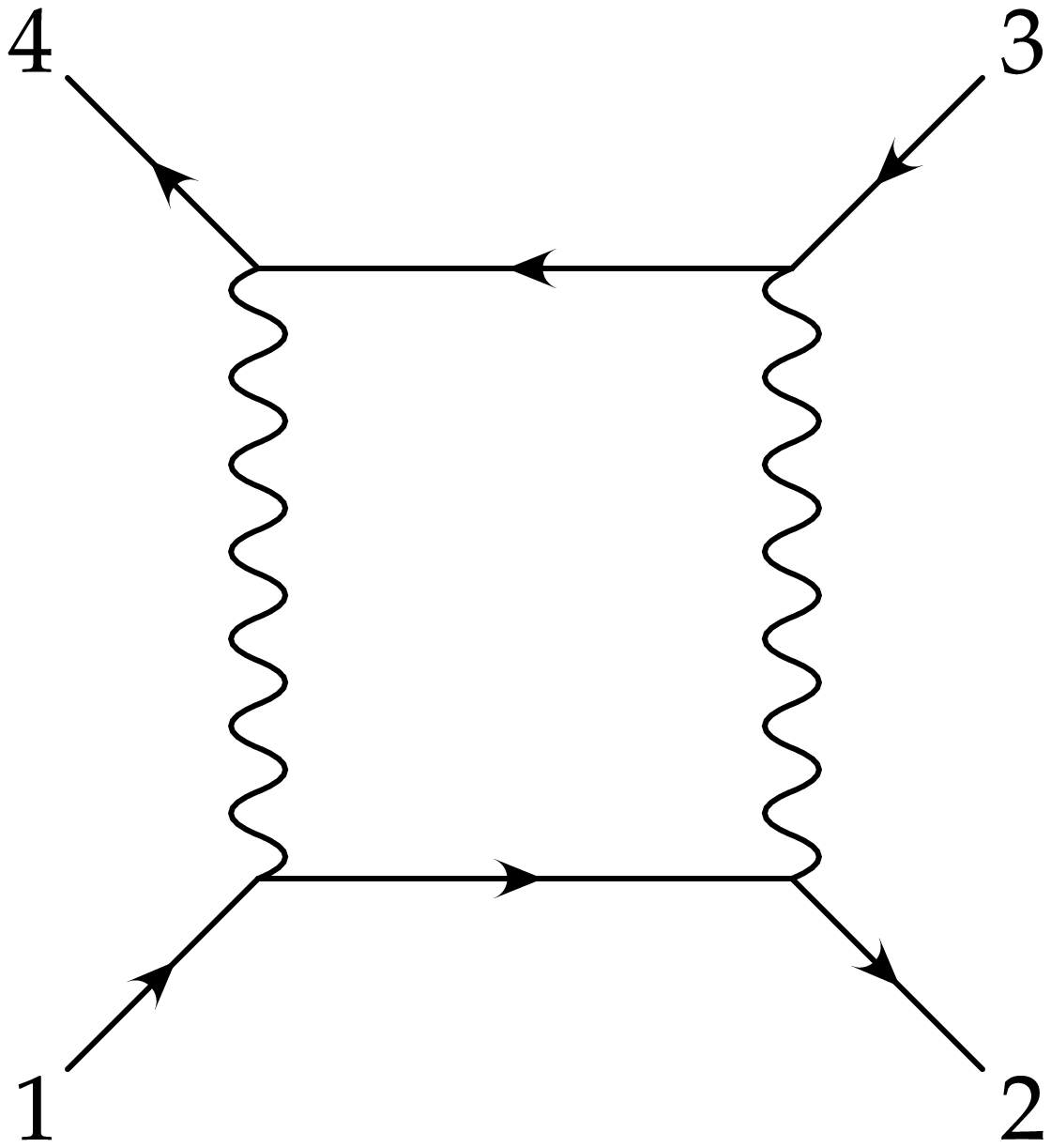}
\end{center}
\caption{\label{fig:app3}}
\end{figure}

The $s$-channel box graph in Fig.~(\ref{fig:app3}) with all fermions left-handed gives
\begin{eqnarray}
V_b &=& \alpha^2 I_2(s,t) A_{LL}\left[C_1\ 1 \otimes 1 +\frac14\left(C_d+C_A\right)T^a \otimes T^a \right]\nn
 &=& \alpha^2 I_2(s,t) A_{LL}\left[\frac{N^2-1}{4N^2} 1 \otimes 1 + \frac{N^2-2}{2N}T^a \otimes T^a \right]\nn
\end{eqnarray}
where
\begin{eqnarray}
I_2(s,t)  &=& I_1(s,t)-\frac{1}{s} f(s,t) \nn
I_1(s,t) &=& \frac{4}{s} \biggl( -\frac{1}{\e_{\rm IR}^2}+\frac{1}{\e_{\rm IR}}\Lt-\frac{1}{2}\Lt^2+\frac{\pi^2}{12}\biggr)+\frac{2}{s}\biggl( \LL_{s/t}^2+\pi^2 \biggr) 
\nn
 f(s,t) 
  &=& \frac{s(s+2t)}{(s+t)^2} \biggl(  \LL_{t/s}^2+\pi^2  \biggr) - \frac{2s}{s+t} \LL_{t/s} 
\end{eqnarray}

\begin{figure}[htb]
\begin{center}
\includegraphics[bb=138 234 474 570,width=4cm]{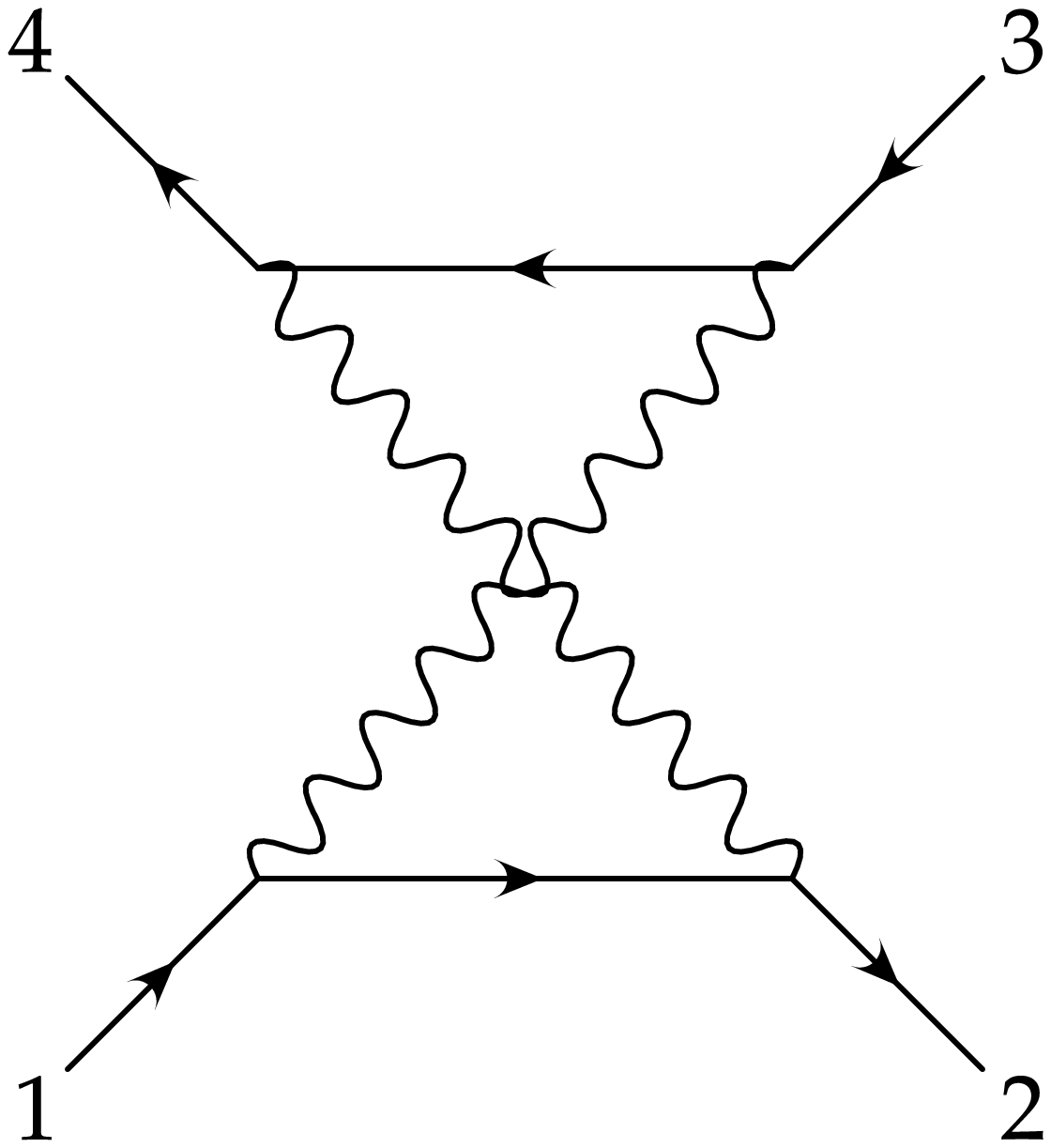}
\end{center}
\caption{\label{fig:4}}
\end{figure}

The $s$-channel crossed-box graph in Fig.~(\ref{fig:4}) with all fermions left-handed gives
\begin{eqnarray}
V_c &=& -\alpha^2 I_1(s,u) A_{LL} \left[C_1\ 1 \otimes 1 +\frac14\left(C_d-C_A\right)T^a \otimes T^a \right]\nn
&=& -\alpha^2 I_1(s,u) A_{LL} \left[\frac{N^2-1}{4N^2} 1 \otimes 1  -\frac{1}{N}T^a \otimes T^a \right]\,.\nn
\end{eqnarray}

The gauge boson self-energy graphs combine to give a contribution of
\begin{eqnarray}
V_{s} 
&=& \frac{\alpha^2}{s}A_{LL}  \biggl\{  C_A  \biggl[  \frac{5}{3\e_{\rm UV}} + \frac{31}{9} - \frac{5}{3} \Ls  \biggr] \nn
&& +  T_F n_F  \biggl[ -\frac{4}{3\e_{\rm UV}} - \frac{20}{9} + \frac{4}{3} \Ls \biggr]  \nn
&&+ T_F n_S  \biggl[ -\frac{1}{3\e_{\rm UV}} - \frac{8}{9} + \frac{1}{3} \Ls \biggr]  \biggr\}
\end{eqnarray}
and the wavefunction graph is 
\begin{equation}
V_w = \frac{\alpha^2}{s}  C_F A_{LL} \biggl[ -\frac{1}{2\e_{\rm UV}} +\frac{1}{2\e_{\rm IR}} \biggr] .
\end{equation}
The sum of all of the diagrams of Fig~\ref{FgA1} including the gauge boson self-energy graphs and the wavefunction graphs is
\begin{eqnarray}
\label{sum}
V_{\rm total} &=& 2V_{v}+2V_{g}+V_{b}+V_{c}+ V_{s}+4V_{w} 
\end{eqnarray}
and gives
\begin{eqnarray}
A_{1LL} &=&  \frac{\alpha^2}{s}  \Biggl\{ \nn
&&2C_F  \biggl[ -\frac{2}{\e_{\rm IR}^2}-\frac{3}{\e_{\rm IR}}+\frac{2}{\e_{\rm IR}}\LL_{s}-\LL_{s}^2 + 3\LL_{s}-8+\frac{\pi^2}{6} \biggl]    \nn
&&  + C_A  \biggl[\frac{2}{\e_{\rm IR}}\LL_{u/s} +  2 \Ls^2-2 \Lu\Ls -\frac{11}{3}\Ls+\pi ^2+\frac{85}{9} \biggr]\nn
&&+  T_F n_F  \biggl[ - \frac{20}{9} + \frac{4}{3} \Ls \biggr]   +T_F n_S  \biggl[ - \frac{8}{9} + \frac{1}{3} \Ls \biggr]    \nn
&&+\biggl[ \frac{4}{\e_{\rm IR}} \LL_{t/u} - 4\LL_{s} \LL_{t/u}  -  f(s,t) \biggr] \frac{(C_d+C_A)}{4}      \Biggr\} \nn
A_{2LL} &=&  \frac{\alpha^2}{s}  \biggl[ \frac{4}{\e_{\rm IR}} \LL_{t/u} - 4\LL_{s} \LL_{t/u}-  f(s,t) \biggr] C_1
\label{A23}
\end{eqnarray}
which are the coefficients of $A_{LL} (T^a \otimes T^a)$ and $A_{LL} (\mathbf{1} \otimes \mathbf{1})$, respectively.

The counterterms of the full theory have been used to cancel to the $1/\e_{\rm UV}$ terms and the remaining poles are all ${1}/{\e_{\rm IR}}$ infrared divergences.  These infrared divergent terms agree with the ultraviolet divergences in the effective theory. The finite parts of  Eq.~(\ref{A23}) give the high scale matching condition at $\mu \sim Q$.

Equation~(\ref{A23}) gives the one-loop matching result for fermions which are distinguishable. If the fermions are identical, then there are also one-loop graphs in the crossed channel, analgous to the crossed channel tree graph Fig.~\ref{fig:app2}. They are obtained by the crossing relation Eq.~(\ref{A9}) used earlier for the tree-level graphs.

The one-loop matching conditions for initial and final fermions of the same chirality (i.e.\ $LL$ or $RR$) is Eq.~(\ref{A23}). If the fermions have opposite chirality, then one can obtain the matching coefficients using charge conjugation. The right handed field $\psi_R$ is replaced by the charge conjugate field $\psi_L^c$. This reverses the sign of the fermion arrow on the fermion line, and exchanges the box and crossed box graphs. One can now use Eq.~(\ref{A23}) for the same-chirality case, and then use charge conjugation on the final amplitude to rewrite the $\psi_L^c$ spinors in terms of the original $\psi_R$ spinors. The result of this procedure is that the matching Eq.~(\ref{A23}) for opposite chirality (i.e.\ $LL$ or $RR$) is given by Eq.~(\ref{A23}) with the replacement $C_1 \to -C_1$, $C_d \to -C_d$ and $t \leftrightarrow u$.

\section{Parameter Integrals}
\label{app:integrals}

The parameter integrals tablulated below arise from vertex and wavefunction graphs where the gauge boson has mass $M$, the external particle has mass $m_{\rm ext}$, and the internal particle has mass $m$. They depend on the variables $w=m_{\rm ext}^2/M^2$ and $z=m^{2}/M^2$. For any function $f(w,z)$ defined below, we define the corresponding function of a single argument by $f(z) \equiv f(z,z)$. In the standard model, where the only fermion with mass comparable to the gauge boson masses is the top quark, we need the integrals $f(z,z)$, $f(z,0)$ and $f(0,z)$, with $z=m_t^2/M_W^2, m_t^2/M_Z^2, m_t^2/M_H^2$.

For $4z \ge 1$, the $f(z,z)$ results can be analytically continued using $\sqrt{1-4 z} \to i\sqrt{4 z-1}$ and $\tanh^{-1}(\sqrt{1-4 z}) \to i\tan^{-1}(\sqrt{4 z-1})$.  In each integral, the factors of $i$ cancel,  and the function remains real. The $f(w,0)$ formul\ae\ are given by using $f(w+i 0^+,0)$ for $w \ge 1$. They have an imaginary part for large values of $w$.

\subsection{Fermions}

The gauge boson vertex graph leads to the integral
\begin{eqnarray}
f_F(w,z) &=& 2 \int_0^{1} {\rd x }\frac{1-x}{x}\log\left(\frac{1-x+z x - w x(1-x)}{1-x}\right) \,.\nn
\label{1}
\end{eqnarray}

\begin{eqnarray}
f_F(z,z) 
&=& 2+\left(\frac{1}{z}-2\right) \log(z) +\frac12 \log^2 (z)\nn
&& + \frac{2\sqrt{1-4z}}{z} \tanh^{-1}\sqrt{1-4z}\nn
&& - 2\left(\tanh^{-1} \sqrt{1-4z}\right)^2\,.
\label{2}
\end{eqnarray}
The function $f_F(z)=f_F(z,z)$ was used in \ptwo.

\begin{eqnarray}
f_F(0,z) &=& \frac{\pi^2}{3}+\frac{2z}{1-z}\log z - 2\, \li{1-z}
\label{1b}
\end{eqnarray}

\begin{eqnarray}
f_F(w,0) &=& 2 + 2 \frac{1-w}{w} \log(1-w)-2\, \li{w}
\label{1c}
\end{eqnarray}

The inverse propagator including the gauge boson wavefunction graph is
\begin{eqnarray}
S^{-1} &=& \slashed{p} \left[1+A(p^2)\right]- m_{\text{ext}}\left[1+ B(p^2)\right].
\end{eqnarray}
The parameter integrals required are
\begin{eqnarray}
a(p^2/M^2,m^2/M^2) &=& A(p^2/M^2,m^2/M^2) \nn
b(p^2/M^2,m^2/M^2) &=& p^2 \frac{\partial B}{\partial p^2}(p^2/M^2,m^2/M^2)\nn
c(p^2/M^2,m^2/M^2) &=&p^2\frac{\partial A}{\partial p^2}(p^2/M^2,m^2/M^2)
\end{eqnarray}
where $m$ is the mass of the internal fermion, and the integrals are evaluated on-shell, with $p^2=m^2_{\text{ext}}$,  where $m_{\text{ext}}$ is the mass of the external fermion.
\begin{eqnarray}
a(w,z) &=& - 2\int_0^1{\rm d}x (1-x) \log\left(\frac{1-x+z x - w x (1-x) }{1-x}\right)\nn
b(w,z) &=&  \int_0^1{\rm d}x\frac{4 \sqrt{wz}\, x(1-x)}{1-x+z x - w x (1-x)}\nn
c(w,z) &=&  \int_0^1{\rm d}x\frac{2wx (1-x)^2}{1-x+z x - w x (1-x) }
\label{3c.6}
\end{eqnarray}

\begin{eqnarray}
a(z,z) &=& \frac52-\frac{1}{z}-\frac{(1-2z)(1-4z)}{z^2\sqrt{1-4z}} \tanh^{-1}\sqrt{1-4z}\nn
&&-\frac{1-4z+2z^2}{2z^2}\log z\nn
a(0,z) &=& -\frac{z}{(1-z)}-\frac{z^2}{(1-z)^2}\log z\nn
a(w,0) &=&\frac32-\frac1{w}-\frac{(1-w)^2}{w^2} \log(1-w)
\end{eqnarray}

\begin{eqnarray}
b(z,z) &=&-4+\frac{4(3z-1)}{z\sqrt{1-4z}} \tanh^{-1}\sqrt{1-4z}\nn
&&+2\left(1-\frac1{z}\right)\log z\nn
b(0,z) &=& 0\nn
b(w,0) &=& 0
\end{eqnarray}

\begin{eqnarray}
c(z,z) &=&\frac2{z}-3+\frac{2(1-5z+5z^2)}{z^2\sqrt{1-4z}} \tanh^{-1}\sqrt{1-4z}\nn
&&+\frac{1-3z+z^2}{z^2}\log z\nn
c(0,z) &=& 0\nn
c(w,0) &=& \frac{2}{w}-1+\frac{2(1-w)}{w^2}\log(1-w)
\end{eqnarray}

The function
\begin{eqnarray}
h_F(z) &=& a(z,z)-2b(z,z)+2c(z,z)
\end{eqnarray}
was used in \ptwo\, and is the wavefunction correction in a vector-like theory.

The corresponding functions for radiative corrections due to a virtual scalar are
\begin{eqnarray}
\tilde a(w,z) &=& \frac 1 2 a(w,z)\nn
\tilde b(w,z) &=& -\frac 1 4 b(w,z)\nn
\tilde c(w,z) &=& \frac 1 2 c(w,z)
\end{eqnarray}
and
\begin{eqnarray}
\tilde h_F(z) &=& \tilde a(z,z)-2\tilde b(z,z)+2 \tilde c(z,z)
\end{eqnarray}
was used in \ptwo.

\subsection{Scalars}

The gauge boson vertex graph for scalar particles leads to the integral
\begin{eqnarray}
f_S(w,z) &=& \int_0^{1} {\rd x }  \frac{(2-x)}{x}  \log\frac{ 1-x+z x - w x(1-x)}{1-x}\,. \nn
\label{7s} 
\end{eqnarray}

\begin{eqnarray}
f_S(z,z) 
&=&1 -\left( 1-\frac{1}{2z} \right) \log(z)+\frac12 \log^2 z \nn
&& +\frac{\sqrt{1-4z} }{z} \tanh^{-1}(\sqrt{1-4z})  \nn
&& - 2\left(\tanh^{-1} \sqrt{1-4z}\right)^2 \nn
f_S(0,z) &=& \frac{\pi^2}{3}+\frac{z}{1-z} \log z - 2\,\li{1-z}\nn
f_S(w,0) &=& 1 + \frac{1-w}{w}\log(1-w)-2\,\li{w}\,.\nn
\label{8}
\end{eqnarray}

Scalar wavefunction renormalization due to gauge boson exchange gives the integral\begin{eqnarray}
h_S(w,z) &=&  \int_0^1{\rm d}x \Biggl\{(3x^2-6x+4)\nn
&&\times \log\left(\frac{1-x+z x - w x(1-x)}{1-x}\right) \nn
&&-\frac{w x(1-x)(2-x)^2}{1-x+z x - w x(1-x)}  \Biggr\} 
\label{9}
\end{eqnarray}

\begin{eqnarray}
h_S(z,z) &=& \frac{3}{2}-\frac{1}{z} + \left[ \frac{3}{2 z}-\frac{1}{2 z^2} \right] \log(z)\nn
&&- \frac{\sqrt{1-4 z} \left(1-z\right) }{z^2} \tanh^{-1}(\sqrt{1-4 z})\nn
h_S(0,z) &=& \frac{z(1-3z)}{2(1-z)^2}-\frac{z(2z^2-2z+1)}{(1-z)^3}\log z\nn
h_S(w,0) &=& -\frac12-\frac1w+\left(2-\frac1{w^2}\right)\log(1-w)
\label{10}
\end{eqnarray}

Scalar wavefunction renormalization due to scalar exchange gives:
\begin{eqnarray}
\tilde h_S(w,z) &=& = - \int_0^1{\rm d}x \frac{z x^3}{1-x+z x - w x(1-x)} \nn
\label{11s}
\end{eqnarray}

\begin{eqnarray}
\tilde h_S(z,z) &=& -\frac{1}{2}-\frac{1}{z}  + \left[\frac{1}{2 z}-\frac{1}{2z^2}\right] \log(z)\nn 
&& + \frac{3z-1}{z^2\sqrt{1-4 z}}  \tanh^{-1}(\sqrt{1-4z}) \nn
\tilde h_S(0,z) &=& \frac{z(2z^2-7z+11)}{6(1-z)^3}+\frac{z}{(1-z)^4}\log z\nn
\tilde h_S(w,0) &=& 0\,.
\label{12}
\end{eqnarray}

\section{Erratum}
\label{app:erratum}

The low-scale matching for the $t$-quark in \ptwo\ is incorrect. The corrected expressions are

\begin{widetext}

\begin{eqnarray}
[\bar \xi_{n,p_2}^{(Q_t)} W_n] \gamma^\mu P_L [W^\dagger_{\bar n} \xi^{(Q_t)}_{\bar n,p_1}]  &\to&  a_1\bar t_{v_2} \gamma^\mu P_L t_{v_1} + a_2 [\bar \xi_{n,p_2}^{(b^\prime)} W_n] \gamma^\mu P_L [W^\dagger_{\bar n} \xi^{(b^\prime)}_{\bar n,p_1}] \, ,\nn
{}[\bar \xi_{n,p_2}^{(t)} W_n] \gamma^\mu P_R [W^\dagger_{\bar n} \xi^{(t)}_{\bar n,p_1}]  &\to&  a_3 \bar t_{v_2} \gamma^\mu P_R t_{v_1} \, ,
\label{t86}
\end{eqnarray}
where the matching coefficients $a_{1-3}$ are given by
\begin{eqnarray}
\log a_1(m_t)&=&
\frac{\aem}{4\pi \sin^2 \theta_W \cos^2 \theta_W} \bigl[g_{Lt}^2 F_g(Q,M_Z,m_t)
+2 U_1\bigr] +\frac{\aem}{4\pi \sin^2 \theta_W}\left(\frac12\right) \left[F_g(Q,M_W,m_t)+2W_1 \right]\nn
&&+ \left(\frac{\alpha_s}{4\pi}\frac{4}{3}+\frac{\aem}{4\pi}\frac{4}{9}\right)\left(\frac{\pi^2}{6}+4\right)+2 H(t_L),\nn
\log a_2(m_t)&=&
\frac{\aem}{4\pi \sin^2 \theta_W \cos^2 \theta_W} g_{Lb}^2 F_g(Q,M_Z,m_t)+\frac{\aem}{4\pi \sin^2 \theta_W}\left(\frac12\right)\left[F_g(Q,M_W,m_t)+2W_2\right] +2H(b^\prime_L) , \nn
\log a_3(m_t)&=&
\frac{\aem}{4\pi \sin^2 \theta_W \cos^2 \theta_W} \bigl[g_{Rt}^2 F_g(Q,M_Z,m_t)
+2U_2\bigr] +\frac{\aem}{4\pi \sin^2 \theta_W}\left(\frac12\right) \left[-c(w_t,0 )\right]\nn
&&+  \left(\frac{\alpha_s}{4\pi}\frac{4}{3}+\frac{\aem}{4\pi}\frac{4}{9}\right)\left(\frac{\pi^2}{6}+4\right)+2H(t_R) ,
\end{eqnarray}
and the required functions $U_{1,2}$, $X_{1,2}$, $H(t_L)$, $H(t_R)$, and $H(b^\prime_L)$ are given in Eqs.~(\ref{90}), (\ref{110}), (\ref{108}).

\end{widetext}

\end{appendix}


\begin{thebibliography}{99}

\bibitem{cgkm1}
  J.-Y.~Chiu, F.~Golf, R.~Kelley and A.~V.~Manohar,
  Phys.\ Rev.\ Lett.\  {\bf 100}, 021802 (2008).
  
\bibitem{cgkm2}
  J.-Y.~Chiu, F.~Golf, R.~Kelley and A.~V.~Manohar,
  Phys.\ Rev.\  D {\bf 77}, 053004 (2008).

  
  
\bibitem{ccc}
  M.~Ciafaloni, P.~Ciafaloni and D.~Comelli,
  Phys.\ Rev.\ Lett.\  {\bf 84}, 4810 (2000).


\bibitem{ciafaloni}
  P.~Ciafaloni and D.~Comelli,
  Phys.\ Lett.\  B {\bf 446}, 278 (1999);
  Phys.\ Lett.\  B {\bf 476}, 49 (2000).


\bibitem{fadin}
  V.~S.~Fadin, L.~N.~Lipatov, A.~D.~Martin and M.~Melles,
  Phys.\ Rev.\  D {\bf 61}, 094002 (2000).
  
\bibitem{kps}
  J.~H.~Kuhn, A.~A.~Penin and V.~A.~Smirnov,
  Eur.\ Phys.\ J.\  C {\bf 17}, 97 (2000).
  
  
\bibitem{fkps}
  B.~Feucht, J.~H.~Kuhn, A.~A.~Penin and V.~A.~Smirnov,
  Phys.\ Rev.\ Lett.\  {\bf 93}, 101802 (2004).
  
  
\bibitem{jkps}
  B.~Jantzen, J.~H.~Kuhn, A.~A.~Penin and V.~A.~Smirnov,
  Phys.\ Rev.\  D {\bf 72}, 051301 (2005)
  [Erratum-ibid.\  D {\bf 74}, 019901 (2006)].
  
\bibitem{jkps4}
  B.~Jantzen, J.~H.~Kuhn, A.~A.~Penin and V.~A.~Smirnov,
  Nucl.\ Phys.\  B {\bf 731}, 188 (2005)
  [Erratum-ibid.\  B {\bf 752}, 327 (2006)].
  

\bibitem{beccaria}
  M.~Beccaria, F.~M.~Renard and C.~Verzegnassi,
  Phys.\ Rev.\  D {\bf 63}, 053013 (2001).

\bibitem{dp1}
  A.~Denner and S.~Pozzorini,
  Eur.\ Phys.\ J.\  C {\bf 18}, 461 (2001).

\bibitem{dp2}
  A.~Denner and S.~Pozzorini,
  Eur.\ Phys.\ J.\  C {\bf 21}, 63 (2001).
 
\bibitem{hori}
  M.~Hori, H.~Kawamura and J.~Kodaira,
  Phys.\ Lett.\  B {\bf 491}, 275 (2000).

\bibitem{beenakker}
  W.~Beenakker and A.~Werthenbach,
  Nucl.\ Phys.\  B {\bf 630}, 3 (2002).

\bibitem{dmp}
  A.~Denner, M.~Melles and S.~Pozzorini,
  Nucl.\ Phys.\  B {\bf 662}, 299 (2003).
  
\bibitem{pozzorini}
  S.~Pozzorini,
  Nucl.\ Phys.\  B {\bf 692}, 135 (2004).
  
\bibitem{js}
  B.~Jantzen and V.~A.~Smirnov,
  Eur.\ Phys.\ J.\  C {\bf 47}, 671 (2006).
  
\bibitem{melles}
  M.~Melles,
  Phys.\ Lett.\  B {\bf 495}, 81 (2000);
  Phys.\ Rev.\  D {\bf 63}, 034003 (2001);
  Phys.\ Rept.\  {\bf 375}, 219 (2003).

   \bibitem{pqcd}
 A.~H.~.~Mueller, {\it Perturbative Quantum Chromodynamics},
 World Scientific, Singapore, 1989.

  \bibitem{BFL}
C.~W.~Bauer, S.~Fleming and M.~E.~Luke,
Phys.\ Rev.\ D {\bf 63}, 014006 (2001).

\bibitem{SCET1}
C.~W.~Bauer, S.~Fleming, D.~Pirjol and I.~W.~Stewart,
Phys.\ Rev.\ D {\bf 63}, 114020 (2001);

\bibitem{SCET2}
C.~W.~Bauer and I.~W.~Stewart,
Phys.\ Lett.\ B {\bf 516}, 134 (2001);

\bibitem{SCET3}
C.~W.~Bauer, D.~Pirjol and I.~W.~Stewart,
Phys.\ Rev.\ D {\bf 65}, 054022 (2002).

\bibitem{ira1}
  C.~W.~Bauer, S.~Fleming, D.~Pirjol, I.~Z.~Rothstein and I.~W.~Stewart,
  Phys.\ Rev.\  D {\bf 66}, 014017 (2002).

 \bibitem{bf}
  M.~Beneke and T.~Feldmann,
  Nucl.\ Phys.\  B {\bf 685}, 249 (2004).

\bibitem{analytic}
  V.~A.~Smirnov and E.~R.~Rakhmetov,
  Theor.\ Math.\ Phys.\  {\bf 120}, 870 (1999)
  [Teor.\ Mat.\ Fiz.\  {\bf 120}, 64 (1999)].

\bibitem{2jet}
  C.~W.~Bauer, A.~V.~Manohar and M.~B.~Wise,
  Phys.\ Rev.\ Lett.\  {\bf 91}, 122001 (2003),
  C.~W.~Bauer, C.~Lee, A.~V.~Manohar and M.~B.~Wise,
  Phys.\ Rev.\  D {\bf 70}, 034014 (2004),
  C.~W.~Bauer and M.~D.~Schwartz,
  Phys.\ Rev.\ Lett.\  {\bf 97}, 142001 (2006),
  M.~Trott,
  Phys.\ Rev.\  D {\bf 75}, 054011 (2007).
  
   \bibitem{lmr}
  M.~E.~Luke, A.~V.~Manohar and I.~Z.~Rothstein,
  Phys.\ Rev.\  D {\bf 61}, 074025 (2000).

\bibitem{collins}
  J.~C.~Collins,
  Phys.\ Rev.\  D {\bf 22}, 1478 (1980).

\bibitem{mueller}
  A.~H.~Mueller,
  Phys.\ Rev.\  D {\bf 20}, 2037 (1979).

\bibitem{sen}
  A.~Sen,
  Phys.\ Rev.\  D {\bf 24}, 3281 (1981).

\bibitem{dis}
  A.~V.~Manohar,
  Phys.\ Rev.\  D {\bf 68}, 114019 (2003).

\bibitem{Bauer:2003pi}
  C.~W.~Bauer and A.~V.~Manohar,
  Phys.\ Rev.\  D {\bf 70}, 034024 (2004).

\bibitem{aybat}
  S.~Mert Aybat, L.~J.~Dixon and G.~Sterman,
  Phys.\ Rev.\  D {\bf 74}, 074004 (2006);
  Phys.\ Rev.\ Lett.\  {\bf 97}, 072001 (2006).

\bibitem{hoang2}
A.H.~Hoang et al., unpublished.

\bibitem{zerobin}
  A.~V.~Manohar and I.~W.~Stewart,
  Phys.\ Rev.\  D {\bf 76}, 074002 (2007).

\bibitem{idilbi1}
  A.~Idilbi and T.~Mehen,
  Phys.\ Rev.\  D {\bf 75}, 114017 (2007).

\bibitem{idilbi2}
  A.~Idilbi and T.~Mehen,
  Phys.\ Rev.\  D {\bf 76}, 094015 (2007).

     
\bibitem{collins:fact}
  J.~C.~Collins, D.~E.~Soper and G.~Sterman,
  Adv.\ Ser.\ Direct.\ High Energy Phys.\  {\bf 5}, 1 (1988)

\bibitem{lee}
  C.~Lee and G.~Sterman,
  Phys.\ Rev.\  D {\bf 75}, 014022 (2007)
  [arXiv:hep-ph/0611061].


\bibitem{hqet}
  A.~V.~Manohar,
  Phys.\ Rev.\  D {\bf 56}, 230 (1997).
  
\bibitem{eft}
  A.~V.~Manohar, {\sl Effective Field Theories},  arXiv:hep-ph/9606222.

\bibitem{Bohm}
  M.~Bohm, H.~Spiesberger and W.~Hollik,
  Fortsch.\ Phys.\  {\bf 34}, 687 (1986).

\bibitem{Roth}
  M.~Roth and A.~Denner,
  Nucl.\ Phys.\  B {\bf 479}, 495 (1996)
  [arXiv:hep-ph/9605420].
  
\bibitem{dorsten}
  C.~W.~Bauer, M.~P.~Dorsten and M.~P.~Salem,
  Phys.\ Rev.\  D {\bf 69}, 114011 (2004),
%
  A.~V.~Manohar,
  Phys.\ Lett.\  B {\bf 633}, 729 (2006).
   

\bibitem{Manohar:2000hj}
  A.~V.~Manohar and I.~W.~Stewart,
  Phys.\ Rev.\  D {\bf 62}, 074015 (2000).

\bibitem{ira2}
  I.~Z.~Rothstein,
  Phys.\ Rev.\  D {\bf 70}, 054024 (2004).

\bibitem{llw}
  A.~K.~Leibovich, Z.~Ligeti and M.~B.~Wise,
  Phys.\ Lett.\  B {\bf 564}, 231 (2003).

\bibitem{book}
  A.~V.~Manohar and M.~B.~Wise,
 {\sl Heavy Quark Physics}, Cambridge University Press (Cambridge, 2000).

\bibitem{korchemsky}
  I.~A.~Korchemskaya and G.~P.~Korchemsky,
  Phys.\ Lett.\  B {\bf 287}, 169 (1992).

\bibitem{kaplan}
  D.~B.~Kaplan and A.~Manohar,
  Nucl.\ Phys.\  B {\bf 310}, 527 (1988);
  A.~V.~Manohar,
  Phys.\ Lett.\  B {\bf 242}, 94 (1990).
  
\bibitem{Melles:2000ia}
  M.~Melles,
  Phys.\ Rev.\  D {\bf 64}, 014011 (2001).

\bibitem{kidonakis}
  N.~Kidonakis, G.~Oderda and G.~Sterman,
  Nucl.\ Phys.\  B {\bf 531}, 365 (1998).
  


\bibitem{MVV}
  S.~Moch, J.~A.~M.~Vermaseren and A.~Vogt,
  Nucl.\ Phys.\  B {\bf 688}, 101 (2004).


\bibitem{Nason:1987xz}
  P.~Nason, S.~Dawson and R.~K.~Ellis,
  Nucl.\ Phys.\  B {\bf 303}, 607 (1988).

  
\bibitem{Beenakker:1988bq}
  W.~Beenakker, H.~Kuijf, W.~L.~van Neerven and J.~Smith,
  Phys.\ Rev.\  D {\bf 40}, 54 (1989).
  
\bibitem{Nason:1989zy}
  P.~Nason, S.~Dawson and R.~K.~Ellis,
  Nucl.\ Phys.\  B {\bf 327}, 49 (1989)
  [Erratum-ibid.\  B {\bf 335}, 260 (1990)].
  

\bibitem{Beenakker:1990maa}
  W.~Beenakker, W.~L.~van Neerven, R.~Meng, G.~A.~Schuler and J.~Smith,
  Nucl.\ Phys.\  B {\bf 351}, 507 (1991).
  
\bibitem{Laenen:1991af}
  E.~Laenen, J.~Smith and W.~L.~van Neerven,
  Nucl.\ Phys.\  B {\bf 369}, 543 (1992).
  
\bibitem{Kidonakis:1995wz}
  N.~Kidonakis and J.~Smith,
  Phys.\ Rev.\  D {\bf 51}, 6092 (1995).
  
  
  
\bibitem{Berger:1997gz}
  E.~L.~Berger and H.~Contopanagos,
  Phys.\ Rev.\  D {\bf 57}, 253 (1998).
  
  
  
\bibitem{Cacciari:2003fi}
  M.~Cacciari, S.~Frixione, M.~L.~Mangano, P.~Nason and G.~Ridolfi,
  JHEP {\bf 0404}, 068 (2004).

\bibitem{Kuhn:2006vh}
  J.~H.~Kuhn, A.~Scharf and P.~Uwer,
  Eur.\ Phys.\ J.\  C {\bf 51}, 37 (2007).
  

\end{thebibliography}
\end{document}